\documentclass[a4paper,oneside,11pt]{article}
\pdfoutput=1
\usepackage{cprform}
\usepackage{amsmath,amstext,amsfonts,amsbsy,amssymb,amscd,slashed}
\usepackage{pdflscape,color,svg,epsfig,cite,hyperref,bbm,multirow}
\usepackage{graphicx}
\usepackage{tabularx}
\usepackage{booktabs}
\usepackage{floatpag}
\usepackage{physics}

\usepackage[makeroom]{cancel}
\usepackage{blkarray}
\usepackage{stackengine}

\usepackage{tikz}
\usetikzlibrary{arrows,shapes}
\usetikzlibrary{trees}
\usetikzlibrary{matrix,arrows}                          
\usetikzlibrary{positioning}                            
\usetikzlibrary{calc,through}                           
\usetikzlibrary{decorations.pathreplacing}  
\usetikzlibrary{decorations.pathmorphing}       
\usetikzlibrary{decorations.markings}
\tikzset{
    vector/.style={decorate, decoration={snake}, draw},
        provector/.style={decorate, decoration={snake,amplitude=2.5pt}, draw},
        antivector/.style={decorate, decoration={snake,amplitude=-2.5pt}, draw},
    fermion/.style={draw=black, postaction={decorate},
        decoration={markings,mark=at position .55 with {\arrow[draw=black]{>}}}},
    fermionbar/.style={draw=black, postaction={decorate},
        decoration={markings,mark=at position .55 with {\arrow[draw=black]{<}}}},
    fermionnoarrow/.style={draw=black},
    gluon/.style={decorate, draw=black,
        decoration={coil,amplitude=4pt, segment length=5pt}},                           
    scalar/.style={dashed,draw=black, postaction={decorate},
        decoration={markings,mark=at position .55 with {\arrow[draw=black]{>}}}},
    scalarbar/.style={dashed,draw=black, postaction={decorate},
        decoration={markings,mark=at position .55 with {\arrow[draw=black]{<}}}},
    scalarnoarrow/.style={dashed,draw=black},
    electron/.style={draw=black, postaction={decorate},
        decoration={markings,mark=at position .55 with {\arrow[draw=black]{>}}}},
        bigvector/.style={decorate, decoration={snake,amplitude=4pt}, draw},
}

\newcolumntype{C}{>{$}c<{$}}
\newcolumntype{R}{>{$}r<{$}}
\newcolumntype{L}{>{$}l<{$}}

\usepackage[normalem]{ulem}
\newcommand\xoutpars[1]{\let\helpcmd\xout\parhelp#1\par\relax\relax}
\newcommand\soutpars[1]{\let\helpcmd\sout\parhelp#1\par\relax\relax}
\long\def\parhelp#1\par#2\relax{%
  \helpcmd{#1}\ifx\relax#2\else\par\parhelp#2\relax\fi%
}

\begin{document}


\begin{titlepage}


\vspace*{-30truemm}
\begin{flushright}
CERN-TH-2025-216
\\
MITP-25-069
\\
KEK-TH-2771
\end{flushright}
\vspace{15truemm}


\begin{center}
{
\Large\bf
Precision lattice calculation of the hadronic contribution$\phantom{\Big|}$
\\ to the running of the electroweak gauge couplings}
\end{center}
\vskip 10 true mm

\vskip -2 true mm
\centerline{\elfrm 
Alessandro~Conigli$^{a,b}$\footnote{aconigli@uni-mainz.de}, 
Dalibor Djukanovic$^{a,b}$,
Georg~von~Hippel$^{c}$,
Simon~Kuberski$^{d}$,
}
\centerline{\elfrm 
Harvey~B.\!~Meyer$^{a,c,d}$,
Kohtaroh~Miura$^{e}$,	
Konstantin~Ottnad$^{c}$,
Andreas~Risch$^{f}$,
Hartmut~Wittig$^{a,b,c}$
}
\vskip 4 true mm
\centerline{\tenit $^a$ Helmholtz Institute Mainz, Johannes Gutenberg-Universit\"{a}t Mainz, 55099 Mainz, Germany}
\vskip 3 true mm
\centerline{\tenit $^b$ GSI Helmholtz Centre for Heavy Ion Research, 64291 Darmstadt, Germany}
\vskip 3 true mm
\centerline{\tenit $^c$ PRISMA$^{++}$ Cluster of Excellence and Institut f\"{u}r Kernphysik,}
\centerline{\tenit Johannes Gutenberg-Universit\"{a}t Mainz, 55099 Mainz, Germany}
\vskip 3 true mm
\centerline{\tenit $^d$ Theoretical Physics Department, CERN, 1211 Geneva 23, Switzerland}
\vskip 3 true mm
\centerline{\tenit $^e$ Institute of Particle and Nuclear Studies, }
\centerline{\tenit  High Energy Accelerator Research Organization (KEK), Tsukuba 305-0801, Japan}
\vskip 3 true mm
\centerline{\tenit $^f$ Department of Physics, University of Wuppertal, Gaussstr. 20, 42119 Wuppertal, Germany}
\vskip 15 true mm


\noindent{\textbf{ Abstract:} We present an update of our lattice QCD
  determination of the hadronic contribution to the running of the
  electromagnetic coupling, $\Delta\alpha_{\mathrm{had}}^{(5)}(-Q^2)$,
  and of the electroweak mixing angle in the space-like momentum
  region up to $Q^2=12 \ \mathrm{GeV}^2$. The calculation is based on
  CLS ensembles with $N_f=2+1$ flavours of $O(a)$-improved Wilson
  fermions, covering five lattice spacings between $0.039$ and $0.085$
  fm and a range of pion masses, including the physical point. A
  refined analysis employing a telescopic window strategy allows for a
  clean separation of systematic effects across Euclidean distance
  scales. Statistical precision is further enhanced through low-mode
  averaging, combined with a spectral reconstruction of the
  vector-vector correlator at long distances on the most chiral
  ensembles. We confirm significant tensions of up to $7\sigma$ at
  space-like virtualities around $Q^2= 1\ \mathrm{GeV}^2$ between our
  lattice results for $\Delta\alpha_{\mathrm{had}}^{(5)}(-Q^2)$ and
  the corresponding data-driven estimates based on $e^+e^-$ cross
  section data. Combining our lattice data with perturbative QCD via
  the Euclidean split technique, we obtain at the $Z$-pole
  $\Delta\alpha_{\mathrm{had}}^{(5)}(M_Z^2) =
  0.027821(34)_{\mathrm{lat}}(35)_{\mathrm{pQCD}}$, which is more than
  two times more precise than recent data-driven estimates. Our result
  deviates slightly, by $1-2\sigma$, from the value produced by global electroweak
  fits. For the electroweak mixing angle, we present the hadronic
  contribution to its running and provide a precise determination of
  the octet-singlet mixing component $\bar\Pi^{(0,8)}$, in good
  agreement with phenomenological models but with significantly higher
  precision.  }

\vspace{10truemm}

\eject
\end{titlepage}

\cleardoublepage

\tableofcontents
\newpage
\section{Introduction}

The electromagnetic coupling and the weak mixing angle are fundamental
parameters in the Standard Model (SM) and play a crucial role in the
world-wide effort to detect signals of beyond-Standard Model (BSM)
physics. Specifically, the value of the electromagnetic coupling at
the $Z$-pole, $\alpha(M_Z^2)^{-1}=127.930\pm0.008$
\cite{ParticleDataGroup:2024cfk} is an important input parameter for
interpreting the results from high-energy colliders. However, the
relatively poor precision with which $\alpha(M_Z^2)$ is known limits
the sensitivity of future SM tests. Hence, to exploit the full
potential of future high-luminosity $e^+e^-$ colliders such as the
FCC-ee \cite{FCC:2018evy} or
CEPC\,\cite{CEPCPhysicsStudyGroup:2022uwl}, it is imperative to drastically
reduce the uncertainty of $\alpha(M_Z)$.

A further sensitive test of the SM is provided by the running of the
electroweak mixing angle, $\sin^2\theta_W$. New precision measurements
in parity-violating electron-proton and electron-positron scattering
are currently prepared by the P2 \cite{Becker:2018ggl} and MOLLER
\cite{MOLLER:2014iki} collaborations, respectively. These measurements
will provide benchmark results for $\sin^2\theta_W$ at low energies
which, when combined with the results obtained directly at the
$Z$-pole, can be compared to the energy dependence predicted by the
SM. Any deviation between the observed and expected energy dependence
will be attributed to BSM physics.

The theoretical knowledge of the running of $\alpha$ and
$\sin^2\theta_W$ is limited by the effects of the strong
interaction. In particular, the contributions to the hadronic vacuum
polarization (HVP) have a sizeable influence on the energy dependence
of both quantities.
The traditional method to determine the hadronic contributions to the
running of $\alpha$ and $\sin^2\theta_W$ employs a data-driven
dispersive approach \cite{Keshavarzi:2018mgv, Davier:2019can,
Jeger_yellow_rep, Jegerlehner:2019alphaQEDc19}, similar to the
evaluation of the HVP contribution to the muon anomalous magnetic
moment, $a_\mu^{\mathrm{hvp}}$.
This method relies on experimental data for 
$e^+e^-\to\hbox{hadrons}$ cross sections, which exhibit strong,
yet to be resolved tensions in the crucial $\pi^+\pi^-$ channel.
As a consequence,
the Muon $g-2$ Theory Initiative refrained from quoting an estimate
for $a_\mu^{\mathrm{hvp}}$ from the data-driven method in their second
White Paper
\cite{Aliberti:2025beg}, switching instead to lattice QCD calculations
of this quantity.
In the case of the weak mixing angle, the data-driven approach for
determining the hadronic contributions to the running is complicated
further by the necessity to perform a potentially model-dependent
flavour separation.

Here we report on our calculation of the running of the
electromagnetic coupling and the weak mixing angle, employing lattice
QCD to compute the offset values used as non-perturbative input in the
Euclidean split technique. A crucial new ingredient is the application
of a telescopic window strategy that has allowed us to reach
significantly higher space-like virtualities. This has the advantage
of a substantial reduction of the uncertainties associated with the
running and matching up to the $Z$~pole, for which we adopt an updated
perturbative framework. These advances, combined with increased statistics and
noise-reduction techniques on our extended set of gauge ensembles, 
have allowed us to improve the precision relative to our previous
calculation~\cite{Ce:2022eix} by a factor of three. Our final result
for the running of $\alpha$ is
(see also Eq.~\eqref{eq:dalpha_z_final_result})
\begin{equation}
  \Delta\alpha_{\mathrm{had}}^{(5)}(M_Z^2) =
  0.027\,821(34)_{\mathrm{lat}}(35)_{\mathrm{pQCD}}\,,
\end{equation}
With its total relative error of 0.17\%, it is thus considerably
more precise than recent data-driven evaluations based on $e^+e^-$
hadronic cross sections \cite{Davier:2017zfy, Keshavarzi:2018mgv,
  Davier:2019can, Keshavarzi:2019abf, Jeger_yellow_rep,
  Jegerlehner:2019alphaQEDc19}, which quote total uncertainties that
are about twice as large as ours. Our main findings and conclusions as well as a detailed summary of the prospects for future improvements are presented in the companion letter~\cite{Conigli:2026xxx}.

In an effort to keep this paper self-contained, we 
present the calculational setup in section~\ref{sec:running_coupling}
and describe the methodological improvements compared to our earlier determination~\cite{Ce:2022eix} in detail in section \ref{subsec:comp_strategy}. The
evaluation of the error budget for our calculation is presented in
section~\ref{sec:lat_res}. Finally, in section \ref{sec:had_run} we
report our best estimate for the running of the electromagnetic
coupling and the weak mixing angle. For the sake of facilitating
comparisons with other lattice calculations as well as
phenomenological studies, we also parametrize the dependence of vacuum
polarization functions on the space-like virtualities considered in
our calculation in terms of rational approximants.

\section{Preliminaries} \label{sec:running_coupling}

\subsection{The electromagnetic coupling}
The electromagnetic coupling $\alpha \equiv e^2/(4\pi)$ in the Thomson
limit ($q^2\to 0$) governs interactions at very low energies, well
below the electron mass. Its precisely measured value $\alpha^{-1} =
137.035\, 999 \, 178(8)$, as reported by the Particle Data Group (PDG)
\cite{ParticleDataGroup:2024cfk}, represents one of the most
accurately known quantities in physics.\footnote{One should be aware,
though, that there are significant tensions between the most precise
measurements of $\alpha$ \cite{Parker:2018sci, Morel:2020dww,
  Fan:2022eto} that are yet to be resolved.} However, at higher
energies -- such as at the electroweak scale probed by the $Z$-boson
-- the relevant coupling increases to $\alpha^{(5)}(M_Z)
=1/127.930(8)$ \cite{ParticleDataGroup:2024cfk}, a roughly $7\%$
enhancement. This energy dependence can either be determined
experimentally (see \cite{Riembau:2025ppc} for a recent proposal) or
predicted from its low-energy value using the Renormalization Group
(RG) evolution. In the on-shell renormalization scheme, one introduces
an effective coupling at any time-like momentum transfer $q^2$,
\begin{equation}
	\alpha(q^2) = \frac{\alpha}{1 - \Delta\alpha(q^2)},
\end{equation}
so that the scale dependence of the coupling is fully described by
$\Delta\alpha(q^2)$.  A precise knowledge of the running is a  key
ingredient in precision electroweak tests, and an accurate first-principles determination of this quantity is therefore crucial for ongoing and future collider programs aiming to probe effects of possible new physics through high-precision electroweak observables.

While perturbation theory provides reliable estimates for the leptonic component of  $\Delta\alpha(q^2)$, the hadronic contribution at low energies is dominated by non-perturbative QCD effects. This contribution can be expressed in terms of the subtracted HVP function  $\bar{\Pi}$,
\begin{equation}\label{eq:space_like_dalpha}
	\Delta\alpha_{\mathrm{had}}(q^2) = 4\pi\alpha 	\Re\bar{\Pi}(q^2), \qquad
	\bar{\Pi}(q^2) = \Pi(q^2) - \Pi(0).
\end{equation}
The HVP can be determined phenomenologically via a dispersion relation that connects it to the experimentally measured $R$-ratio, $R(s) = \sigma(e^+e^+\to \mathrm{hadrons}) / \sigma(e^+e^-\to \mu^+\mu^-)$. Integrating  $R(s)$ over the hadronic spectrum  \cite{Erler:2017knj, Proceedings:2019vxr, Davier:2019can, Keshavarzi:2019abf}, and combining it  with our previous lattice determination \cite{Ce:2022eix}, yields  the current world average $\Delta\alpha_{\mathrm{had}}^{(5)}(M_Z^2)=0.02783(6)$ \cite{ParticleDataGroup:2024cfk}.

Alternatively, the hadronic contribution to the running of $\alpha$
can be determined directly from first principles using lattice QCD, thereby avoiding reliance on experimental data. On the lattice, computations are naturally performed at space-like momenta $Q^2=-q^2$, where the HVP function $\Pi(Q^2)$ is obtained from the two-point correlation function  of the  electromagnetic currents  $j_\mu^\gamma$ \cite{Burger:2015lqa,Francis:2015grz, Budapest-Marseille-Wuppertal:2017okr}
\begin{equation}
	j_\mu^\gamma = \frac{2}{3} \bar{u}\gamma_\mu u - \frac{1}{3} \bar{d}\gamma_\mu d -\frac{1}{3}\bar{s}\gamma_\mu s + \frac{2}{3}\bar{c}\gamma_\mu c + \ldots \ ,
\end{equation}
according to
\begin{equation}
	(Q_\mu Q_\nu - \delta_{\mu\nu} Q^2)\Pi(-Q^2) = \Pi_{\mu\nu}^{(\gamma,\gamma)}(Q) = 
	\int\dd^4 x \;e^{iQx} \langle j_\mu^\gamma(x) j_\nu^\gamma(0)\rangle.
\end{equation}
As discussed in \cite{Eidelman:1998vc, Jegerlehner:2008rs, Jegerlehner:1999hg}, the connection between the HVP and $\Delta\alpha_{\mathrm{had}}$  is established through  the Adler function $D(Q)$ \cite{Adler:1974gd}. This quantity  is defined as the derivative of the HVP with respect to $Q^2$, and can equivalently  be expressed as a dispersion integral over the experimentally measured $R$-ratio,
	\begin{equation}
	D(Q^2) = 12\pi^2Q^2
	\dv{\Pi(-Q^2)}{Q^2} = Q^2 \int_{0}^{\infty} \dd{s} \frac{R(s)}{(s+Q^2)^2}.
\end{equation}
Thus, by computing  the correlation function of the  electromagnetic current $j_\mu^\gamma$, one can access $\bar\Pi^{(\gamma,\gamma)}$ and thereby obtain  $\Delta\alpha_{\mathrm{had}}$ directly from QCD.

\subsection{The electroweak mixing angle}
The electroweak mixing angle (or Weinberg angle) $\theta_W$ uniquely
defines the relation between the electromagnetic and weak interactions
in the SM. It relates the $g$ and $g'$ couplings of the $\rm SU(2)_L$
weak isospin and $\rm U(1)_Y$ weak hypercharge interactions, respectively, with the electromagnetic coupling $\alpha = e^2/(4\pi)$ through \cite{ParticleDataGroup:2020ssz, Glashow:1961tr}
\begin{equation}
	e = g\sin\theta_W = g'\cos\theta_W, \qquad \sin^2\theta_W = \frac{g'^2}{g^2+g'^2}.
\end{equation}
Beyond tree-level the value of $\sin^2\theta_W$ depends on the renormalization scheme and energy scale, and while precise determinations exist at the Z-pole \cite{ParticleDataGroup:2024cfk},  there is a growing interest in its low-energy value, currently known at the percent level \cite{Dzuba:2012kx, SLACE158:2005uay, NuTeV:2001whx}, especially from experiments probing precision electroweak observables at $q^2\ll M_Z^2$. In this regime, the electroweak mixing angle is sensitive to hadronic effects and can be used to probe BSM physics \cite{Kumar:2013yoa, Ramsey-Musolf:1999qyv, Erler:2013xha, Chang:2009yw}.  

A $q^2$-dependent definition of the electroweak mixing angle, linking $\sin^2\theta_W(q^2)$ to its value in the Thomson limit takes the form \cite{Jeger1986, Jeger_yellow_rep, Jegerlehner:2011mw, Jegerlehner:2017zsb}
\begin{equation}
	\sin^2\theta_W(q^2) = \bigg(
	\frac{1-\Delta\alpha_2(q^2)}{1-\Delta\alpha(q^2)} + \Delta\kappa_b(q^2) - \Delta\kappa_b(0)
	\bigg)\sin^2\theta_W(0).
\end{equation}
In this work we adopt $\sin^2\theta_W(0) = 0.23857(5)$, quoted in~\cite{ParticleDataGroup:2020ssz} as the average of different evaluations~\cite{Czarnecki:2000ic, Kumar:2013yoa, Erler:2004in, Erler:2017knj}.
Here $\Delta\kappa_b$ denotes the bosonic contribution as given in \cite{Czarnecki:2000ic}, while $\Delta\alpha_2$ is the energy running contribution to the coupling $g^2 = 4\pi\alpha_2$, defined as
 \begin{equation}
 	\alpha_2(q^2) = \frac{\alpha_2}{1 - \Delta\alpha_2(q^2)}.
 \end{equation}
 Similarly to Eq.~\eqref{eq:space_like_dalpha}, the hadronic contribution at low energies is computed from the HVP function
 \begin{equation}
		(\Delta\sin^2\theta_W)_{\mathrm{had}}(q^2) = 
		\Delta\alpha_{\mathrm{had}}(q^2) - \Delta\alpha_{2,\mathrm{had}}(q^2)
		= -\frac{4\pi\alpha}{\sin^2\theta_W(0)}\bar{\Pi}^{(Z,\gamma)}(q^2),
\end{equation}
where  $\bar{\Pi}^{(Z,\gamma)}(q^2)$ denotes the HVP mixing of the electromagnetic current $j_\mu^\gamma$ and the vector component of the neutral weak current
\begin{equation}
	j_\mu^Z|_{\mathrm{vector}} = 	j_\mu^{T_3}|_{\mathrm{vector}} - \sin^2\theta_Wj_\mu^\gamma.
\end{equation}

\subsection{The TMR method}
As described above, the correlation function $G_{\mu\nu}(x)= \langle
j_\mu(x)j_\nu(0)\rangle$ of two vector currents computed on the
lattice is the primary quantity for evaluating the hadronic running
for both $\Delta\alpha_{\mathrm{had}}$ and $(\Delta\sin^2\theta_W)_{\mathrm{had}}$. The relevant HVP $\bar{\Pi}^{(\gamma,\gamma)}$ and $\bar{\Pi}^{(Z,\gamma)}$ can therefore be computed as a function of $Q^2$ from the vector correlators by providing the currents $j_\mu^\gamma$ and $j_\mu^Z$, respectively.
We adopt the Time-Momentum Representation (TMR) \cite{Bernecker:2011gh, Francis:2013fzp} to compute the subtracted HVP in the Euclidean theory through the integral
\begin{equation} \label{eq:PibarTMR}
	\bar{\Pi}^{(\alpha,\gamma)}(-Q^2) = 
	\int_{0}^{\infty} \dd x_0\; G^{(\alpha,\gamma)}(x_0)\bigg[
	x_0^2 - \frac{4}{Q^2}\sin[2](\frac{Qx_0}{2})	\bigg], \qquad \alpha=Z,\gamma.
\end{equation}
Here $G^{(\alpha,\gamma)}(x_0)$ represents the zero-momentum projected correlator
\begin{equation}
	G^{(\alpha,\gamma)}(x_0) = -\frac{1}{3}\int\dd^3x \sum_{k=1}^{3}\langle
	j_k^\alpha(x_0,\vec{x}) j_k^\gamma(x_0,\vec{0})
	\rangle\,.
\end{equation}
In this work, we update our previous result \cite{Ce:2022eix} with the inclusion of an extended set of ensembles with improved coverage in lattice spacing and quark masses. Furthermore, inspired by techniques developed for the non-perturbative calculations of the HVP contribution to the muon anomalous magnetic moment $a_\mu$, we introduce a new family of kernel functions designed to better control  systematics across different Euclidean regions. A detailed description of our computational strategy is given in  the following section.

\subsection{Flavour decomposition}
Following the notation of \cite{Ce:2022eix,Kuberski:2024bcj, Djukanovic:2024cmq}, we conveniently express the electromagnetic current through a matrix  $T^a$ acting in flavour space,
\begin{equation}
	j_\mu^a = \bar{\psi} T^a \gamma_\mu \psi, \qquad \bar{\psi} = \big(
	\bar{u}\ \bar{d} \ \bar{s} \ \bar{c} \ \bar{b}
	\big),
\end{equation}
such that the generic flavour-specific vector correlator takes the form
\begin{equation}
	\delta_{kl}\, G^{(m,n)}(x_0) = - \int\dd^3 x\langle
	j_k^m(x_0,\mathbf{x}) j_l^n(0)\rangle.
\end{equation}
We set $T^a = \frac{\lambda^a}{2} \oplus 0_c \oplus 0_b$  for $a=1,\ldots, 8$ to describe the $(u,d,s)$ sector through the corresponding Gell-Mann matrix $\lambda^a$ and define $T^c=\mathrm{diag}(0,0,0,1,0)$, $T^b=\mathrm{diag}(0,0,0,0,1)$ to describe the charm and bottom sector, respectively. We also introduce the $SU(3)$-singlet generator $T^0=\frac{1}{2}\mathbf{1}_3 \oplus 0_c \oplus 0_b$ corresponding to the singlet current $j_\mu^0$.   The vector correlators of interest then read
\begin{eqnarray}\label{eq:corr_building_blocks_gg}
	G^{(\gamma,\gamma)} = G^{(3,3)} + \frac{1}{3}G^{(8,8)} + \frac{4}{9}G^{(c,c)} + \frac{4}{9}G^{(c,c)}_{\mathrm{disc}} + 
	\frac{2}{3\sqrt{3}}G^{(c,8)}_{\mathrm{disc}} + \frac{1}{9}G^{(b,b)}+ \ldots,
	\\
	\label{eq:corr_building_blocks_zg}
	G^{(Z,\gamma)} = \bigg(
	\frac{1}{2} - \sin^2\theta_W
	\bigg)G^{(\gamma,\gamma)} - \frac{1}{6\sqrt{3}}G^{(0,8)} - \frac{1}{18}G^{(c,c)} - \frac{1}{18}G^{(c,c)}_{\mathrm{disc}} + \ldots,
\end{eqnarray}
in terms of the flavour-specific building blocks, which can be computed separately.
We treat the quark-connected and disconnected contributions in the
heavy quark sector individually, while the ellipses represent
contributions that are too small to be relevant at the current
statistical precision, such as the disconnected diagrams involving bottom and top quarks.

\section{General computational strategy}\label{subsec:comp_strategy}
One of the main challenges in achieving a lattice determination of the
electroweak couplings with controlled and conservative uncertainties
is to perform a reliable extrapolation to the physical point while
accounting for the systematics involved. To address this, we follow a
strategy that separates the dominant sources of uncertainty. Inspired
by the splitting in Short (SD), Intermediate (ID) and Long (LD)
Distance windows introduced in the calculation of the muon $g-2$
\cite{RBC:2018dos}, we propose the following decomposition for the
subtracted HVP\footnote{In sections \ref{subsec:comp_strategy} and
\ref{sec:lat_res}, we omit the minus sign in the argument of the
vacuum polarization that reminds us that we consider space-like virtualities. The minus sign is restored in section \ref{sec:had_run}.}
\begin{align}	\label{eq:hvp_splitting}
	  \bar{\Pi}(Q^2)
	&= \widehat{\Pi}(Q^2) + \widehat{\Pi}(Q^2/4) + \bar{\Pi}(Q^2/16),
\end{align}
where we have defined
\begin{equation}
   \widehat{\Pi}(Q^2) \equiv     \Pi(Q^2) - \Pi(Q^2/4).
  \end{equation}
We refer to the three terms $\widehat{\Pi}(Q^2)$, $
\widehat{\Pi}(Q^2/4)$  and $ \bar{\Pi}(Q^2/16)$ as High- (HV), Mid-
(MV) and Low-Virtuality (LV) regions, respectively. The above
telescopic sum offers a clear separation of the different Euclidean
regions, making it possible to disentangle the strong cutoff effects
at short times from the sizeable chiral dependence expected at larger
distances. This separation allows us to tailor the fit models to the
specific behaviour of each term in Eq.~\eqref{eq:hvp_splitting}. In
particular, perturbation theory becomes applicable in the HV region,
where the short distance nature of the observable makes discretization
effects most prominent. As a result, we can reliably achieve higher
values of $Q^2$ compared to \cite{Ce:2022eix}, thereby increasing the
threshold energy at which the running to the $Z$-pole is continued
using QCD perturbation theory. This is a crucial aspect for improving
the overall precision in $\Delta\alpha_{\mathrm{had}}^{(5)}(M_Z^2)$, as in our previous determination \cite{Ce:2022eix} the perturbative piece accounted for approximately $60\%$ of the total variance. 

\begin{figure}[h!]
	\centering
	\includegraphics[scale=0.48]{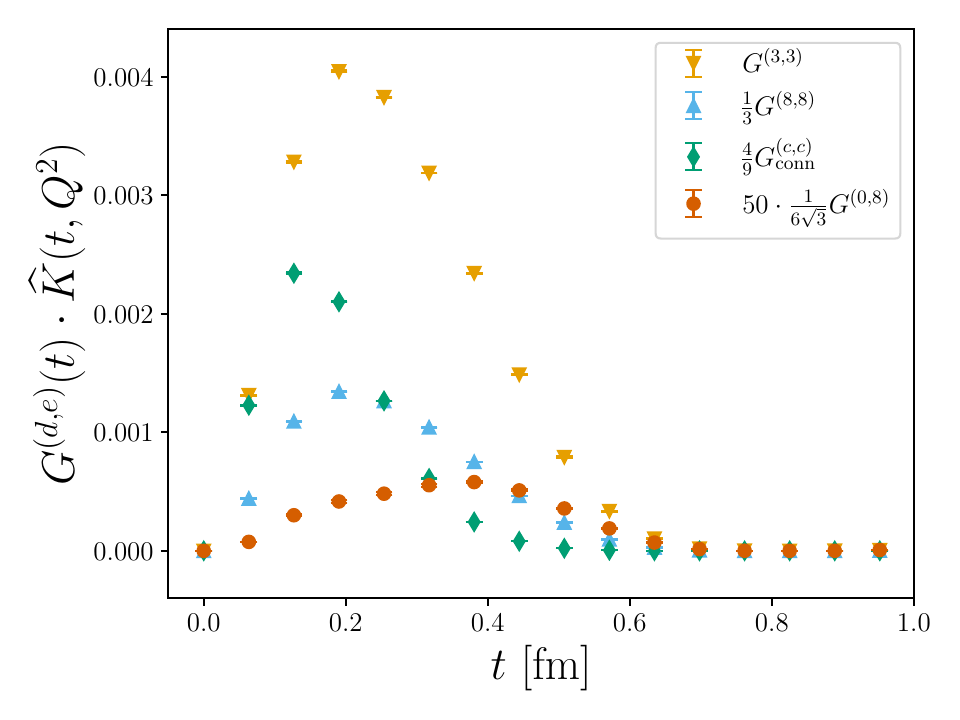}
	\includegraphics[scale=0.48]{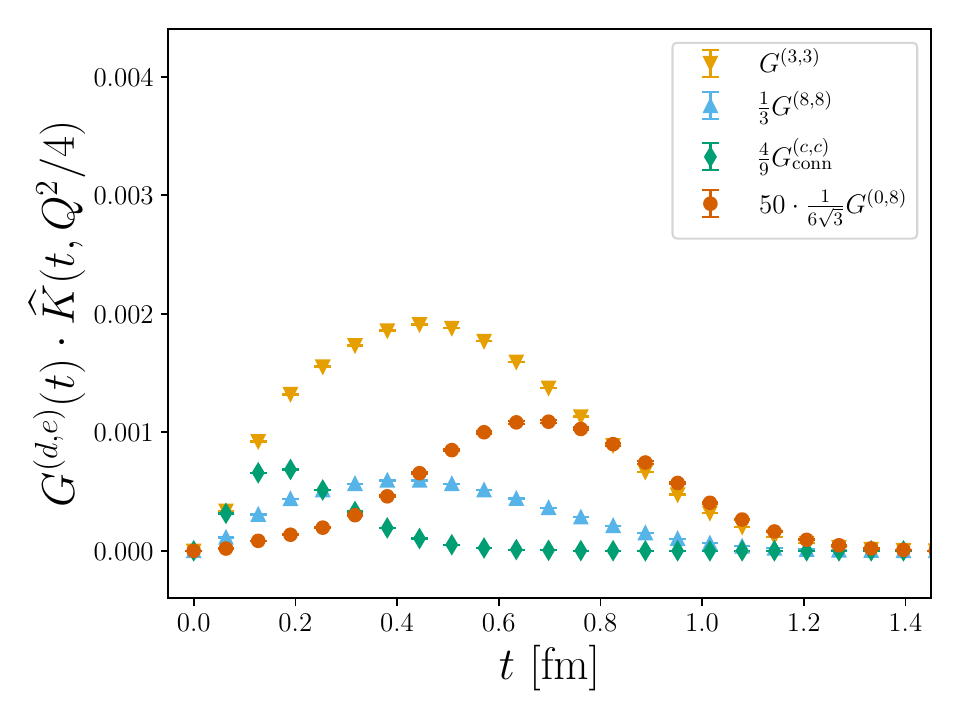}
	\includegraphics[scale=0.48]{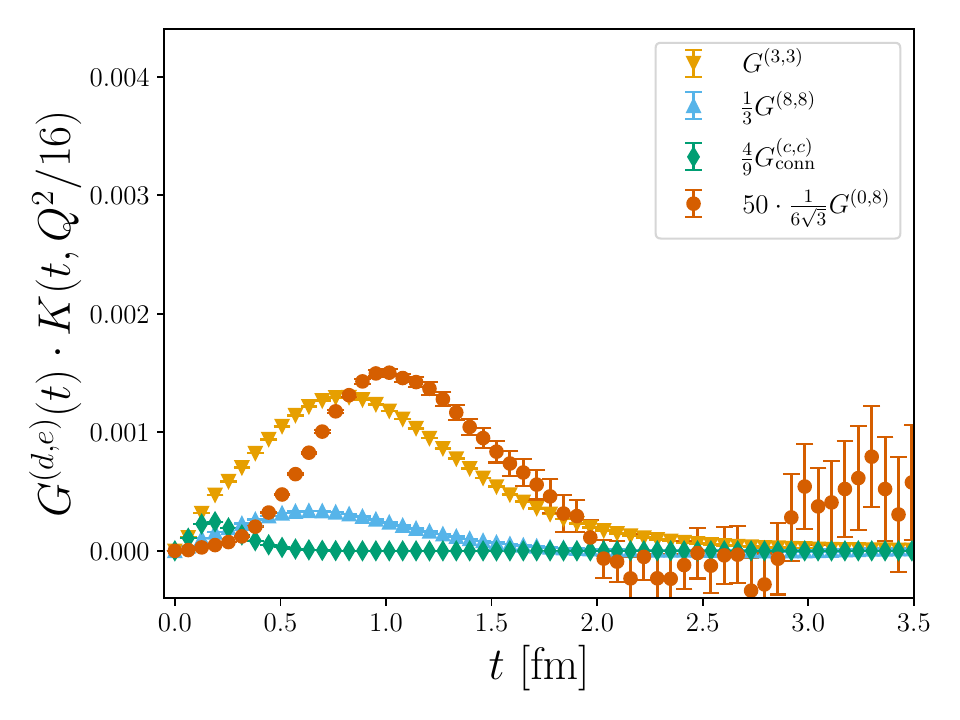}
	\caption{ Illustration of the integrands for the various
          contribution entering the computation of the electroweak
          couplings. \textit{Upper left}: HV region integrands
          using the non-subtracted kernel in
          Eq.~\eqref{eq:non_sub_kernel_QSD_QID}. \textit{Upper right:}
          MV region integrands
          using the non-subtracted kernel in
          Eq.~\eqref{eq:non_sub_kernel_QSD_QID}.  \textit{Bottom:} LV
          region integrands
          from the non-subtracted kernel in Eq.~\eqref{eq:non_sub_kernel_QLD}. Results are shown for physical pion mass ensemble E250 with $a \approx 0.064\ \mathrm{fm}$ at the virtuality $Q^2=9\ \mathrm{GeV}^2$. }
	\label{fig:kernel_comparison}
\end{figure}

The finite differences of the HVP, namely  $\widehat{\Pi}(Q^2)$ and $\widehat{\Pi}(Q^2/4)$  in the HV and MV regions are estimated via the TMR integral by convoluting the correlators of interest with the $Q^2$-dependent function
\begin{align}
	\widehat{\Pi}(Q^2) &=
	\int_{0}^{\infty} \dd x_0\; G(x_0)\widehat{K}(x_0, Q^2), 
\\ 
  \label{eq:non_sub_kernel_QSD_QID}
	\widehat{K}(x_0, Q^2) &= \frac{16}{Q^2}\sin[4](\frac{Qx_0}{4}),
\end{align}
while the LV term $\bar{\Pi}(Q^2/16)$ is extracted from
Eq.~\eqref{eq:PibarTMR}, i.e.\ with the standard kernel
\begin{equation}\label{eq:non_sub_kernel_QLD}
	K(x_0, Q^2) =  x_0^2 - \frac{4}{Q^2}\sin[2](\frac{Qx_0}{2}).
\end{equation}
In figure~\ref{fig:kernel_comparison} we plot the contributions to the
integrand for the three different virtuality regions considered
throughout this work. The integrands in the HV, MV and LV regions show
a behaviour analogous to the SD, ID and LD windows in the $g-2$
calculations. The separation of these regions makes it clear that
cutoff effects are most prominent in the HV regime, while the LV one
is dominated by a strong chiral dependence. The LV region also provides the largest absolute contribution to the final result and, due to the long-distance nature of the observable, the signal there is more susceptible to signal-to-noise degradation. In all three regions, the isovector channel gives the dominant contribution. In what follows we describe the computational strategy adopted for each term contributing to the HVP.

\subsection{The isovector contribution}
Following the approach introduced in \cite{Kuberski:2024bcj}, we make use of  the decomposition
\begin{equation}\label{eq:Pi33_decompo}
	\bar{\Pi}^{(3,3)}(Q^2) = \bar{\Pi}^{(3,3)}_{\mathrm{sub}}(Q^2) + \bar{b}^{(3,3)}(Q^2, Q_m^2),
\end{equation}
where  $ \bar{\Pi}^{(3,3)}_{\mathrm{sub}}(Q^2)$ is computed from the isovector correlator $G^{(3,3)}$,
\begin{equation}
	\bar{\Pi}^{(3,3)}_{\mathrm{sub}}(Q^2) =  \int_{0}^{\infty}
	\dd x_0\; G^{(3,3)}(x_0)\, K_{\mathrm{sub}}(x_0, Q^2, Q_m^2)
\end{equation}
with the help of a subtracted kernel function
\begin{equation}\label{eq:sub_kernel_QLD} 
	K_{\mathrm{sub}}(x_0, Q^2, Q_m^2) = x_0^2 - \frac{4}{Q^2}\sin[2](\frac{Qx_0}{2})
	-
	\frac{4}{3}\frac{Q^2}{Q_m^4}\sin[4](\frac{Q_mx_0}{2}),
\end{equation}
and
\begin{equation}\label{eq:b33_def0}
	\bar{b}^{(3,3)}(Q^2, Q_m^2) = \frac{Q^2}{3Q_m^2} \widehat{\Pi}^{(3,3)}(4Q_m^2)
\end{equation}
is to be computed in continuum perturbation theory.
The above decomposition suppresses the $x_0^4$ behaviour from the
kernel function, thus reducing higher-order cutoff effects and
cancelling completely the potentially dangerous lattice artefacts of
$O(a^2\log(a))$ in the very short Euclidean distance region \cite{Ce:2021xgd, Sommer:2022wac}.

Similarly,
we compute  the quantity $\widehat{\Pi}(Q^2)$ in the HV and MV regions from the subtracted kernel function,
\begin{align}\label{eq:Pihat33_decompo}
	\widehat{\Pi}^{(3,3)}(Q^2) &= \widehat{\Pi}^{(3,3)}_{\mathrm{sub}}(Q^2) + {b}^{(3,3)}(Q^2, Q_m^2),
        \\
        \widehat{\Pi}^{(3,3)}_{\mathrm{sub}}(Q^2) &= \int_{0}^{\infty}
        \dd x_0\; G^{(3,3)}(x_0)\, \widehat{K}_{\mathrm{sub}}(x_0, Q^2, Q_m^2),
        \\  \label{eq:sub_kernel_QSD_QID}
	\widehat{K}_{\mathrm{sub}}(x_0, Q^2, Q_m^2) &= \frac{16}{Q^2}\sin[4](\frac{Qx_0}{4})
	-
	\frac{Q^2}{Q_m^4}\sin[4](\frac{Q_mx_0}{2}),
\end{align}
where the second term in Eq.~\eqref{eq:Pihat33_decompo} is defined as
\begin{equation}\label{eq:b33_def}
  {b}^{(3,3)}(Q^2, Q_m^2) = \frac{Q^2}{4Q_m^2} \widehat{\Pi}^{(3,3)}(4Q_m^2).
\end{equation}
The quantities $b^{(3,3)}$ and $\bar{b}^{(3,3)}$ can be computed reliably using massless perturbation theory through the Adler function at virtualities $Q\geq 2.5 \ \mathrm{GeV}$, where good convergence is expected. We further  decompose the HV isovector contribution  into a SD and Intermediate plus Long Distance (ILD) part using the standard Euclidean time windows as defined in \cite{RBC:2018dos}, performing separate chiral-continuum extrapolations for each. This is beneficial, as tree-level improvement (see section \ref{subsubsec:tree_level_imp}) is applied solely to the SD term, leading to an effective reduction of cutoff effects where they are most pronounced.

Throughout this work, we set
\begin{equation}
  Q_m^2=9 \ \mathrm{GeV}^2
\end{equation}
as the space-like momentum value at which we perform the subtraction. The HV, MV and LV terms  are computed in the energy ranges $4.0 \ \mathrm{GeV}^2 \leq Q^2 \leq 12.0 \ \mathrm{GeV}^2 $,  $1.0\ \mathrm{GeV}^2 \leq Q^2/4 \leq 3.0  \ \mathrm{GeV}^2$ and $0.25 \ \mathrm{GeV}^2 \leq Q^2/16 \leq 0.75 \ \mathrm{GeV}^2$, respectively.

\subsection{The isoscalar contribution }\label{subsec:isoscalar_contrib}
Similarly to the isovector contribution, we evaluate the isoscalar HVP  through the decomposition in Eq.~\eqref{eq:hvp_splitting}.  Knowing $\widehat{\Pi}^{(3,3)}(Q^2)$ in the HV region,  we  estimate  the corresponding  $\widehat{\Pi}^{(8,8)}(Q^2)$    via the breakdown \cite{Kuberski:2024bcj}
\begin{equation}
	\widehat{\Pi}^{(8,8)}(Q^2) = \widehat{\Pi}^{(3,3)}(Q^2) + \widehat{\Delta}_{ls}(Q^2),
	\label{eq:isoscalar_decomposition}
\end{equation}
where only the term $\widehat{\Delta}_{ls}$ has to be computed anew. The key observation is that  isovector and isoscalar correlation functions at short distances are nearly identical. As a result, their difference is  parametrically suppressed, leading to a significant reduction of the statistical uncertainty and to a cancellation of log-enhanced cutoff effects. In practice, we evaluate $\widehat{\Delta}_{ls}$ through the integral 
\begin{equation}
	\widehat{\Delta}_{ls}(Q^2) = \int_{0}^{\infty} \dd x_0 \big[
	G^{(8,8)}(x_0) - G^{(3,3)}(x_0)
	\big] \widehat{K}(x_0, Q^2),
	\label{eq:delta_ls}
\end{equation} 
using the non-subtracted kernel function $\widehat{K}(x_0, Q^2)$ defined in Eq.~\eqref{eq:non_sub_kernel_QSD_QID}.
On the other hand, we extract the MV  $\widehat{\Pi}^{(8,8)}(Q^2/4)$ and LV   $\bar{\Pi}^{(8,8)}(Q^2/16)$ contributions through a straightforward computation by employing the non-subtracted kernel introduced in Eq.~\eqref{eq:non_sub_kernel_QLD}.

\subsubsection{$Z\gamma$ mixed isoscalar contribution}
The mixed contribution $\bar{\Pi}^{(0,8)}$, relevant exclusively to
the electroweak mixing angle, includes the $\rm SU(3)$-singlet
current and vanishes linearly in $m_s-m_l$ when approaching the $\rm SU(3)$-symmetric point, ensuring  that log-enhanced cutoff effects are not relevant. Therefore, no perturbative subtraction of the latter is required.  Specifically, we compute this contribution in the three virtuality regions defined in Eq.~\eqref{eq:hvp_splitting} through the non-subtracted kernels  in Eq.~\eqref{eq:non_sub_kernel_QSD_QID} and Eq.~\eqref{eq:non_sub_kernel_QLD}.

\subsection{Charm-connected contribution}
The charm-connected contribution is computed analogously to the isovector HVP by employing the subtracted kernels introduced in Eq.~\eqref{eq:sub_kernel_QSD_QID} and Eq.~\eqref{eq:sub_kernel_QLD} for the (HV,\,MV) and LV regions, respectively. In the (HV,\,MV) cases, the subtraction function is determined according to
\begin{equation}
	b_{\mathrm{conn}}^{(c,c)}(Q^2) = 2b^{(3,3)}(Q^2, Q_m^2) + \Delta_{lc}b(Q^2,Q_m^2).
	\label{eq:charm_bcc}
\end{equation}
Here, the first term $b^{(3,3)}$ is evaluated in massless perturbation theory, while $\Delta_{lc}b$, defined as
\begin{equation}
  \Delta_{lc}b(Q^2, Q_m^2) =  \frac{Q^2}{4Q_m^2} (2 \widehat\Pi^{(3,3)}(4Q_m^2) - \widehat\Pi^{(c,c)}(4Q_m^2)),
\end{equation}
is computed on the lattice using the non-subtracted kernels in Eq.~\eqref{eq:non_sub_kernel_QSD_QID}  convoluted with the difference between the light-connected and charm-connected correlators. A completely analogous procedure is used for the LV part, the subtraction function being obtained
as the sum of the perturbatively computed $2\bar{b}^{(3,3)}(Q^2,Q_m^2)$ and the charm-mass effect $\Delta_{lc}\bar b$ computed on the lattice.

\subsection{Lattice setup}
\subsubsection{Gauge ensembles}
Our calculation is performed on  Coordinated Lattice Simulations (CLS)
ensembles with $N_f=2+1$ flavours \cite{Bruno:2014jqa,
  Bali:2016umi, Mohler:2017wnb, Mohler:2020txx, Kuberski:2023zky} of
non-perturbatively $O(a)$-improved Wilson quarks \cite{Bulava:2013cta}
and a tree-level improved L\"{u}scher-Weisz gauge action. The analysis
is focused on a subset of CLS ensembles in which the sum of the bare
quark masses is held constant as we approach the physical pion mass,
ensuring that the improved bare coupling
$\tilde{g}_0$~\cite{Luscher:1996sc} also remains constant along the
chiral trajectory. In addition, we include four ensembles on a different chiral trajectory where the strange quark mass is close to its physical value and with a pion mass of roughly $220\ \mathrm{MeV}$. This helps to account for a small mistuning in $m_K^{\mathrm{phys}}$ when approaching the physical pion mass \cite{Bruno:2016plf}. 
Compared to our 2022 publication \cite{Ce:2022eix}, we have
substantially expanded the set of gauge ensembles, which now covers five values of the lattice spacing in the range $0.039\ \rm{fm} \leq \mathit{a} \leq 0.087 \ \rm{fm}$ and pion masses  in the range $130 \ {\rm MeV}  \leq m_\pi \leq 420 \ {\rm MeV}$. Notably, we employ three ensembles with physical pion mass, allowing us to firmly constrain the chiral dependence. An overview of the ensembles entering this analysis is provided in table \ref{tab:CLS}. Further details on the set of ensembles can be found in \cite{Ce:2022eix, Ce:2022kxy, Kuberski:2024bcj, Djukanovic:2024cmq}.

\begin{table}[h!]
	\centering
		\begin{tabular}{l c c c c c c c c c } 
			\toprule
			\text{Id} & $\beta $ & bc & $L/a$ & $T/a$ & $a \ [\mathrm{fm}]$  & $m_\pi$ [MeV] & $m_K$ [MeV] & $m_\pi L $ & $L \ [\mathrm{fm}]$\\
			\noalign{\smallskip}\hline\noalign{\smallskip}
			A654& 3.34 & p & 24 & 48 & 0.0978(5) &  338 &  462 & 4.0 & 2.3  \\
			\noalign{\smallskip}\hline\noalign{\smallskip}
			H101 & 3.40 & o & 32 & 96 & 0.0847(5) &  424 &  424 & 5.8 & 2.7  \\
			H102 &   & o & 32 & 96 & &  358 & 445 & 4.9 & 2.7  \\
			N101 &   & o & 48 & 128 & &  282 & 468 & 5.8 & 4.1  \\
			C101 &   & o & 48 & 96 &  &  222 & 478 & 4.6 & 4.1  \\
			C10$2^\dagger$ &   & o & 48 & 96 & &  225 & 506 & 4.6 & 4.1  \\
			D15$0^\dagger$ &   & p & 64 & 128 & &  131 & 484 & 3.6 & 5.4  \\
			\noalign{\smallskip}\hline\noalign{\smallskip} 
			
			B450 & 3.46 & p & 32 & 64 & 0.0752(5)  &  422  & 422 & 5.1 & 2.4  \\
			N452 &          & p & 48 & 128 &                  &  356  & 447 & 6.5 & 3.6  \\
			N451 &          & p & 48 & 128 &                  &  291  & 468 & 5.3 & 3.6  \\
			D450 &          & p & 64 & 128 &                  &  219  & 483 & 5.3 & 4.8  \\
			D45$1^\dagger$ &          & p & 64 & 128 &                &  220  & 510 & 5.3 & 4.8  \\
			D452 &          & p & 64 & 128 &                  &  156  & 490 & 3.8 & 4.8  \\
			\noalign{\smallskip}\hline\noalign{\smallskip}
			N202 & 3.55 & o & 48 & 128 & 0.0633(4)  &  418 & 418 & 6.5 & 3.0  \\
			N203 &  & o & 48 & 128 &    & 349   & 447 & 5.4 & 3.0  \\
			N200 &  &  o & 48 & 128 &     &  286  & 468 & 4.4 & 3.0  \\
			D251 &  &  p & 64 & 128 &     &  286  & 468 & 5.9 & 4.1  \\
			D200 &  & o & 64 & 128 &    &  202    & 486 & 4.2 & 4.1  \\
			D20$1^\dagger$ &  & o & 64 & 128 &    &  202    & 507 & 4.2 & 4.1  \\
			E25$0^\dagger	$ &  & p & 96 & 192 &    &  132    & 495 & 4.1 & 6.1  \\
			\noalign{\smallskip}\hline\noalign{\smallskip}
			J307 & 3.70 & o & 64 & 192 &  0.0490(3)  &  425   & 425 & 6.7 & 3.1 \\
			J306 &  & o & 64 & 192 &    &  350   & 456 & 5.6 & 3.1 \\
			J303 &  & o & 64 & 192 &     &  260  &  480 & 4.1 & 3.1  \\
			J30$4^\dagger$ &  & o & 64 & 192 &     &  263  &  530 & 4.3 & 3.1  \\
			E300 &  & o & 96 & 192 &     &  177  &  497 & 4.2 & 4.7  \\
			F30$0^\dagger$ &  & o & 128 & 256 &     &  136  &  496 & 4.3 & 6.3  \\
			\noalign{\smallskip}\hline\noalign{\smallskip}
			J500 & 3.85 & o & 64 & 192 & 0.0385(3)     &  417  &  417 & 5.2 & 2.5  \\
			J501 &  & o & 64 & 192 &     &  337  &  450 & 4.2 & 2.5  \\
			\bottomrule
		\end{tabular}
	\caption{List of CLS $N_f=2+1$ ensembles used in this work. Columns show the following parameters: the bare coupling, the temporal boundary conditions, open (o) or anti-periodic (p), the lattice dimensions, the lattice spacing $a$ in physical units based on \cite{Bussone:2025wlf}, approximate pion and kaon masses  and the physical size of the lattice. Ensembles marked by a dagger belong to the second chiral trajectory where $m_s\approx m_s^{\mathrm{phys}}$. Ensemble A654 is  used exclusively for estimating isospin-breaking effects.}
	\label{tab:CLS}
\end{table}

\subsubsection{Renormalization and $O(a)$-improvement}
We employ two discretizations of the vector current, the local $(L)$   and point-split $(C)$ currents, defined as
\begin{eqnarray}
	J_\mu^{(L),a}(x) &=& \overline{\psi}(x)\gamma_\mu T^a\psi(x),
	\\ \nonumber
	J_\mu^{(C),a}(x) &=& \frac{1}{2}\big(
	\overline{\psi}(x+a\hat{\mu})(1+\gamma_\mu) U_\mu^\dagger(x)T^a\psi(x)
	-
	\overline{\psi}(x)(1-\gamma_\mu)U_\mu(x)T^a\psi(x+a\hat{\mu})
	\big),
\end{eqnarray}
with $U_\mu(x)$ denoting the gauge link at site $x$ with direction
$\hat{\mu}$. The $O(a)$-improved versions of the currents in the
massless theory are defined by
\begin{equation}
	J_\mu^{(\alpha),a,I}(x) = 	J_\mu^{(\alpha),a}(x) + a c_{\mathrm{V}}^{(\alpha)(g_0)} \partial_\nu \Sigma_{\mu\nu}^a (x),
	\qquad
	\alpha=L, C,
\end{equation}
where the local tensor current is given by $\Sigma_{\mu\nu}^a(x)=-\frac{1}{2}\overline{\psi}(x)\comm{\gamma_\mu}{\gamma_\nu}T^a\psi(x)$. Adopting the prescription of \cite{Kuberski:2024bcj}, we replace the commonly used symmetric derivative acting on $\Sigma_{\mu\nu}^a(x)$ with
\begin{equation}
	\tilde{\partial}_0 \Sigma_{\mu 0}^a(x) \rightarrow
	\frac{1}{x_0^2} \big[
	\tilde{\partial}\big(
	x_0^2\Sigma_{\mu 0}^a(x)
	\big)
	-
	2x_0 \Sigma_{\mu 0}^a(x)
	\big],
\end{equation}
such that cutoff effects arising from the discrete derivative in the very-short Euclidean region are substantially reduced.
 In line with our previous works, we use two independent sets of non-perturbatively determined improvement coefficients and renormalization constants. By set 1 we denote  the improvement coefficients from large-volume simulations  as determined in \cite{Gerardin:2018kpy}, while we refer to set 2 when using $c_V$ and $Z_V$ from \cite{Heitger:2020zaq}  and $b_V,\bar{b}_V$ from \cite{Fritzsch:2018zym}, determined in the Schr\"{o}dinger Functional (SF) setup. In practice, an update on the value of $\tilde{b}_A$ \cite{Bali:2023sdi}, used as input for the extraction of $c_V$, with respect to the one entering in \cite{Gerardin:2018kpy}, leads to a change in the set 1  improvement coefficients. As a result in this work we consider the updated~\cite{Harris:2025xvk} (yet unpublished) values from set 1 exclusively as a cross-check of our continuum extrapolation, while the final results are based only on set 2.  However, as a conservative choice, we still include set 1 in our estimate of systematic uncertainties from the model variation. Details on the renormalization pattern for the electromagnetic currents employed in this work are discussed  in \cite{Gerardin:2019rua,Ce:2022kxy, Ce:2022eix}.

\subsubsection{Tree-level improvement}\label{subsubsec:tree_level_imp}
On top of subtracting the perturbative evaluation of
$b^{(d,e)}(Q^2,Q_m^2)$ to reduce lattice artefacts in short-distance
observables, we may achieve an additional reduction of cutoff effects
in the isovector channel by evaluating the correlators for non-zero~$a$ in massless perturbation theory at leading order. Given the tree-level computation $\mathcal{O}^{\mathrm{tl}}(a)$ of an observable $\mathcal{O}(a)$, we perform the tree-level improvement by means of 
\begin{equation}
	\mathcal{O}(a) \rightarrow \mathcal{O}(a)
	\frac{\mathcal{O}^{\mathrm{tl}}(0)}{\mathcal{O}^{\mathrm{tl}}(a)}.
\end{equation}
We also apply the same improvement schemes to the vector correlators computed in massless perturbation theory as those used in the non-perturbative calculations.  The implementation of tree-level improvement significantly reduces cutoff effects and the associated systematic uncertainties, while leaving the continuum-extrapolated result unchanged within errors. For further details, we refer the reader to \cite{Kuberski:2024bcj}.

\subsection{Finite-volume correction}\label{sec:fvc}
To obtain reliable estimates for the electroweak couplings it is
crucial to correct for finite-size effects in the isovector channel,
arising from the finite spatial volume $L^3$  used in lattice QCD calculations. Following our previous studies \cite{Ce:2022kxy, Ce:2022eix, Djukanovic:2024cmq}, we employ two correction schemes: the Hansen-Patella (HP) approach \cite{Hansen:2019rbh, Hansen:2020whp}, based on the pion's electromagnetic form factor, is particularly effective in the relatively short Euclidean distances. On the other hand, in the long-distance domain, we apply the Meyer-Lellouch-L\"{u}scher (MLL) formalism \cite{Lellouch:2000pv, Meyer:2011um}, which relies on the time-like pion form factor. 

In practice, our final estimate for finite-volume effects is obtained
by applying the HP correction for times $t<t^\star = (m_\pi L/4)^2 /
m_\pi$, and then switching to the MLL formalism beyond this threshold.
The HP corrections are computed from the vector-meson dominance
parametrization of the pion form factor, while the Gounaris-Sakurai
formalism is used in the MLL method. In addition, we also include
corrections for kaon propagation in finite volumes, non-negligible for
ensembles near the $\rm SU(3)$-symmetric point along the chiral trajectory $\Tr(M_q)=\mathrm{const}$. Conversely, the leading finite-volume effects cancel in the isoscalar channel and we only include contributions from $K\bar{K}$ states, treated via the HP method. Further details of our implementation of the finite-volume corrections are given in \cite{Djukanovic:2024cmq}. 

As in \cite{Djukanovic:2024cmq}, we follow the procedure first proposed in \cite{Borsanyi:2020mff} by correcting all ensembles to a common reference value of $m_\pi L$ before performing the chiral-continuum extrapolation to the physical point. To minimize the correction applied in the vicinity of the physical point, we select a reference volume that closely matches our physical pion ensembles, defined by
\begin{equation}
	(m_\pi L)_{\mathrm{ref}} = (m_{\pi^0})_{\mathrm{phys}} \cdot 6.272 \ \mathrm{fm} \approx 4.290.
\end{equation}
Eventually, we compute the correction from the reference to infinite volume in the continuum following the same strategy outlined in \cite{Djukanovic:2024cmq}. We arrive at
\begin{equation}
	(\bar{\Pi}^{(3,3)})(L=\infty) - (\bar{\Pi}^{(3,3)})(L_{\mathrm{ref}}) = 28.6(1.6) \times 10^{-5}
\end{equation}
at $Q^2=9.0\ \mathrm{GeV}^2$. As in \cite{Djukanovic:2024cmq}, the
quoted error of $\sim10$\% has been estimated from the variation of
the resulting correction for different input parameters in the HP and
MLL formalisms, noting that contributions from higher channels to the
finite-volume correction are subleading.
In table~\ref{tab:fvc_Lref_to_Linf} we collect results for the finite-volume corrections to $L=\infty$  at various values of $Q^2$.

\begin{table}[t]
	\centering
	\renewcommand{\arraystretch}{1.1}
	\begin{minipage}{0.47\linewidth}
		\flushright
		\begin{tabular}{c | c}
			\toprule
			$Q^2\,[\mathrm{GeV}^2]$ & $\Delta L$ \\
			\hline
			0.25   & 22.1(1.3) \\
			0.3125 & 23.2(1.4) \\
			0.375  & 24.0(1.4) \\
			0.4375 & 24.7(1.5) \\
			0.5    & 25.1(1.5) \\
			0.75   & 26.3(1.5) \\
			1.0    & 26.9(1.6) \\
			1.25   & 27.3(1.6) \\
			1.5    & 27.5(1.6) \\
			1.75   & 27.7(1.6) \\
			\bottomrule
		\end{tabular}
	\end{minipage}\hfill
	\begin{minipage}{0.47\linewidth}
		 \flushleft
		\begin{tabular}{c | c}		
			\toprule
			$Q^2\,[\mathrm{GeV}^2]$ & $\Delta L$ \\
			\hline
			2.0   & 27.9(1.6) \\
			2.25  & 28.0(1.6) \\
			3.0   & 28.2(1.6) \\
			4.0   & 28.4(1.6) \\
			5.0   & 28.5(1.6) \\
			6.0   & 28.5(1.6) \\
			7.0   & 28.6(1.6) \\
			8.0   & 28.6(1.6) \\
			9.0   & 28.6(1.6) \\
			12.0  & 28.7(1.6) \\
			\bottomrule
		\end{tabular}
	\end{minipage}
	\caption{Values for the isovector finite-volume correction in the continuum from $L_\mathrm{ref}$ to $L=\infty$ at different values of $Q^2$ employed in this work. Results are presented in units of $10^{-5}$. }
\label{tab:fvc_Lref_to_Linf}
\end{table}

\section{Lattice results}
\label{sec:lat_res}

\subsection{Physical point extrapolation}\label{sec:phys_point_extrap}
The proxies to describe the light and strange quark masses are defined in terms of the dimensionless hadronic combinations
\begin{equation}
	\phi_2 = 8t_0 m_\pi^2, \qquad \phi_4 = 8t_0\left(m_K^2+ \frac{1}{2} m_\pi^2\right).
\end{equation}
The scale setting is performed using  the  gradient flow parameter $t_0/a^2$ \cite{Luscher:2010iy}, with its physical value $t_0^{\mathrm{phys}} = 0.1440(7)\ \mathrm{fm}$ taken from the determination based on $f_{\pi K}$, a combination of pion and kaon decay constants, in Ref.~\cite{Bussone:2025wlf}. 
We define the physical point  in the isospin-symmetric limit by imposing $m_\pi = (m_{\pi^0})_{\mathrm{phys}}$ and $2m_K^2 - m_\pi^2 = (m_{K^+}^2 + m_{K^0}^2 - m_{\pi^+}^2)_{\mathrm{phys}}$ \cite{Urech:1994hd, Neufeld:1995mu}, leading to the physical masses  $m_\pi = 134.9768(5) \ \mathrm{MeV}$ and $m_K = 495.011(10)\ \mathrm{MeV}$ for the pion and kaon, respectively, while the tuning  of the charm quark mass is performed by fixing  $m_{D_s}=1968.47 \ \mathrm{MeV}$.

The extrapolation to the physical point adopts a similar strategy as
our previous works \cite{Kuberski:2024bcj, Djukanovic:2024cmq} and
proceeds by fitting concurrently the chiral and cutoff dependence,
using Symanzik effective theory to guide the continuum behaviour. Denoting by $X_a^2 = a^2/(8t_0)$ the proxy for the lattice spacing, our most general fit ansatz reads
\begin{equation}
\begin{split}
\mathcal{O}(X_a) &= \beta_2 X_a^2 + \beta_3X_a^3 + \beta_4X_a^4 + \delta_2X_a^2 \left(
\phi_2 - \phi_2^{\mathrm{phys}}
\right) \\
&+ \delta_3X_a^3 \left(
\phi_2 - \phi_2^{\mathrm{phys}}
\right)
+ \epsilon_2 X_a^2 \left(
\phi_4 - \phi_4^{\mathrm{phys}}
\right).
\end{split}
\end{equation}
In our fits we always include the leading term proportional to $X_a^2$
and check for the significance of higher order effects by selectively
dropping one or more of the terms multiplied by the parameters
$\beta_i, \delta_i, \epsilon_2$. Following \cite{Husung:2019ytz,
  Husung:2021mfl, Husung:2024cgc}, we also include terms describing
logarithmic enhancements of cutoff effects, modelling the leading lattice artefacts as $X_a^2[\alpha_s(1/a)]^{\hat{\Gamma}}$. We consider two values of the anomalous dimension $\hat{\Gamma}\in [0, 0.395]$, where for each fit both choices of $\hat{\Gamma}$ are tested to assess for the stability of the extrapolation.  Similarly, the chiral dependence of observable $\mathcal{O}$ is modelled by including a linear term in $\phi_2$ along with higher-order corrections, 
\begin{equation}
	\begin{split}
		\mathcal{O}\big(\phi_2\big)  = \mathcal{O}\big(\phi_2^{\mathrm{phys}}\big) &+  \gamma_1 \left(
	\phi_2 - \phi_2^{\mathrm{phys}} 	\right)
	\\
	 & + \gamma_2\left(
	 f_{\chi,1}(\phi_2) - f_{\chi,1}(\phi_2^{\mathrm{phys}})
	 \right)
	 \\
	 &+ \gamma_3\left(
	 f_{\chi,2}(\phi_2) - f_{\chi,2}(\phi_2^{\mathrm{phys}})
	 \right),
\end{split}
\end{equation}
where
\begin{equation}
	f_{\chi,1} \in \{ 1/\phi_2; \ \log(\phi_2); \ \phi_2\log(\phi_2); \  \phi_2^2 \}, \qquad 
	f_{\chi,2} \in \{ 1/\phi_2;  \  \phi_2^2 \}.
\end{equation}
In practice,  the inclusion of  higher-order terms is evaluated individually based on their statistical significance,  as the different HV, MV and LV virtuality regions may not require the full set of terms.

 Since the strange quark proxy $\phi_4$ is close to its physical value on all ensembles, we describe it via 
\begin{equation}
\mathcal{O}\big(\phi_4\big) = \mathcal{O}\big(\phi_4^{\mathrm{phys}}\big) + \delta_0 \left(
\phi_4  - \phi_4^{\mathrm{phys}}
\right) 
+
\delta_1 \left(
\phi_4^2  - \phi_4^{2,\mathrm{phys}}
\right)
,
\end{equation} 
where the higher-order term is only included in the isoscalar MV region.

The charm-quark mass dependence is treated separately by interpolating the data to the physical point using $\sqrt{8t_0}m_{D_s}$ as a proxy, exploiting the approximately linear behaviour in the simulated mass range, prior to the global fit.

To assess systematic uncertainties arising from the chiral-continuum
extrapolations, we apply  cuts to the data sets,  either by excluding
the coarsest lattice spacing or by removing all ensembles with $m_\pi
> 400 \ \mathrm{MeV}$. For the final estimate and systematic error
analysis we follow our previous work and perform a weighted model
average\,\cite{Jay:2020jkz}, with weights assigned according to the Akaike Information Criterion (AIC) \cite{Akaike:1998zah}. The quality of the fits is assessed by monitoring the ratio $\chi^2/\chi^2_{\mathrm{exp}}$, where $\chi^2_{\mathrm{exp}}$ denotes the expected value of the $\chi^2$ as defined in Ref.~\cite{Bruno:2022mfy}, and by verifying the corresponding $p$-values. Error analysis and propagation are performed using the $\Gamma$-method of Ref.~\cite{Wolff:2003sm}, as implemented in the \texttt{ADerrors} package~\cite{Ramos:2018vgu}. 
The split into statistical and systematic errors is performed according to Eq.~(18) of \cite{Jay:2020jkz}, where in the estimate of the statistical error we set the derivatives of the AIC weights to zero. The chiral-continuum extrapolation plots and fit quality indicators for all channels entering the analysis are collected in the Supplemental Material~\cite{SuppMat}. 
  
\subsection{Noise reduction strategies}\label{sec:noise_redduction}
The rapid degradation of signal quality in both light-quark connected
and disconnected correlation functions in the LD  Euclidean region is
one of the limiting factors for a high precision determination of the
electroweak couplings. To overcome this, we employ several noise
reduction strategies, focusing primarily on the isovector
contribution, which accounts for roughly $50\%$ of the total
HVP. Following our recent publication \cite{Djukanovic:2024cmq}, we
adopt the improved estimator computed via low-mode averaging (LMA)
\cite{Giusti:2004yp,DeGrand:2004qw} for the light-quark connected correlation function.

Additionally, in both isovector and isoscalar channels we utilize the bounding method, an established technique to tackle the signal-to-noise problem in HVP calculations \cite{RBC:2018dos, Gerardin:2019rua, Borsanyi:2016lpl,Lehner2016}. The method consists of replacing the correlation function at $t>t_c$ with appropriate lower and upper bounds according to
\begin{equation}\label{eq:bounding_method}
	0 \leq G^{(\alpha,\gamma)}(t_c) e^{-E_{\mathrm{eff}}(t-t_c)} \leq G^{(\alpha,\gamma)}(t)
	\leq
	G^{(\alpha,\gamma)}(t_c) e^{-E_0(t-t_c)},
\end{equation}
where $E_0$ is the ground-state energy level contributing to the
vector correlation function. In practice, when available we estimate
$E_0$ from a dedicated spectroscopy study by solving a Generalized
Eigenvalue Problem (GEVP) in the isovector channel. On the other
ensembles we estimate the ground-state energy through a
Gounaris-Sakurai parametrization of the time-like pion form factor. In
line with our previous work  \cite{Ce:2021xgd, Djukanovic:2024cmq}, we
apply the same estimate for $E_0$ to the isoscalar correlation function, where the near degeneracy  $m_\rho \lesssim m_\omega$ makes this a conservative approximation.

For the lower bound, we determine $E_\mathrm{eff}$ at some time $t_{\mathrm{eff}} < t_c$ from the logarithmic derivative of the vector correlation function. For each ensemble, $t_{\mathrm{eff}}$  is set to ensure that the effective mass at that point is larger than the region where the bounding method takes over.  

Eventually, we estimate the corresponding HVP contribution by averaging both bounds over an interval of roughly $0.4 \ \mathrm{fm}$, starting from the value of $t_c$ where the two bounds are compatible within $0.5\sigma$ uncertainty. 

Besides the $\bar{\Pi}^{(3,3)}$ and $\bar{\Pi}^{(8,8)}$ contributions,
we also apply the bounding method to the mixed isoscalar
$\bar{\Pi}^{(0,8)}$ term, as detailed in \cite{Ce:2022eix}. Unlike
$G^{(3,3)}$ and $G^{(8,8)}$, the correlator $G^{(0,8)}$ lacks a
strictly positive-definite spectral decomposition, which invalidates
Eq.~\eqref{eq:bounding_method} in this case. However, it is known that $G^{(0,8)}$ shares the same ground-state energy $E_0$ as the isoscalar contribution, and its associated amplitude $Z_0$ is positive. In addition, the correlator approaches its asymptotic behaviour $\sim Z_0e^{-E_0t}$ from below, and the effective energy  $E_{\mathrm{eff}}$ also closes in on $E_0$ from below. Consequently, the role of the upper and lower bounds for $G^{(0,8)}$ are inverted compared to Eq.~\eqref{eq:bounding_method}.

\begin{figure}
	\centering
	\includegraphics[scale=0.305]{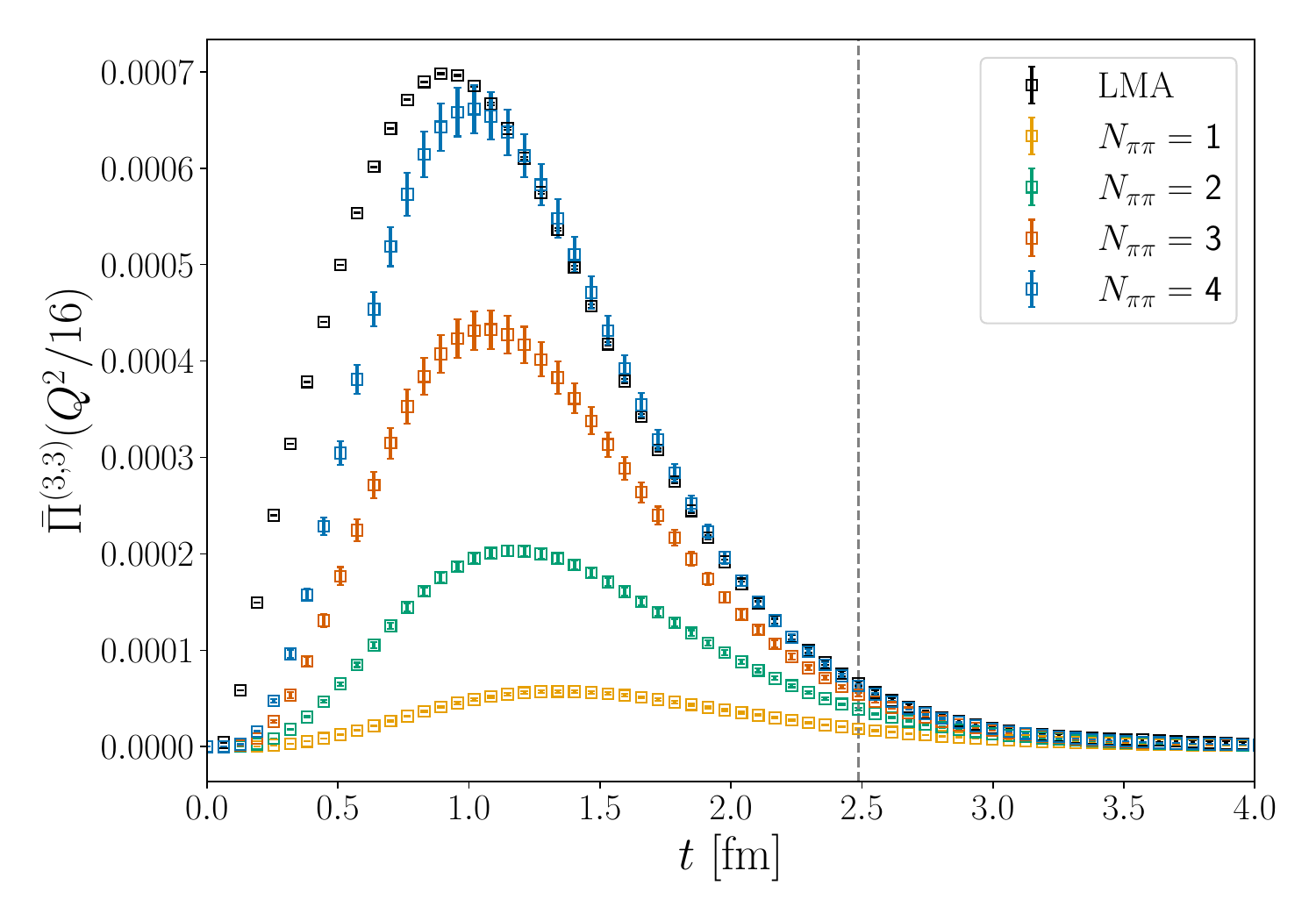}
	\includegraphics[scale=0.305]{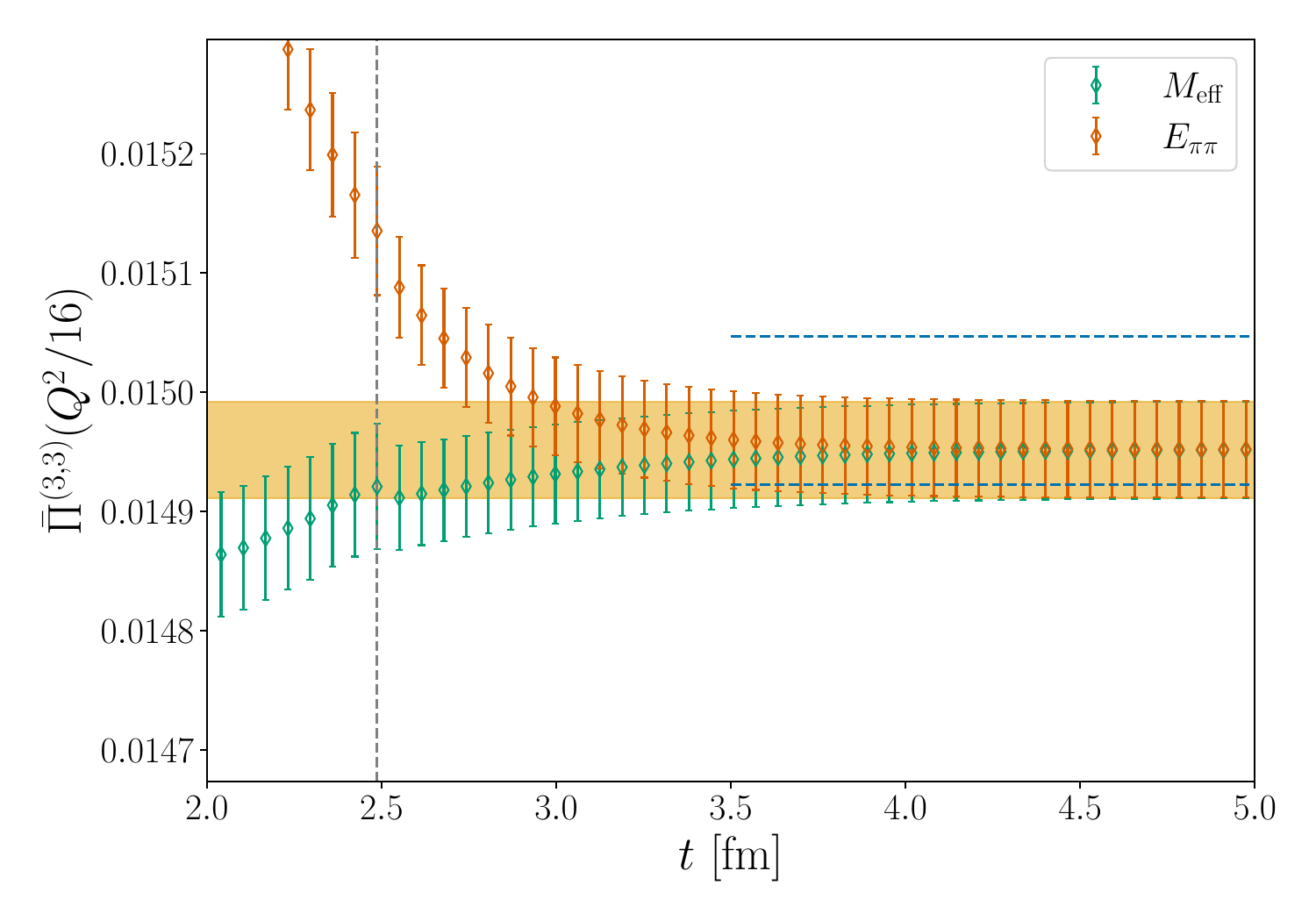}
	\caption{Illustration of noise-reduction techniques for the  $\bar\Pi^{(3,3)}$ channel in the LV momentum region. \textit{Left:} integrand contribution on the physical-mass ensemble E250. Black squares correspond to the correlation function obtained using LMA, while the coloured points show the reconstructed integrand from $N_{\pi\pi}$ states. The vertical dotted line marks the transition between LMA and spectroscopy data sets. \textit{Right:} application of the bounding method on reconstructed correlator. The dashed vertical lines indicates where the correlator is replaced by spectroscopy data. Green and orange points represent the lower and upper bound, respectively, with the yellow band illustrating our final estimate. The dashed horizontal lines indicate the results we would have obtained based solely on the LMA data. }
	\label{fig:bm_spectroscopy}
\end{figure}

We further improve the statistical accuracy in the  isovector LD
region for the two ensembles D200 and E250, with pion mass of
approximately $200\ \mathrm{MeV}$ and $ 130\ \mathrm{MeV}$,
respectively, by supplementing the LMA calculation with a dedicated spectroscopy calculation to explicitly reconstruct the tail of the $G^{(3,3)}$ correlator \cite{Gerardin:2019rua}. This involves computing the lowest energy levels $E_n$ and amplitudes $Z_n$ through a dedicated GEVP study. Full computational details for the physical point ensemble are provided in Appendix D of \cite{Djukanovic:2024cmq}. 

We find that for E250 the isovector correlator is saturated by the
four lowest-lying energy states from approximately $1.2\ \mathrm{fm}$
onward, while for D200 the two lowest states are sufficient to achieve
this, due to the larger pion mass on this ensemble. However, LMA continues to provide better precision at shorter time separations (below roughly $2.5\ \mathrm{fm}$), therefore we switch to the reconstructed correlator only when it yields lower statistical uncertainty.

An illustration of the noise-reduction strategies applied throughout this work for the physical point ensemble E250 is shown in figure~\ref{fig:bm_spectroscopy}. The left panel illustrates the reconstruction of the isovector TMR integrand when including an increasing number of two-pion states in the correlator, with the signal being saturated beyond $1.2 \ \mathrm{fm}$. The  dashed vertical line represents the source-sink separation where we switch from LMA to the spectroscopy data set. On the right panel we show the bounding method for the isovector channel, with the green and orange points representing the lower and upper bounds, respectively. The yellow band shows the final estimate, while the dashed horizontal lines depict the result we would have obtained without the spectroscopy data.

\subsection{The isovector contribution}
The isovector channel contributes by far the biggest share to the HVP,
particularly on ensembles close to the physical pion mass.
As explained in section~\ref{sec:noise_redduction}, LMA together with
the bounding method and spectral reconstruction allow us to achieve
high precision on the most important set of data close to the physical
point. For instance, on ensemble E250, we were able to reduce the uncertainty from $0.6\, \%$ in \cite{Ce:2022eix} to $ 0.2\, \%$ in the current analysis at $Q^2=1.0 \ \mathrm{GeV}^2$.

At any fixed value of $Q^2$ in the range $0.25 \ \mathrm{GeV}^2 \leq Q^2 \leq 12 \ \mathrm{GeV}^2$, we extrapolate the isovector contribution to the physical point by exploring several functional forms, as described in section \ref{sec:phys_point_extrap}. During the chiral-continuum extrapolation, higher-order cutoff and chiral effects are treated independently in each Euclidean window, reflecting the different dominant sources of uncertainty in the various regions. In the HV region, we find that, even after applying tree-level improvement to the SD correlator to suppress discretization effects, higher-order terms in the lattice spacing are still required for a precise description of the data. The effect of tree-level improvement can be assessed quantitatively: at $Q^2=9 \ \mathrm{GeV}^2$, without it, we obtain $\Pi_{\mathrm{sub, SD}}^{(3,3)}(Q^2)=1038(2)_{\rm stat}(12)_{\rm syst}$, while applying tree-level improvement yields $\Pi_{\mathrm{sub, SD}}^{(3,3)}(Q^2)=1043.9(1.4)_{\rm stat}(3.4)_{\rm syst}$. These results are fully consistent with each other, with the tree-level improvement leading to a significant reduction of the systematic uncertainty.  At the same time, the light quark mass dependence  is found to be mild, indicating that chiral effects are subdominant in this region. 
 
In contrast the LV region, dominating the total isovector contribution, approaches the continuum limit with a significantly flatter behaviour,  consistent with what was observed in \cite{Djukanovic:2024cmq} for the LD contribution to the muon anomalous magnetic moment.
We note that, within this region, models incorporating only leading-order lattice artefacts tend to be favoured in the model average, compared to those including higher-order cutoff effects. Fits including $a^3$-terms generally lead to lower central values due to the curvature introduced in the extrapolation, but the resulting shift remains well within the final systematic uncertainty assigned. However, in this region we observe an increased sensitivity to the light quark mass. In particular, to get a good description of the data close to the physical point, it is crucial to include fits with chirally divergent terms. We find that models including $1 / \phi_2$ and $\log(\phi_2)$ terms dominate the  average in this virtuality region, while the inclusion of a third parameter describing the chiral dependence results in a negligible shift.

\begin{figure}
	\centering
	\includegraphics[scale=0.3]{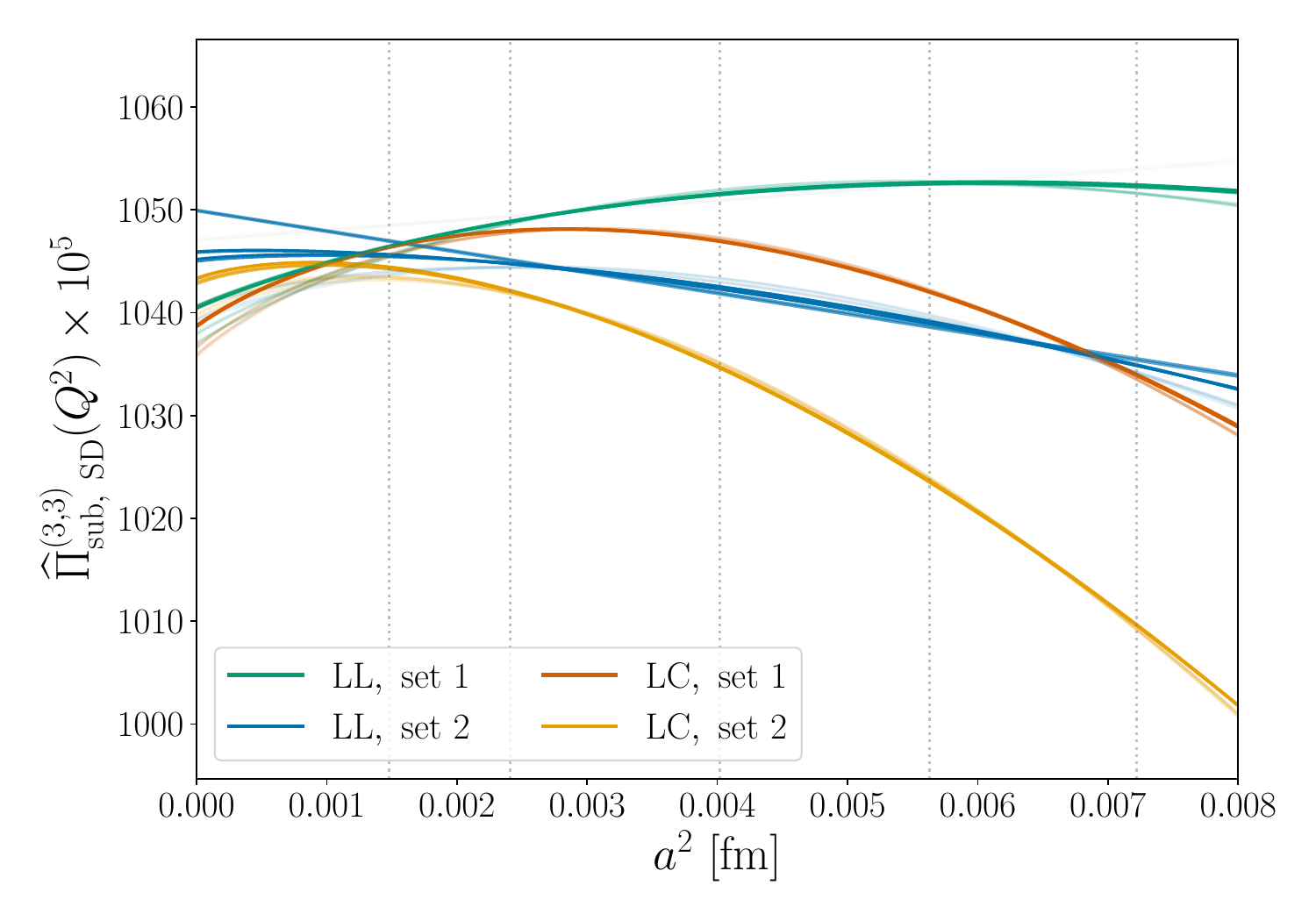}
	\includegraphics[scale=0.31]{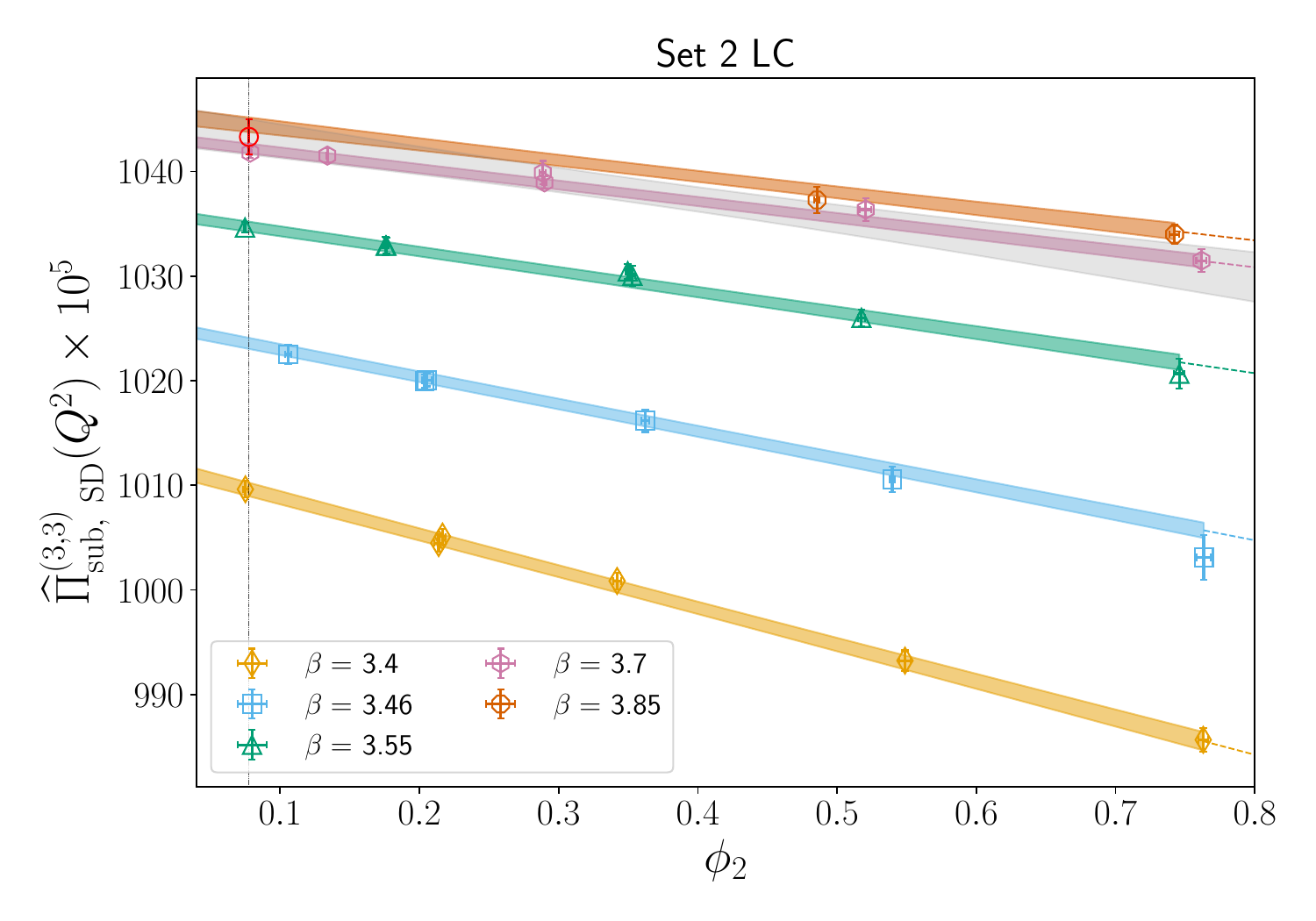}
	\caption{Illustration of fits to the  $\widehat{\Pi}_{\mathrm{sub}}^{(3,3)}(Q^2)$ isovector contribution in the HV  region. Specifically, we show results for the SD Euclidean window, where tree-level improvement is applied. \textit{Left:} continuum limit behaviour for four sets of data based on different improvement schemes and discretizations of the vector current. Each line corresponds to a single fit, with the opacity associated to the weights as given by our model average prescription. Dashed vertical lines denotes the lattice spacings used in this work. \textit{Right:} chiral approach to the physical pion mass for one of the  fits with the highest weight, and with $\chi^2/\chi^2_{\mathrm{exp}}=0.78$. Data points are projected to $\phi_4^{\mathrm{phys}}$. Coloured lines denote the chiral trajectories at finite lattice spacing, while the grey band shows the dependence on $\phi_2$ in the continuum. Results are shown  for $Q^2= 9\ \mathrm{GeV}^2 $ and $Q_m^2= 9\  \mathrm{GeV}^2$.  }
	\label{fig:isov_high_q_cl}
\end{figure}

\begin{figure} 
	\centering 
	\includegraphics[scale=0.3]{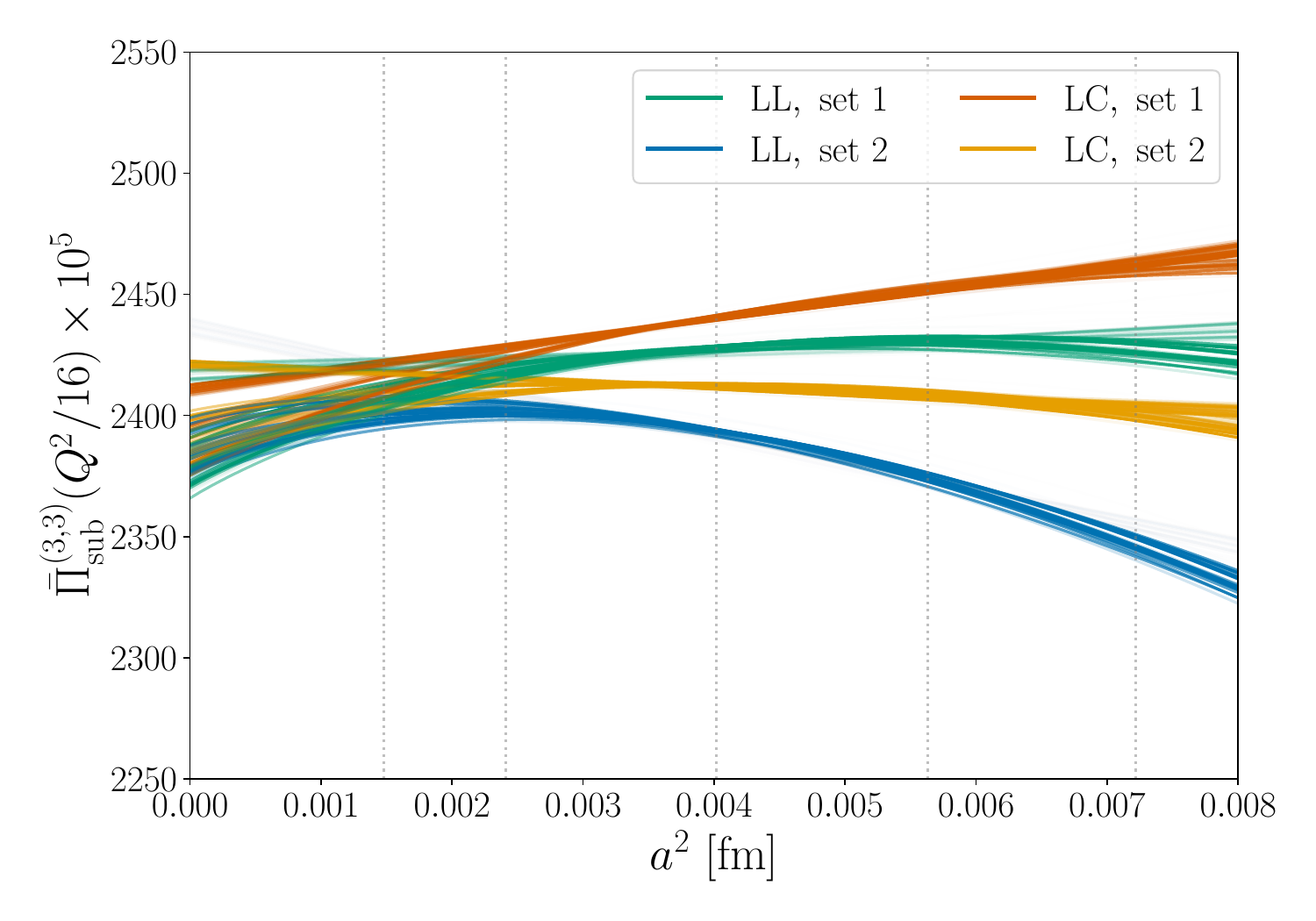}
	\includegraphics[scale=0.31]{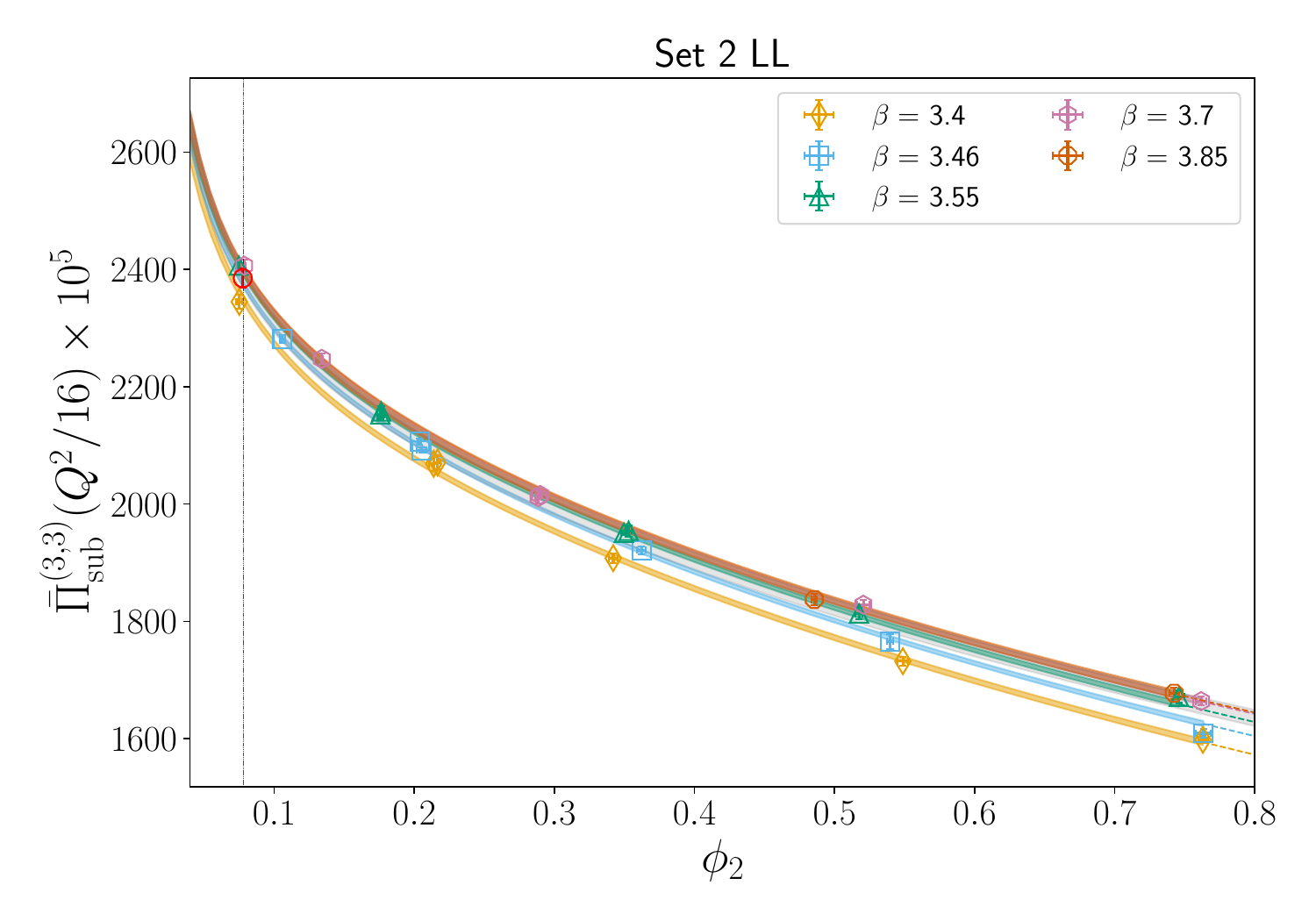}
	\caption{ Same as figure~\ref{fig:isov_high_q_cl} for the isovector LV contribution $\bar\Pi^{(3,3)}_{\mathrm{sub}}(Q^2/16)$ as determined from the subtracted kernel in Eq.~\eqref{eq:sub_kernel_QLD} . Results are shown  for $Q^2= 9\ \mathrm{GeV}^2 $ and $Q_m^2= 9\  \mathrm{GeV}^2$, with $\chi^2/\chi^2_{\mathrm{exp}}=1.1$ for the best fit model.  }  
	\label{fig:isov_low_q_cl}	
\end{figure}

In figures~\ref{fig:isov_high_q_cl} and~\ref{fig:isov_low_q_cl} we
show a summary of our fits for
$\widehat{\Pi}^{(3,3)}_{\mathrm{sub,SD}}(Q^2)$ and
$\bar{\Pi}^{(3,3)}_{\mathrm{sub}}(Q^2/16)$, in the HV and LV regions,
respectively. The left panels show the approach to the continuum
limit, projected to physical pion mass, for the two discretizations
of the vector current and each set of improvement coefficients. Each line depicts a single fit, with the opacity given by the corresponding weight as determined from our information criterion. As mentioned above, we observe that higher-order lattice artefacts are required to properly describe the continuum approach for $\widehat{\Pi}^{(3,3)}_{\mathrm{sub,SD}}(Q^2)$ in the HV region, while the LV contribution $\bar{\Pi}^{(3,3)}_{\mathrm{sub}}(Q^2/16)$ presents a flatter behaviour well described by leading-order lattice artefacts. 

The right panels show the light quark mass dependence for the best
fits according to our model average prescription, for two choices of
improvement coefficients and current discretization. Data points are
projected to $\phi_4^{\mathrm{phys}}$. Coloured lines represent the
chiral trajectories at finite lattice spacing, with the grey bands
showing the light quark mass dependence in the continuum limit. We
note that the chiral behaviour is well constrained over the full range
of light quark masses, with three ensembles in the vicinity of the
physical point. For the HV term (see figure~\ref{fig:isov_high_q_cl})
we observe a mild scaling on $\phi_2$, while
$\bar{\Pi}^{(3,3)}_{\mathrm{sub}}(Q^2/16)$ shows a stronger dependence
on the light quark mass (see figure~\ref{fig:isov_low_q_cl}), with fits including chirally divergent terms dominating the model average.

\subsubsection{Perturbative evaluation of $b^{(3,3)}(Q^2,Q_m^2)$}
The subtracted piece $b^{(3,3)}$ defined in Eq.~\eqref{eq:b33_def} can be computed reliably using massless perturbation theory. Similarly to \cite{Kuberski:2024bcj}, we evaluate the Adler function using the perturbative coefficients from \cite{Baikov:2010je} up to $\alpha_s^3$  and from \cite{Baikov:2008jh} for $O(\alpha_s^4)$. The computation is performed in the $N_f=3$ massless theory with $\Lambda_{\overline{\mathrm{MS}}} = 0.338(10) \ \mathrm{GeV}$ \cite{FlavourLatticeAveragingGroupFLAG:2024oxs}.
We observe a fast convergence of perturbation theory: taking $Q^2=5 \ \mathrm{GeV}^2$  and $Q_m^2= 9 \ \mathrm{GeV}^2$ we obtain, from the tree-level prediction to the highest perturbative order
\begin{equation}
	b^{(3,3)}(5\ \mathrm{GeV}^2,9\ \mathrm{GeV}^2) = 
	\{243.85, 
	\ 260.65, 
	\ 262.55, 
	\ 263.07,
	\ 263.34
	\}
	\ \times 10^5.
\end{equation}
To estimate the total uncertainty of the perturbative calculation we add in quadrature the error from $\Lambda_{\overline{\mathrm{MS}}}$ and the truncation error arising from the size of the last $O(\alpha_s^4)$ perturbative term. Results for several values of $Q^2$ employed in this work are listed in table~\ref{tab:b33_bcc_perturb_res}.

Eventually, after combining the lattice results for $\bar{\Pi}_{\mathrm{sub}}^{(3,3)}$ with the perturbative evaluation of $b^{(3,3)}$, we obtain for the isovector contribution at $Q^2=9\ \mathrm{GeV}^2$ in units of $10^{-5}$
\begin{eqnarray}
	\widehat{\Pi}^{(3,3)}(Q^2) &=& 1963.8 (2.1)_\mathrm{stat}(3.2)_{\mathrm{syst}}(0)_\mathrm{scale}(0)_{\mathrm{FV}}[3.8],
	\\
	\widehat{\Pi}^{(3,3)}(Q^2/4) &=& 1961.5  (3.6)_\mathrm{stat}(3.5)_{\mathrm{syst}}(0.7)_\mathrm{scale}(0)_{\mathrm{FV}}[5.1],
	\\
	\bar{\Pi}^{(3,3)}(Q^2/16) &=&2465 (12)_\mathrm{stat}(14)_{\mathrm{syst}}(10)_\mathrm{scale}(3)_{\mathrm{FV}}[21] ,
\end{eqnarray} 
for the HV, MH and LV regions, respectively. Here the first error is
statistical, the second is the systematic uncertainty arising from the
model variation, and the third is the uncertainty due to the physical
scale $t_0^{\mathrm{phys}}$. The fourth error is associated with the finite volume correction from $L_{\mathrm{ref}}$ to $L=\infty$ as computed in section \ref{sec:fvc}. 

\begin{table}		
			\centering
			\renewcommand{\arraystretch}{1.3}
		\begin{tabular}{c | c c c}
			\toprule
			$Q^2 \ [\mathrm{GeV}^2]$ & $b^{(3,3)}_{\mathrm{HV}}(Q^2, Q_m^2)$ & $b^{(3,3)}_{\mathrm{MV}}(Q^2/4, Q_m^2)$  & $\bar{b}^{(3,3)}_{\mathrm{LV}}(Q^2/16, Q_m^2)$\\
			\hline
			
			$4.0$ & 210.68(20)(22)& 52.669(50)(55) & 17.557(16)(19) \\
			
			$5.0$ &263.35(24)(28) & 65.837(61)(69) & 21.946(20)(24) \\
			
			$6.0$ & 316.02(29)(33) & 79.004(73)(83) &  26.335(24)(28) \\
			
			$7.0$ & 368.69(34)(39)& 92.172(86)(97) & 30.724(29)(32) \\
			
           $8.0$ & 421.36(39)(44) & 105.34(10)(11)& 35.113(33)(37)\\
			
			$9.0$ & 474.03(44)(50)& 118.51(11)(12) & 39.502(37)(42) \\
	\bottomrule
		\end{tabular}
	\caption{The subtraction term $b^{(3,3)}(Q^2, Q_m^2)$ for the isovector correlator (in units of $10^{-5}$) as obtained perturbatively from the $N_f=3$ massless Adler function computed up to $O(\alpha_s^4)$. Results are shown for the three virtuality regions HV, MV and LV. The first error denotes the uncertainty induced by $\Lambda_{\overline{\mathrm{MS}}}$, while the second error represents the absolute size of the last $O(\alpha_s^n)$ included in the computation and it provides an estimate of the truncation error of the perturbative series.}
	\label{tab:b33_bcc_perturb_res}
\end{table}

\subsection{The isoscalar contribution}
\begin{figure}
	\centering
	\includegraphics[scale=0.3]{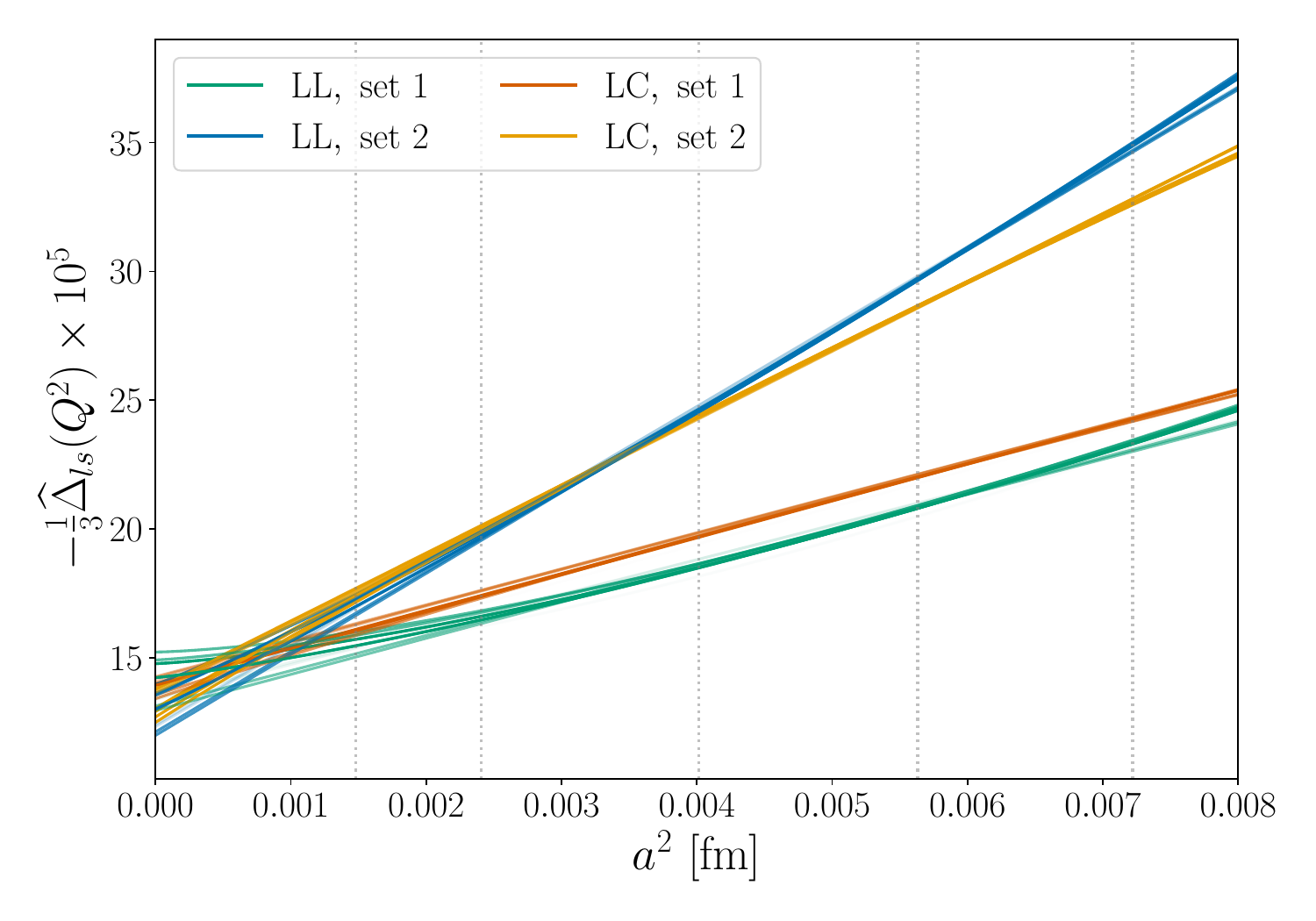}
	\includegraphics[scale=0.31]{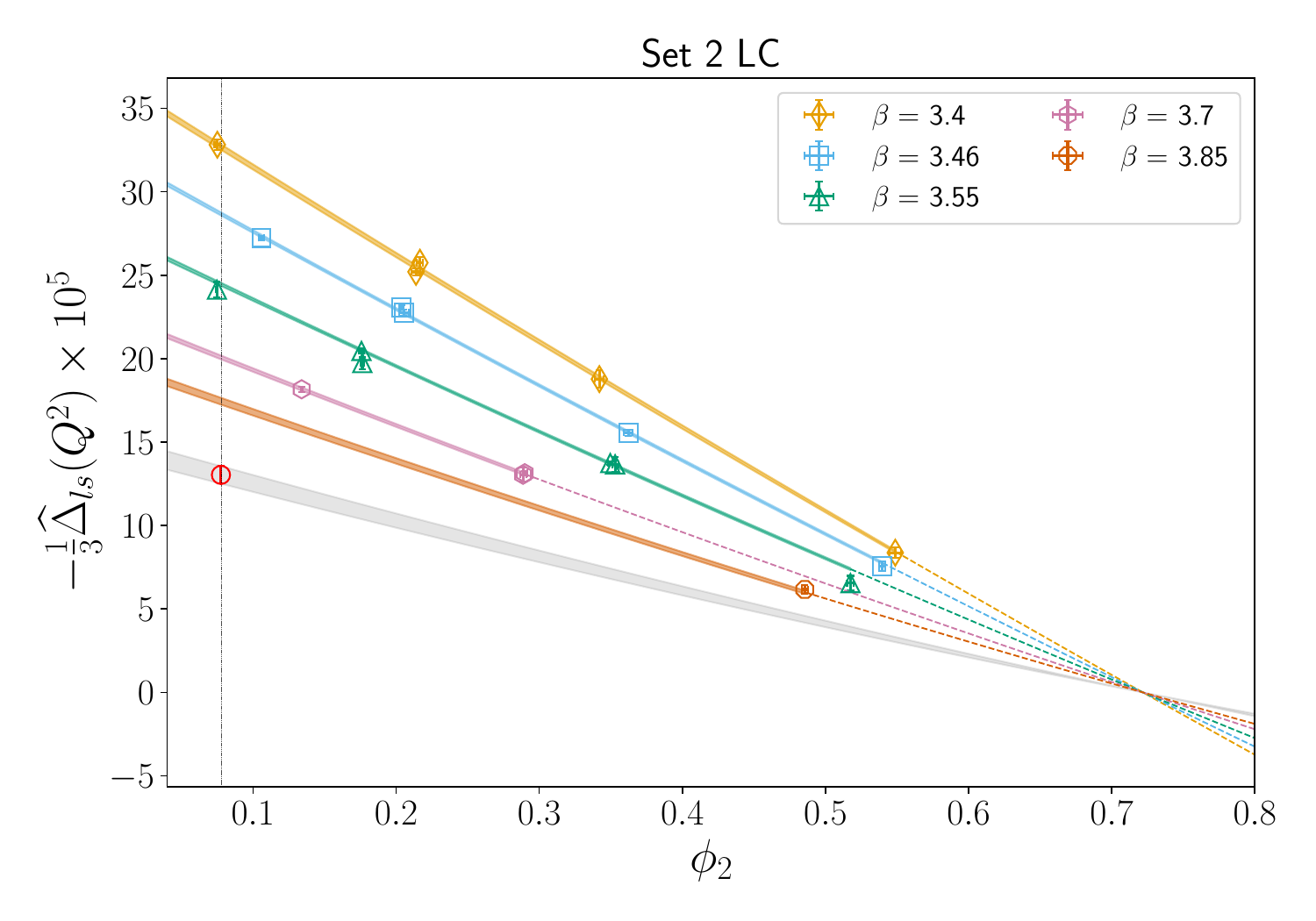}
	\caption{Same as figure~\ref{fig:isov_high_q_cl} for the term $\hat{\Delta}_{ls}(Q^2)$ used to extract the HV isoscalar contribution.  Results are shown  for $Q^2= 9\ \mathrm{GeV}^2$, with $\chi^2/\chi^2_{\mathrm{exp}}=1.16$ for the best fit model.}
	\label{fig:isoscal_SD}
\end{figure}

The  isoscalar contribution is  computed  through the telescopic series  introduced in Eq.~\eqref{eq:hvp_splitting}.
 In the HV piece, we employ the decomposition as given in Eq.~\eqref{eq:isoscalar_decomposition}  to extract  $\widehat{\Pi}^{(8,8)}(Q^2)$. In practice, only the quantity $\widehat{\Delta}_{ls}$ has to be computed anew following Eq.~\eqref{eq:delta_ls}.
 This quantity vanishes by definition at the $\rm SU(3)$-symmetric point and is expected  to scale proportionally to $m_s - m_l$ at leading order. The chiral-continuum extrapolation is therefore performed using the following parametrization
\begin{equation}
	\widehat{\Delta}_{ls}(Q^2)(\phi_\delta, \phi_4, X_a) = \phi_\delta\left(
	\gamma_1 + \gamma_2\phi_\delta + \beta_2X_a^2 + \beta_3X_a^3 + \gamma_0 \phi_4
	\right),
	\label{eq:su3_fit_ansatz}
\end{equation}
where $\phi_\delta = \phi_4 - \frac{3}{2}\phi_2$. This  ensures that
all cutoff effects are suppressed by $\phi_\delta$ in the proximity of
the $\rm SU(3)$-symmetric point. To describe the lattice spacing
dependence we explore the same functional forms introduced in section
\ref{sec:phys_point_extrap}, multiplied by the parameter $\phi_\delta$
to describe the suppression at the $\rm SU(3)$-symmetric point.  Multiple fit ansätze are then formed by selectively setting  some of the coefficients in the parametrization to zero.
 An illustration of the chiral-continuum behaviour  for
 $\widehat{\Delta}_{ls}$ is shown in figure~\ref{fig:isoscal_SD}. The
 approach to the continuum for all the functional forms explored in
 this work and evaluated at the physical light and strange quark
 masses is given in the left panel, while  the right panel shows the
 chiral dependence at finite lattice spacing. The intersection point
 of the curves reflects the $\widehat{\Delta}_{ls}=0$ constraint at
 the $\rm SU(3)$-symmetric point. As we move further away from this point the suppression is removed and cutoff effects grow. Given the result at $Q^2=9 \ \mathrm{GeV}^2$ in units of $10^{-5}$
\begin{equation}
	\frac{1}{3}\widehat{\Delta}_{ls}(Q^2) = - 13.16(40)_\mathrm{stat}(74)_{\mathrm{syst}}(10)_\mathrm{scale}[85] ,
\end{equation} 
obtained from our model average prescription, together with the result for the isovector contribution in the HV region, we quote for the HV isoscalar at the same value of $Q^2$
\begin{equation}
	\frac{1}{3}\widehat{\Pi}^{(8,8)}(Q^2) = 641.5 (0.9)_\mathrm{stat}(1.2)_{\mathrm{syst}}(0)_\mathrm{scale}[1.5] .
\end{equation} 

 On the other hand, the MV and LV isoscalar contributions are computed
 in a straightforward manner using the non-subtracted kernels in
 Eq.~\eqref{eq:non_sub_kernel_QSD_QID} and
 Eq.~\eqref{eq:non_sub_kernel_QLD}, respectively.  At the $\rm SU(3)$-symmetric point, where the quark-disconnected contribution vanishes, the isoscalar and  isovector terms coincide.   As outlined in section \ref{sec:noise_redduction}, we apply the bounding method to mitigate the signal-to-noise problem.  This is found to be particularly effective in the LV region, where the signal extends to larger Euclidean times. In contrast, in the MV region, the bounding method leads only to a moderate reduction of the statistical uncertainty. Contrary to the isovector channel,  we find that the  data  show a milder chiral dependence as the divergent behaviour of light-connected and disconnected pieces cancels in the isoscalar channel. In both regions, we explore similar functional forms as the ones considered in the isovector channel. We observe that the  chiral dependence is well described by fits that do not include chirally divergent terms, and the inclusion of additional parameters describing the light quark mass dependence results in a subleading shift  well within the statistical uncertainty. Fits including mass-dependent cutoff effects lead to better fit quality and dominate our model average, particularly in the MV region. Additionally we find that, in order to accurately describe ensembles along the chiral trajectory with fixed strange quark mass $m_s = m_s^{\mathrm{phys}}$, it is necessary to include a  higher-order term in $\phi_4$. Following our model average prescription, we arrive at the results
 \begin{eqnarray}
 	\frac{1}{3}\widehat{\Pi}^{(8,8)}(Q^2/4) &=& 575.7 (2.5)_\mathrm{stat}(2.0)_{\mathrm{syst}}(0.8)_\mathrm{scale}[3.3],
 	\\
 	\frac{1}{3}\bar{\Pi}^{(8,8)}(Q^2/16) &=& 519.3 (5.4)_\mathrm{stat}(5.0)_{\mathrm{syst}}(1.7)_\mathrm{scale}[7.5],
 \end{eqnarray} 
 at $Q^2=9\ \mathrm{GeV}^2$ in units of $10^{-5}$.
 
\begin{figure}
	\centering
	\includegraphics[scale=0.3]{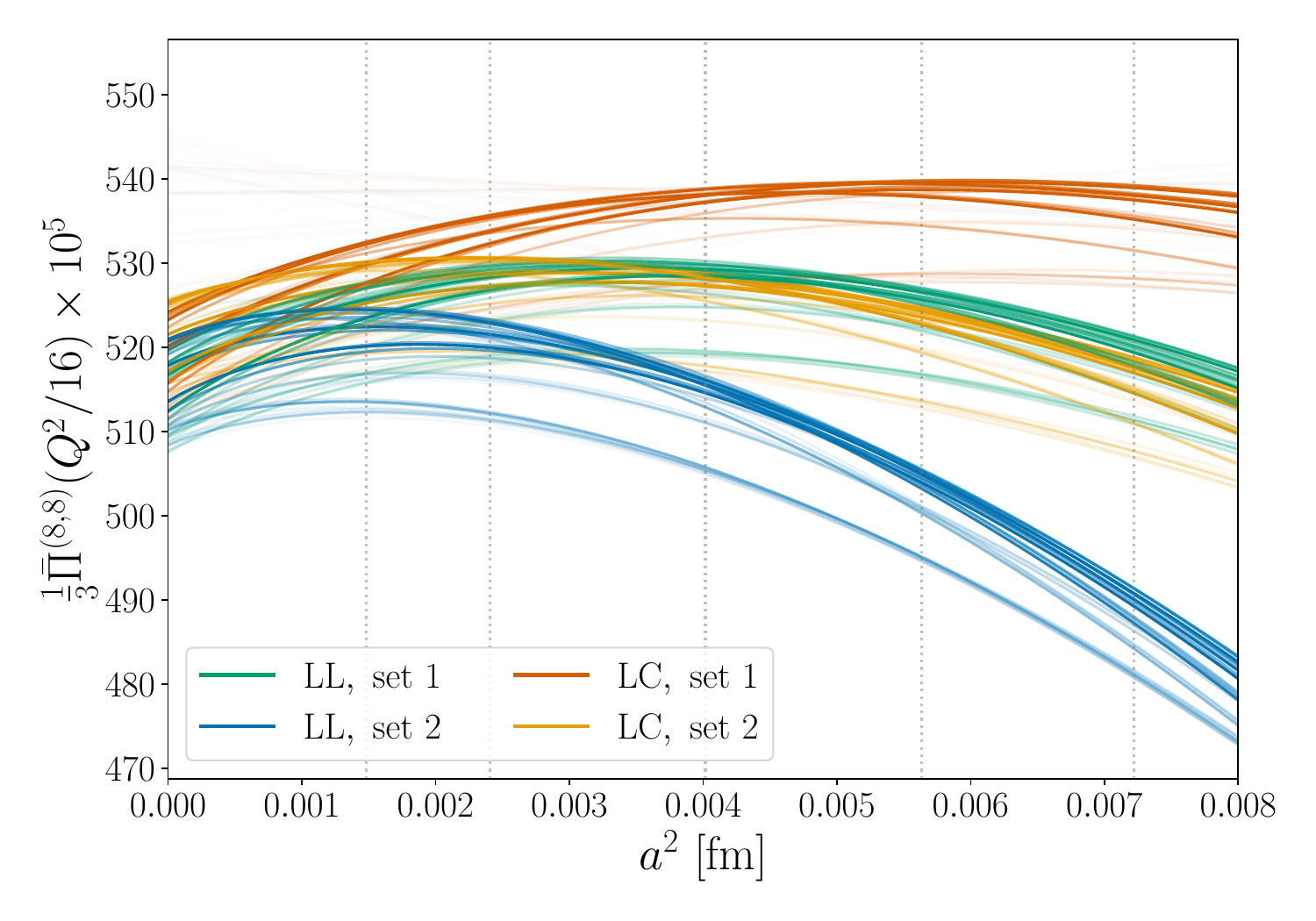}
	\includegraphics[scale=0.31]{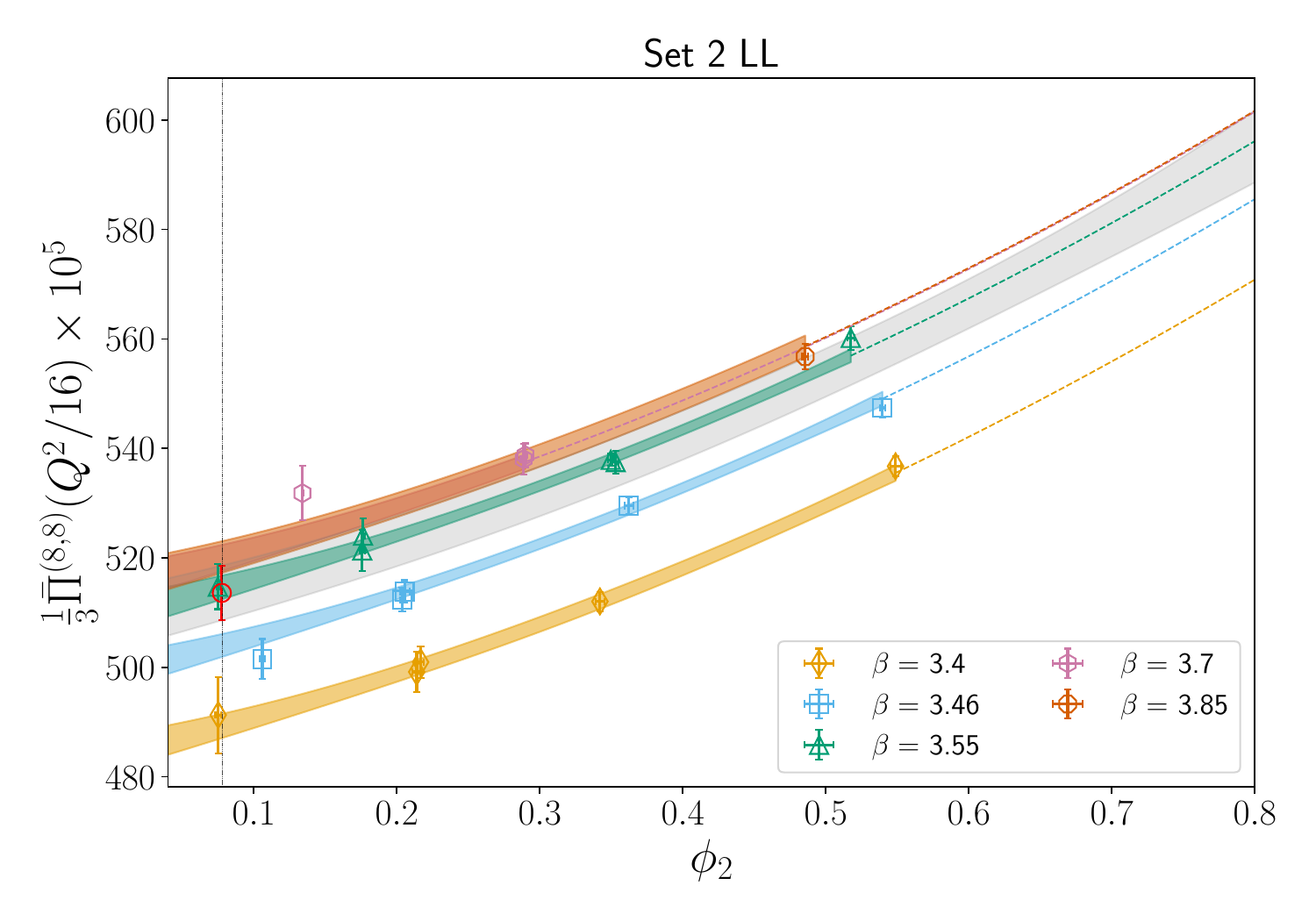}
	\caption{ Same as figure~\ref{fig:isov_high_q_cl} for the isoscalar LV contributions $\bar{\Pi}^{(8,8)}(Q^2/16)$. Results are shown  for $Q^2= 9\ \mathrm{GeV}^2$, with $\chi^2/\chi^2_{\mathrm{exp}}=0.82$ for the best fit model.  }
	\label{fig:isoscal_LD} 
\end{figure}

 The left-hand side of figure~\ref{fig:isoscal_LD} shows the approach
 to the continuum in the LV region for the four set of data
 considered, while the right panel illustrates one of the best fit for
 set 2 and $LL$ discretization of the vector current.

\subsection{The charm-connected contribution}
The charm-connected contribution on the lattice is evaluated employing the subtracted kernel as introduced in  Eq.~\eqref{eq:sub_kernel_QSD_QID} for the HV and MV regions, and the subtracted kernel in Eq.~\eqref{eq:sub_kernel_QLD} in the LV region. The tuning of the charm quark mass to match the physical $m_{D_s}=1968.47 \ \mathrm{MeV}$ meson  mass is described in detail in \cite{Ce:2022kxy, Gerardin:2019rua}. Following the chiral-continuum extrapolation, a small shift is applied to adapt the tuning of the charm quark  hopping parameter to the updated value of $t_0^\mathrm{phys}$ used in this work. The charm-connected contribution shows a strong curvature as approaching the continuum that requires the inclusion of higher-order terms beyond $a^2$ together with mass-dependent cutoff effects. In line with our previous studies, we only include the $LC$ discretization of the current in the model average as it shows significantly milder cutoff effects.  When evaluating the charm contribution, we choose to use $t_0^{\mathrm{sym}}$ from \cite{Bussone:2025wlf}  instead of the ensemble-specific $t_0$ to set the $Q^2$ input in the TMR kernel and to determine the value of $\phi_2$ for each ensemble entering the chiral continuum extrapolation. While using $t_0^{\mathrm{sym}}$ introduces correlations among ensembles at the same lattice spacing, it significantly reduces the light-quark mass dependence of the charm-connected contribution.
 
Quark mass effects in the subtraction function $b^{(c,c)}$, defined in Eq.~\eqref{eq:charm_bcc}, are also computed on the lattice by evaluating non-perturbatively $\Delta_{lc}b$ using the non-subtracted kernel in Eq.~\eqref{eq:non_sub_kernel_QSD_QID}. The same quantity can also be extracted in massive perturbation theory. As previously observed in \cite{Kuberski:2024bcj}, the perturbative and lattice results yield similar central values, but the perturbative prediction carries a significantly larger uncertainty, dominated by the absolute size of the last $O(\alpha_s^n)$ term included in the massive perturbative expansion. In contrast, the uncertainty in the lattice result, driven by the continuum extrapolation, is found to be smaller at the value of $Q_m$ considered in this work.
Eventually, combining the non-perturbative calculations of $\Pi_{\mathrm{sub}}^{(c,c)}$ and $\Delta_{lc}b$ with the perturbative evaluation for $b^{(3,3)}$, we obtain at $Q^2=9 \ \mathrm{GeV}^2$
 \begin{eqnarray}
\frac{4}{9} 	\widehat{\Pi}^{(c,c)}(Q^2) &=& 711.9 (9.4)_\mathrm{stat}(9.2)_{\mathrm{syst}}(1.8)_\mathrm{scale}[13],
 	\\
\frac{4}{9}	\widehat{\Pi}^{(c,c)}(Q^2/4) &=& 262.5 (4.1)_\mathrm{stat}(3.0)_{\mathrm{syst}}(0.6)_\mathrm{scale}[5.1],
	\\
\frac{4}{9}	\bar{\Pi}^{(c,c)}(Q^2/16) &=&99.5 (1.7)_\mathrm{stat}(1.8)_{\mathrm{syst}}(0.3)_\mathrm{scale}[2.5],
\end{eqnarray} 
for the HV, MV and LV regions, respectively, in units of $10^{-5}$

\begin{figure} 
	\centering
	\includegraphics[scale=0.3]{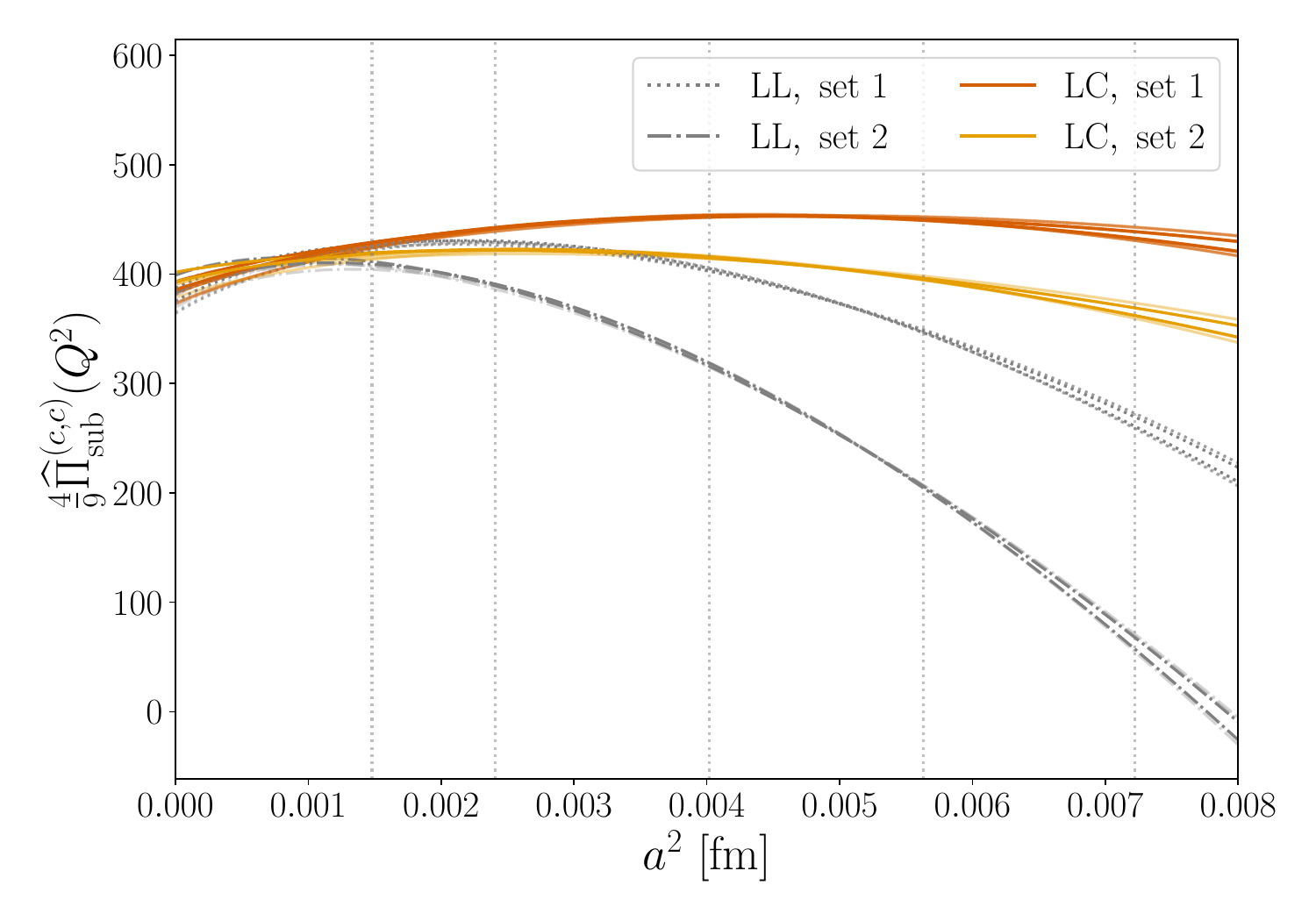}
	\includegraphics[scale=0.3]{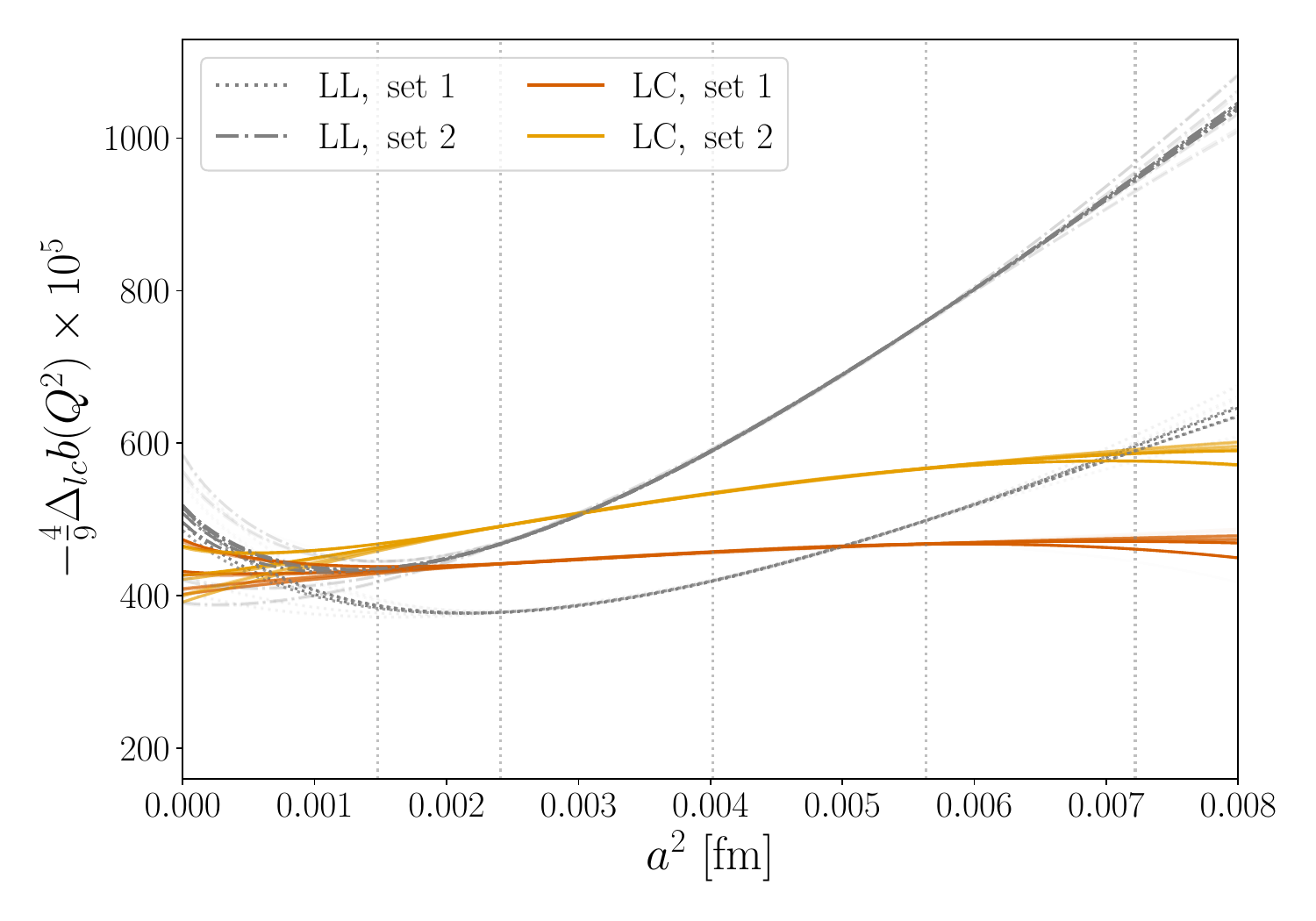}
	\caption{ Illustration of fits to the charm-connected contribution.  We show the continuum limit behaviour for the four sets of data based on different improvement schemes and discretizations of the vector current. Each line corresponds to a single fit, with the opacity associated to the weights as given by our model average prescription. \textit{Left}: fits to the $\widehat{\Pi}^{(c,c)}(Q^2)$ contribution in the HV region with   $Q^2= Q_m^2 = 9\ \mathrm{GeV}^2$, with $\chi^2/\chi^2_{\mathrm{exp}}=0.92$ for the best fit model. \textit{Right:} fits for the subtraction function $\Delta_{lc}b$  computed non-perturbatively at $4Q_m^2 = 36 \ \mathrm{GeV}^2$, with $\chi^2/\chi^2_{\mathrm{exp}}=0.82$ for the best fit model.  }
	\label{fig:charm_conn_cl}
\end{figure}

An illustration of the continuum behaviour for
$\widehat{\Pi}^{(c,c)}(Q^2)$ in the HV region and $\Delta_{lc}b$ is
shown on the left and right panels of figure~\ref{fig:charm_conn_cl},
respectively. We observe a good agreement in the continuum among the
different data sets for the two quantities, despite having
substantially different cutoff effects. Contrary to the other
contributions, we notice that the charm-connected HVP has only a mild
dependence on the light quark mass. 

\subsection{Z$\gamma$ mixing isoscalar contribution}
\begin{figure}
	\centering
	\includegraphics[scale=0.3]{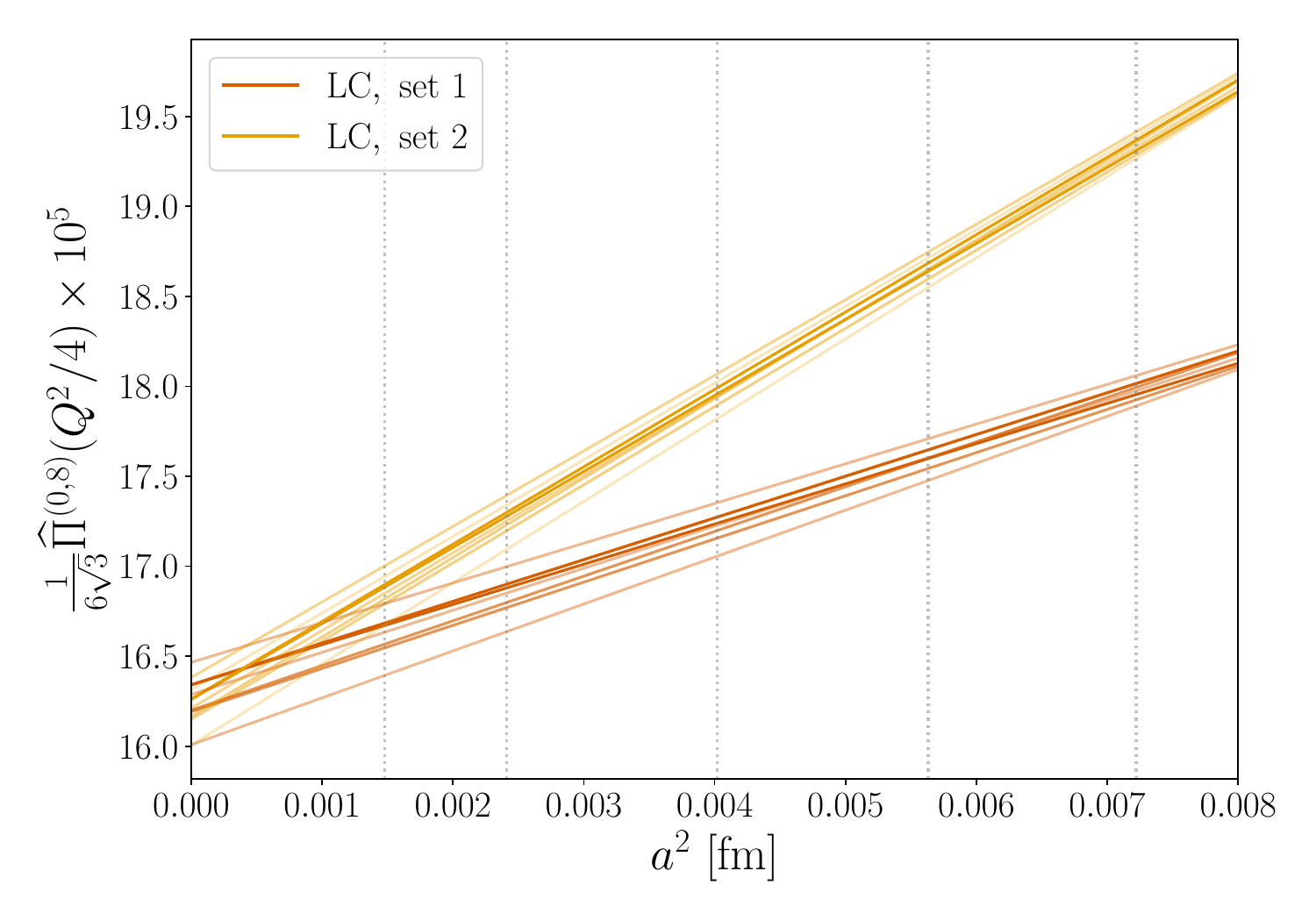}
	\includegraphics[scale=0.31]{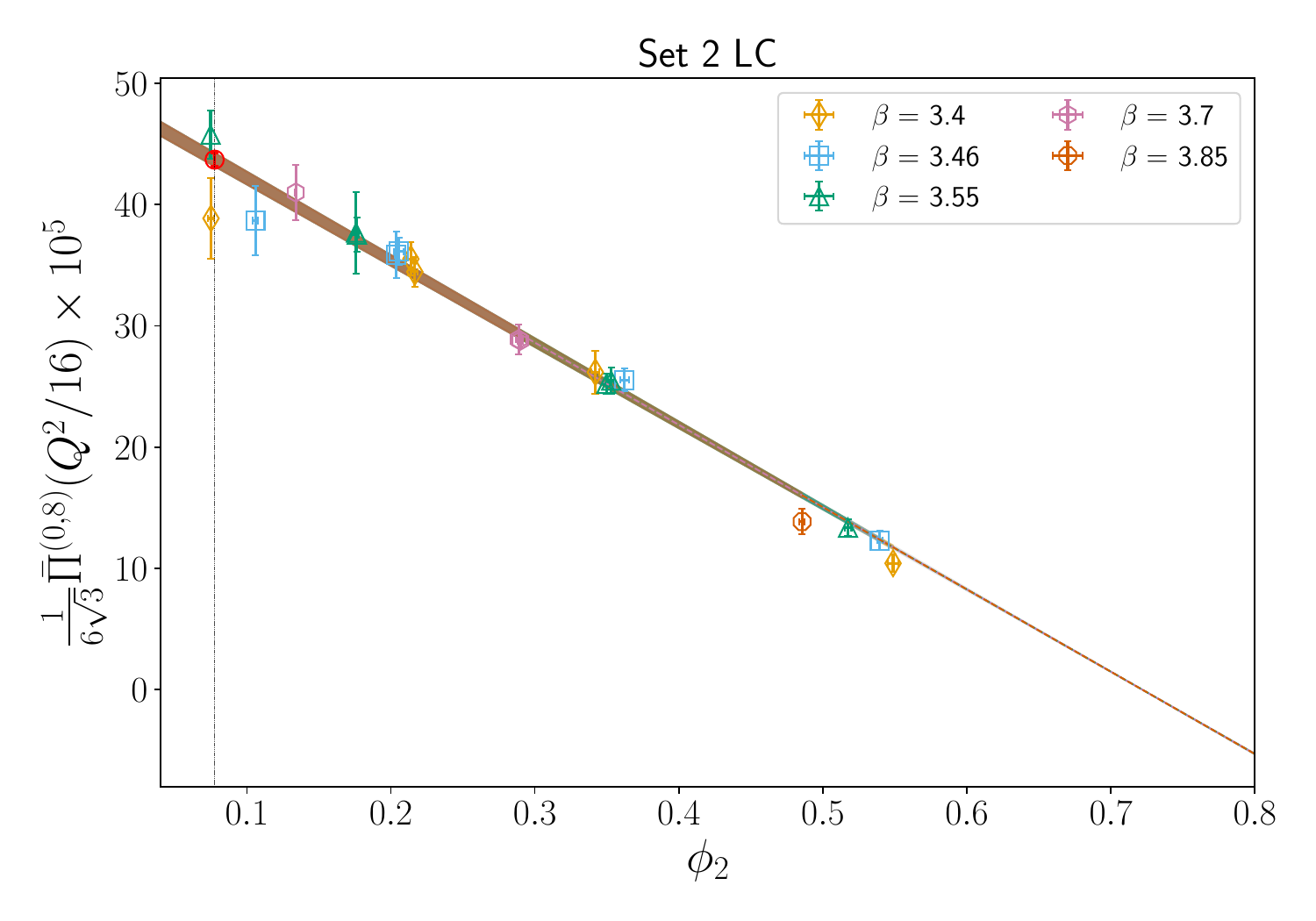}
	\caption{Illustration of fits to the mixing isoscalar contribution at $Q^2=9\ \mathrm{GeV}^2$. \textit{Left:} continuum limit behaviour for two available $LC$ discretizations for $\widehat{\Pi}^{(0,8)}(Q^2/4)$ in the MV region. Each line corresponds to a single fit, with the opacity associated to the weights as given by our model average prescription. \textit{Right}: chiral approach of $\bar{\Pi}^{(0,8)}(Q^2/16)$ in the LV region to the physical pion mass for one of the fits with the highest weight, with $\chi^2/\chi^2_{\mathrm{exp}}=0.68$ for the best fit model. 
	 }
	\label{fig:pi08_cl}
\end{figure}

Finally, the mixing isoscalar contribution $\bar{\Pi}^{(0,8)}$
entering the electroweak mixing angle is computed through
the splitting in Eq.~\eqref{eq:hvp_splitting}. At leading order this
term is proportional to $m_s - m_l$, thus vanishing linearly toward
the $\rm SU(3)$-symmetric point and ensuring that log-enhanced cutoff effects are not present. The bounding method is applied as outlined in section \ref{sec:noise_redduction} in the MV and LV regions to mitigate the uncertainty arising from the large Euclidean distances where the disconnected contribution entering $\bar{\Pi}^{(0,8)}$ suffers from a strong signal-to-noise problem.
We describe the chiral-continuum dependence along the same
lines as for the HV isoscalar $\Delta_{ls}$, by considering the fit ansatz
\begin{equation}
	\widehat{\Delta}_{ls}(Q^2)(\phi_\delta, \phi_4, X_a) = \phi_\delta\left(
	\gamma_1 + \gamma_2\phi_\delta + \beta_2X_a^2  + \gamma_0 \phi_4
	\right).
\end{equation}
In the LV region, we do not include an explicit dependence on the
lattice spacing in the base fit model, as no discretization effects
are visible within the current statistical precision. Nonetheless, we
allow for fits including $O(a^2)$ terms to probe potential systematic
effects. For this observable, we fit only the local-conserved
discretization, which avoids relying on the renormalized singlet local
current. Figure~\ref{fig:pi08_cl} summarizes the extrapolation to the physical point for this quantity: the left panel shows the continuum limit approach in the MV region for the two available data sets, while the right panel illustrates the chiral behaviour for one of the best fits in the LV region. The absence of leading $O(a^2)$ terms is consistent with the observation that data from different lattice spacings cannot be distinguished within uncertainties.

From the model average, we eventually quote for the mixing isoscalar contribution  at $Q^2 =9 \ \mathrm{GeV}^2$ in units of $10^{-5}$
\begin{eqnarray}
	\frac{1}{6\sqrt{3}}\widehat{\Pi}^{(0,8)}(Q^2) &=& 3.20 (8)_\mathrm{stat}(6)_{\mathrm{syst}}(0)_\mathrm{scale}[10],
	\\
	\frac{1}{6\sqrt{3}}\widehat{\Pi}^{(0,8)}(Q^2/4) &=& 16.20 (22)_\mathrm{stat}(11)_{\mathrm{syst}}(4)_\mathrm{scale}[25],
	\\
	\frac{1}{6\sqrt{3}}\bar{\Pi}^{(0,8)}(Q^2/16) &=& 44.08 (76)_\mathrm{stat}(50)_{\mathrm{syst}}(34)_\mathrm{scale}[98],
\end{eqnarray} 
for the HV, MV and LV regions, respectively.

\subsection{Isospin-breaking effects}\label{sec:ib_effects}
In this section, we describe our determination of the correction to $\bar \Pi(Q^2)$
due to electromagnetic and strong isospin-breaking (IB) effects.

We have performed a lattice calculation of the strong and electromagnetic
	isospin-breaking corrections to the quark-connected contribution to
	$\bar{\Pi}$, taking into account only effects in the valence sector.
	The setup follows Refs.~\cite{Ce:2022eix, Ce:2022kxy, Kuberski:2024bcj,
		Djukanovic:2024cmq}, where QCD+QED effects are included perturbatively
	around the isospin-symmetric theory using Monte Carlo reweighting combined
	with a leading-order expansion in the electromagnetic coupling $e^{2}$ and
	the quark-mass shifts $\Delta m_{u}$, $\Delta m_{d}$, and $\Delta m_{s}$
	relative to their isosymmetric values~\cite{deDivitiis:2013xla, Risch:2021hty, Risch:2019xio,Risch:2018ozp, Risch:2017xxe}.
	The additional correlation functions corresponding to quark-mass insertions and
	photon-exchange diagrams are computed in the non-compact formulation of
	lattice QED with the ${\rm QED}_{L}$ prescription~\cite{Hayakawa:2008an} to
	regularize infrared divergences.
	The physical point of QCD+QED is defined through the meson mass
	combinations $m_{\pi^{0}}^{2}$,
	$m_{K^{+}}^{2}+m_{K^{0}}^{2}-m_{\pi^{+}}^{2}$, and
	$m_{K^{+}}^{2}-m_{K^{0}}^{2}-m_{\pi^{+}}^{2}+m_{\pi^{0}}^{2}$ together with
	the fine-structure constant $\alpha$, following the hadronic
	renormalization scheme of Ref.~\cite{Risch:2021hty}.
	
	\begin{figure}
		\centering
		\includegraphics[width=.7\textwidth]{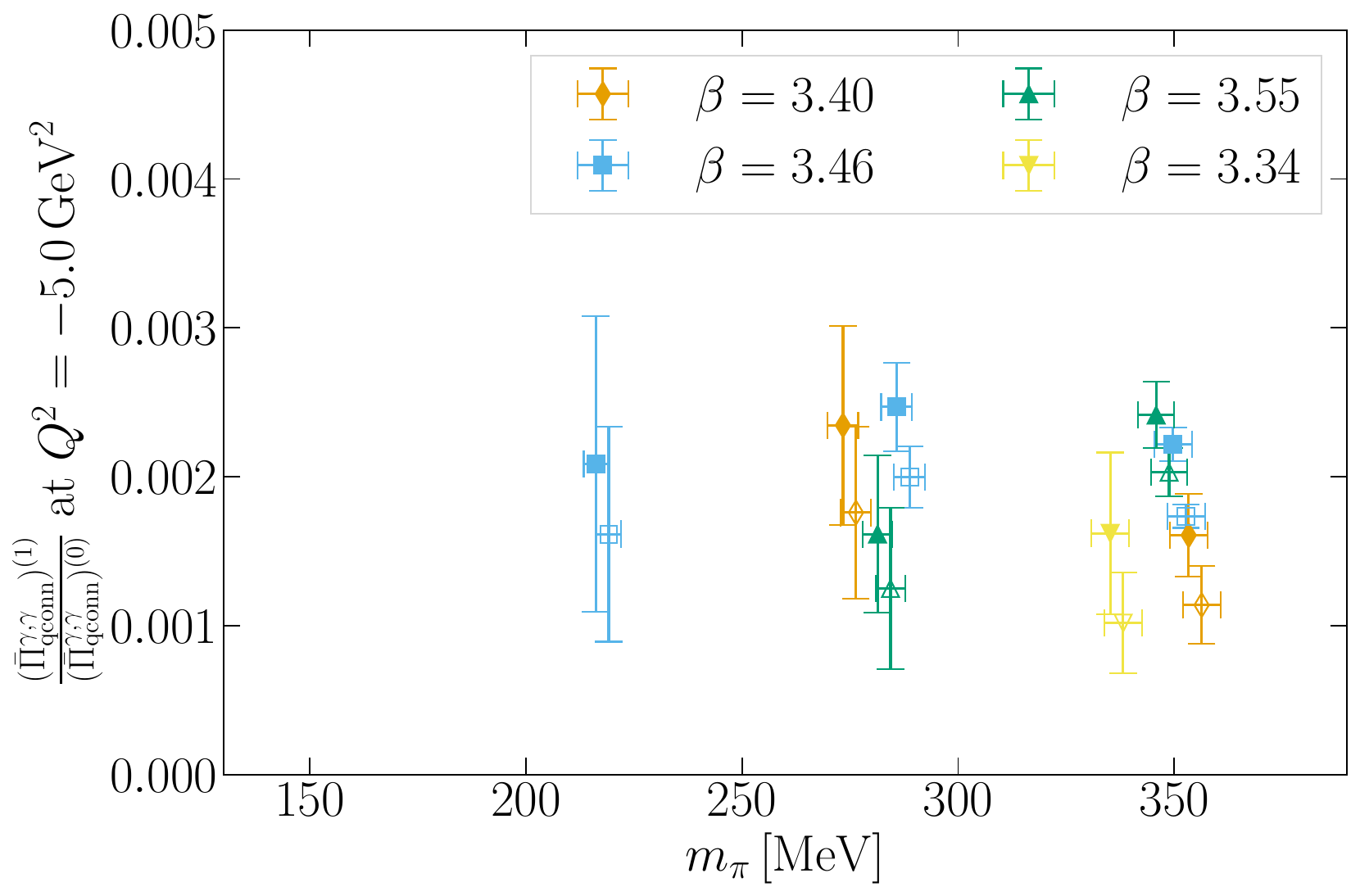}
		\caption{Relative effect from isospin-breaking on the quark-connected
		contribution to $\bar\Pi$ at $Q^2=-5\,\mathrm{GeV}^2$ as described in the
		text.
		Filled symbols correspond to the local-local discretization and open symbols
		to the local-conserved current. 
		The points are slightly displaced horizontally for better visibility. }
		\label{fig:isospin_breaking} 
	\end{figure}
	
	We have performed the computation on eight gauge ensembles
	(A654, H102, N101, N452, N451, D450, N203, and N200),
	covering four values of the lattice spacing down to $0.064\,\mathrm{fm}$
	and pion masses in the range from $350$ to $220\,\mathrm{MeV}$.
	In figure~\ref{fig:isospin_breaking} we show the relative size of the
	isospin-breaking correction to the quark-connected contribution to
	$\bar{\Pi}(Q^2=-5\,\mathrm{GeV}^2)$,
	where, as in the isosymmetric theory, both local–local and
	local–conserved discretizations of the vector current are employed.
	Within the statistical uncertainties, the data exhibit only a mild
	dependence on the pion mass and the lattice spacing in the explored
	parameter range and point to a four per~mil correction.

        In addition, we pursue a complementary approach to the lattice calculations described above.
	A phenomenological estimate of the correction to $a_\mu^{\rm hvp}$
	from electromagnetic effects was given in~\cite{Parrino:2025afq}, based on the formalism set up in~\cite{Biloshytskyi:2022ets},
	and an estimate of the strong isospin-breaking correction in~\cite{Erb:2025nxk}.
        Here we adapt those predictions to the subtracted vacuum polarization itself. We
	focus on two space-like virtualities, 9\,GeV$^2$ and 5\,GeV$^2$.
	Since the hadronic models used are meant to describe low-to-moderate energies,
	we will use their predictions at 5\,GeV$^2$ and rely on QCD perturbation theory for the subsequent 
	IB effects from 5 to 9\,GeV$^2$ in the vacuum polarization.
	Indeed, for the difference $\Pi(9{\,\rm GeV}^2) - \Pi(5{\,\rm GeV}^2)$, which is short-distance dominated,
	the perturbative prediction for e.m.\ effects amounts to a relative correction of about half a per~mil
	(see ~\cite{Kuberski:2024bcj}, Eq.\ (54), which is based on the perturbative calculation~\cite{Kataev:1992dg}).
	In absolute terms, this amounts to
	\begin{equation}\label{eq:9m5IB}
		\Delta_{\rm IB}\{\Pi(9{\,\rm GeV}^2) - \Pi(5{\,\rm GeV}^2)\} \simeq 0.7 \times 10^{-5}.
	\end{equation}
	We find strong isospin-breaking in this interval of virtualities to be entirely negligible, as they
	are of order $\Delta_{\rm SIB} R(s) = 18({\alpha_s}/{\pi}) (4/9 - 1/9) (m_u^2-m_d^2)/s$ in the $R$-ratio~\cite{Chetyrkin:1997qi}.

	\begin{table}
		\renewcommand{\arraystretch}{1.2}
		\centerline{\begin{tabular}{c c c c}
			\toprule
 contribution   &    $\overline\Pi(9{\,\rm GeV}^2)$  &  $\overline\Pi(5{\,\rm GeV}^2)\phantom{\Big|}$ &  $\overline\Pi(5{\,\rm GeV}^2)$ \\
		$(\times10^5)$	& $M_\pi=135\,$MeV       &   $M_\pi=135\,$MeV    & $M_\pi=291\,$MeV         \\
				\hline
				$\pi^0$ pole &     1.20    &  1.20  & 1.10 \\
				$\eta$ pole &     0.79    &   0.79 & 0.62 \\
				$\eta'$ pole &     1.16    & 1.15  & 1.58 \\
				$\eta_c$ pole &     0.51    & 0.40  & 0.40 \\
				$\pi^\pm$ loop &  $-9.25$    &  $-9.25$ & $-1.7$ \\
				\hline
				strong IB      &    13.3    &  13.2    &  12.0 \\
				\hline
				Total:           &     7.7      &   7.5   &  14.0    \\
				\bottomrule
		\end{tabular}}
		\caption{Corrections from IB effects to the subtracted vacuum polarization for two different virtualities,
			and for two different pion masses. Results based on a hadronic model, see the main text.\label{tab:IBhadmodel}}
	\end{table}
	
	Table \ref{tab:IBhadmodel} provides the quantitative predictions of
	the hadronic model for $M_\pi=291\,$MeV and for the physical pion
	mass. At the latter mass, the table also indicates the results for
	$Q^2=9$\,GeV$^2$, showing very little change in the prediction as
	compared to 5\,GeV$^2$. The single largest contribution is the strong
	isospin-breaking one. The (negative) charged pion loop contribution
significantly reduces the size of the correction at the physical point.
It is computed with vector-meson-dominance (VMD) pion form factors and
renormalized according to the FLAG scheme.\footnote{Unlike in the case of $a_\mu^{\rm hvp}$~\cite{Parrino:2025afq},
we observe little difference between the FLAG and the on-mass-shell renormalization scheme.}
Similarly, the pseudoscalar meson exchanges are computed using the
master formula derived in~\cite{Biloshytskyi:2022ets}, with a VMD ansatz for the transition form factor.

	A direct comparison between the hadronic model and our lattice data for the quark-connected contributions is possible
	at $M_\pi=291\,$MeV, corresponding to gauge ensemble N451. Here, according to the model calculations,
	the charged pion loop contribution is still small in magnitude, the strong IB contribution dominates and disconnected diagrams are relatively small.
        The lattice result at finite lattice spacing,  $11(3)\times 10^{-5}$ in the $(u,d,s)$ sector, is compatible with the model prediction\footnote{The model for the electromagnetic corrections can approximately be adapted~\cite{Parrino:2025afq}
        to the fact that the lattice results contain the quark-connected diagrams only,
        ensuring that the leading chiral behaviour of the connected diagrams is reproduced.  At $M_\pi=291\,$MeV, the model prediction
        for the electromagnetic effects would increase by about $1\times 10^{-5}$.}, $13.6\times 10^{-5}$ leaving out the $\eta_c$ pole contribution.
	
	Our final estimate for the correction to $\bar\Pi(9\,{\rm GeV}^2)$ from IB effects in the FLAG scheme is
	the sum of the entry in table \ref{tab:IBhadmodel} for the correction to $\bar\Pi(5\,{\rm GeV}^2)$ and of Eq.\ (\ref{eq:9m5IB}),
	\begin{equation}\label{eq:ib_effects}
		\Delta_{\rm IB} \bar\Pi(9\,{\rm GeV}^2) = (8.2\pm14.0) \times 10^{-5}.
	\end{equation}
	The generous uncertainty we have assigned to this correction is based 
	on the size of the (dominant) strong IB correction. It also covers the typical size of the quark-connected IB correction
        obtained in the lattice calculation.

\begin{table}[h!]	
	\centering
			\renewcommand{\arraystretch}{1.1}
		\begin{tabular}{c c c  }
			\hline\noalign{\smallskip}
			$Q^2 \ [\mathrm{GeV}^2]$ & $\bar\Pi^{(3,3)}$ & $\frac{1}{3}\bar\Pi^{(8,8)}$  \\
			\noalign{\smallskip}\hline\noalign{\smallskip}
			
			$0.5$ &  2310(11)(14)(9)[20] & 479.1(5.2)(4.9)(1.6)[7.3]  \\
			
			$1.0$ &3258(12)(11)(10)[19] & 740.4(6.4)(6.1)(2.9)[9.3]  \\ 
			
			$2.0$ & 4256(14)(14)(10)[23] & 1041.5(6.7)(5.3)(2.4)[9.0]  \\ 
			
			$3.0$ & 4835(15)(17)(10)[25] & 1228.2(7.0)(6.0)(2.5)[9.5] \\ 
			
			$4.0$ & 5257(14)(11)(10)[20] & 1364.3(6.8)(6.2)(2.9)[9.6]  \\
			
			$5.0$ &5572(14)(12)(10)[21] & 1468.5(6.6)(5.8)(2.0)[9.1]  \\
			
			$6.0$ & 5829(14)(13)(10)[21] & 1552.6(6.7)(5.5)(1.9)[8.9] \\ 
			
			$7.0$ & 6043(14)(14)(10)[22] & 1622.2(6.8)(5.4)(2)[8.9] \\ 
			
			$8.0$ & 6228(14)(15)(10)[23] & 1682.1(6.8)(5.4)(2.4)[9.0]  \\ 
			
			$9.0$ & 6390(15)(15)(10)[24] & 1736.5(6.9)(5.5)(2.5)[9.2]  \\ 
			
			$12.0$ &  6784(16)(17)(10)[25] & 1869.9(7.1)(6.2)(2.6)[9.7] \\ 
			\noalign{\smallskip}\hline \noalign{\smallskip}			
			 $Q^2 \ [\mathrm{GeV}^2]$   & $\frac{4}{9}\bar\Pi^{(c,c)}$ & $\frac{1}{6\sqrt{3}}\bar\Pi^{(0,8)}$  \\
			 \noalign{\smallskip}\hline\noalign{\smallskip}
			 
			 $0.5$  &  88.8(1.6)(1.6)(0.2)[2.3] & 41.85(76)(48)(34)[96] \\
			 
			 $1.0$  &  173.1(3.0)(1.7)(0.5)[3.5] & 52.42(87)(47)(44)[1.1] \\ 
			 
			 $2.0$  &  326.3(5.8)(2.7)(0.8)[6.4] & 59.44(81)(49)(39)[1.02] \\ 
			 
			 $3.0$  & 463.5(7.2)(5.1)(1.2)[8.9] & 61.94(82)(60)(41)[1.09] \\ 
			 
			 $4.0$  & 585.7(9.3)(6.8)(1.4)[12.0] & 61.91(91)(47)(43)[1.11]  \\
			 
			 $5.0$  & 699(11)(8)(2)[13] & 62.66(88)(42)(44)[1.07] \\ 
			 
			 $6.0$  & 803(12)(9)(2)[15] & 63.25(87)(44)(40)[1.05] \\ 
			 
			 $7.0$  & 900(14)(10)(2)[17] & 63.24(85)(47)(38)[1.04] \\ 
			 
			 $8.0$  & 990(14)(11)(2)[18] & 63.24(83)(49)(38)[1.04] \\ 
			 
			 $9.0$  & 1074(15)(12)(3)[19] & 63.48(83)(52)(38)[1.05] \\ 
			 
			 $12.0$ & 1298(18)(13)(3)[22] & 63.96(84)(60)(40)[1.10] \\ 
			 \bottomrule
		\end{tabular}
	\caption{Contribution to the running for the various channel at the physical point, in units of $ 10^{-5}$. The first quoted uncertainty corresponds to the statistical error, the second to systematics from model exploration, and the third to the scale-setting error. The final uncertainty in squared brackets is the sum in quadrature of the previous ones. }
	\label{tab:hvp_various_channel}
\end{table}

\begin{table}[h!]	
		\centering
		\renewcommand{\arraystretch}{1.1}
		\begin{tabular}{c c c  }
			\toprule
			$Q^2 \ [\mathrm{GeV}^2]$ & $\Delta\alpha_{\mathrm{had}}^{(5)}$ & $(\Delta\sin^2\theta_W)_{\mathrm{had}}$  \\
			\noalign{\smallskip}\hline\noalign{\smallskip}
			
			$0.5$ & 264.0(1.4)(1.4)(1.0)[2.2] & $-268.8(1.5)(1.5)(1.0)[2.3]$  \\
			
			$1.0$ & 382.9(1.5)(1.1)(1.3)[2.3] & $-390.7(1.6)(1.3)(1.2)[2.4]$  \\ 
			
			$2.0$ & 516.4(1.7)(1.5)(1.2)[2.6] & $-526.7(1.8)(1.6)(1.2)[2.7]$  \\ 
			
			$3.0$ & 599.4(1.9)(1.7)(1.3)[2.9] & $-609.8(2.0)(1.8)(1.2)[2.9]$ \\ 
			
			$4.0$ & 662.1(1.8)(1.4)(1.3)[2.7] & $-672.3(1.8)(1.4)(1.2)[2.6]$   \\
			 
			$5.0$ & 711.2(1.9)(1.4)(1.3)[2.7] & $-720.1(1.8)(1.5)(1.2)[2.6]$  \\
			
			$6.0$ & 752.3(2.0)(1.5)(1.3)[2.8] & $-759.6(1.9)(1.5)(1.2)[2.7]$ \\ 
			
			$7.0$ & 787.6(2.1)(1.6)(1.3)[3.0] & $-793.2(2.0)(1.6)(1.2)[2.8]$ \\ 
			
			$8.0$ & 818.5(2.2)(1.7)(1.3)[3.1] & $-822.4(2.0)(1.7)(1.2)[2.9]$ \\ 
			
			$9.0$ & 846.4(2.3)(1.8)(1.4)[3.2] & $-848.5(2.1)(1.8)(1.3)[3.0]$  \\ 
			
			$12.0$ & 916.2(2.4)(2.0)(1.5)[3.5] & $-913.1(2.2)(1.9)(1.3)[3.2]$ \\ 
			\bottomrule
		\end{tabular}
		\caption{Total HVP contribution to the running of $\alpha$ and $\sin^2\theta_W$ in isospin-symmetric QCD. The first quoted uncertainty is the  statistical error, followed by the systematic error arising from the model exploration and the scale setting error.  The final uncertainty in squared brackets is the sum in quadrature of the previous ones. Results are shown in units of $ 10^{-5}$.}
		\label{tab:dalpha_sin2_space_like_resl}
\end{table}

\section{Hadronic running of the couplings}\label{sec:had_run}
In this work, we evaluate the HVP functions $\bar\Pi^{(3,3)}, \bar\Pi^{(8,8)}, \bar\Pi^{(0,8)}, \bar\Pi^{(c,c)}$ at the physical point for several discrete values of the squared momentum transfer in the range $0.25\  \mathrm{GeV}^2\leq Q^2 \leq 12 \ \mathrm{GeV}^2$, as reported in table~\ref{tab:hvp_various_channel}. Compared to our previous work~\cite{Ce:2022eix}, we find good agreement for the isovector, isoscalar, and charm-connected contributions. A tension at the level of about $2\sigma$ is instead observed for the octet-singlet mixing contribution $\bar{\Pi}^{(0,8)}$, primarily driven by the inclusion of an additional fine lattice spacing and a substantially improved treatment of lattice artefacts. In addition, we determine the strange-quark contribution $\bar\Pi^{(s,s)}$, which does not directly enter our main analysis; the corresponding results are presented in  Appendix~\ref{app:strange_contribution}.  From these quantities, we construct  the  $\bar{\Pi}^{(\gamma,\gamma)}$ and $\bar{\Pi}^{(Z,\gamma)}$ HVP 	functions, which provide direct access to the  electroweak couplings at space-like momenta. 

Our determination of $\Delta\alpha_{\mathrm{had}}(-Q^2)$ and
$(\Delta\sin^2\theta_W)_\mathrm{had}$ includes the contributions from
$u,d,s$ and $c$ quarks.  To incorporate the missing $b$-quark effects,
we use results from the HPQCD collaboration for the lowest four time
moments of the HVP \cite{Colquhoun:2014ica}. The bottom contribution
is obtained by constructing  Padé approximants from these moments,
yielding  a small correction -- at most $0.4\%$ at $Q^2=12\ \mathrm{GeV}^2$ -- to the total hadronic running of the coupling. This contribution is added  to our lattice results to obtain a complete $N_f=5$ flavour  determination of the electroweak couplings. The final results for  $\Delta\alpha_{\mathrm{had}}^{(5)}(-Q^2)$ and $(\Delta\sin^2\theta_W)_\mathrm{had}$ in the space-like region are listed in table~\ref{tab:dalpha_sin2_space_like_resl}.
Though we have performed our lattice calculation with $N_f=2+1$ flavours of quarks,
we have estimated the charm quenching effects along the lines of~\cite{Ce:2022eix,Kuberski:2024bcj}
  and found them to be negligible.

\begin{figure}[h!]
	\centering
	\includegraphics[scale=0.42]{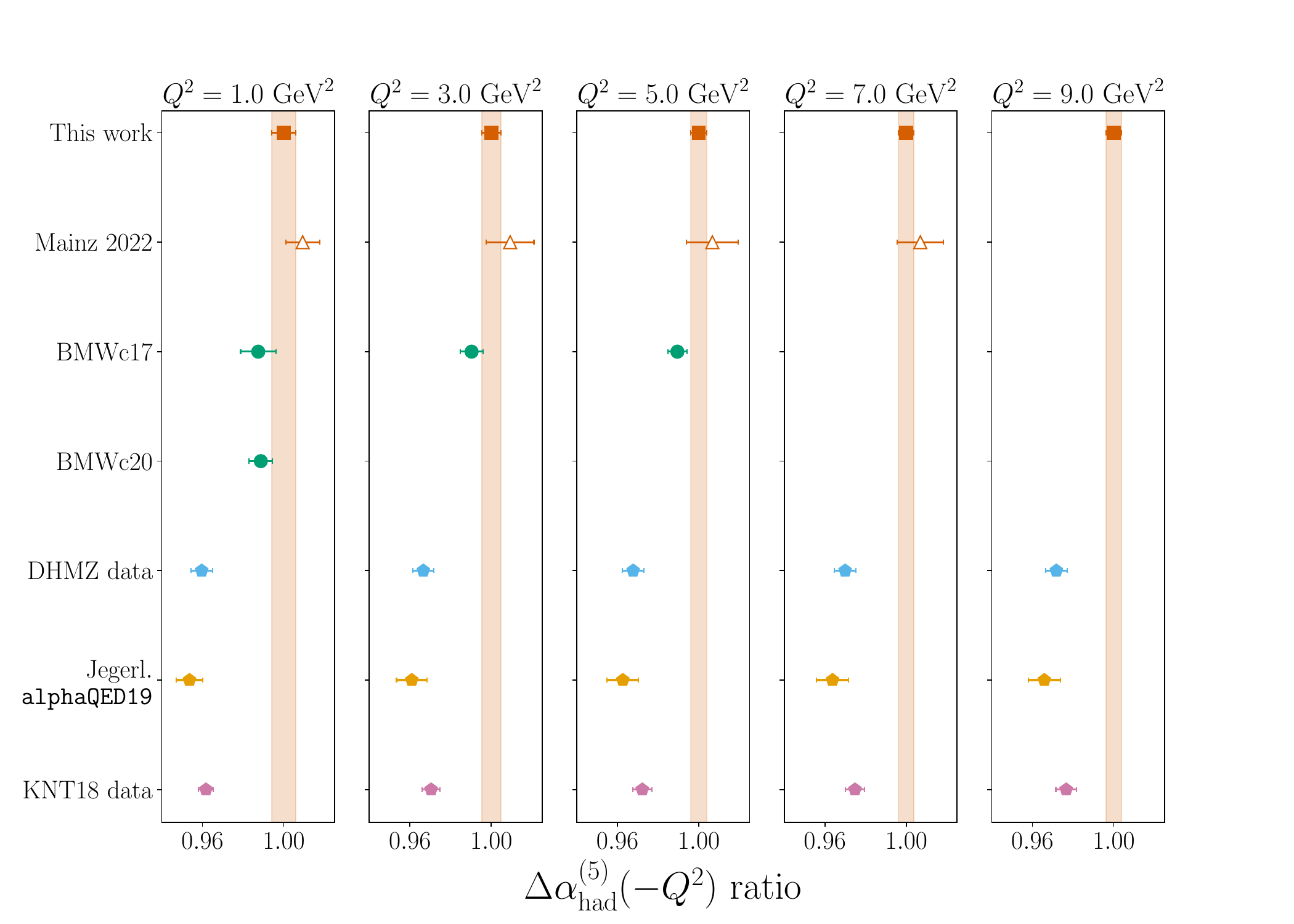}	
	\caption{Summary of results for the determination of $\Delta\alpha_{\mathrm{had}}^{(5)}(-Q^2)$ from lattice and phenomenological estimates. We display the ratio of these determinations to the central value of our results, highlighting relative deviations. For our calculation, the orange vertical  band represents the total uncertainty, including the small bottom-quark contribution. The subleading isospin-breaking effects are not included here.}
	\label{fig:comparison_space_like_region}
\end{figure}

Our estimates for $\Delta\alpha_{\mathrm{had}}^{(5)}(-Q^2)$ can be
directly compared with existing determinations. In
figure~\ref{fig:comparison_space_like_region} we plot ratios between
our new results and several other lattice and phenomenological
estimates. We note that our previous results \cite{Ce:2022eix} are up
to $0.6\%$ larger compared to this analysis, but in good agreement
within errors. With respect to the BMW calculation from
2017~\cite{Budapest-Marseille-Wuppertal:2017okr}, we observe tensions
of up to $2.3\sigma$, our results being larger by $1\%-2\%$. A similar
tension is observed with the more recent BMW determination
\cite{Borsanyi:2020mff}, available for $Q^2= 1 \ \mathrm{GeV}^2$
only. Although the first lattice calculation of the quark-connected
HVP contribution to the running of the electroweak couplings was
reported by \cite{Burger:2015lqa}, we do not include this result in
our comparison, since the corresponding disconnected contribution was
not evaluated in that work. The phenomenological results shown in
figure~\ref{fig:comparison_space_like_region}, labelled by ``DHMZ
data'' \cite{Davier:2019can}, ``Jegerl. $\mathtt{alphaQED19}$ ''
\cite{Jeger_yellow_rep, Jegerlehner:2019alphaQEDc19} and ``KNT18
data'' \cite{Keshavarzi:2018mgv}, are in good agreement within each
other but lie significantly below the lattice determinations. At $Q^2
= 1 \ \mathrm{GeV}^2$, we find a discrepancy of up to $7$ standard
deviations between the phenomenological estimates and our lattice
calculation. Although the tension decreases at larger space-like
momenta, it remains sizeable, reaching about $4.5$ standard deviation
at $Q^2 = 9 \ \mathrm{GeV}^2$. To further investigate the origin of
the tension with data-driven results, we have explored whether modifications
to the experimental $R$-ratio $R(s)$ in some specific intervals of the
centre-of-mass energy $\sqrt{s}$ could reconcile the two
determinations. The analysis and its implications are discussed in
Appendix~\ref{app:modified_Rratio}.

\subsection{Rational approximation of the running}
To describe the dependence of the HVP on the space-like virtuality in a continuous and analytically tractable way, we adopt a Padé approximant, which provides a rational function representation already employed in previous studies \cite{Aubin2012,Ce:2022eix}. The general form of the Padé approximant used is 
\begin{equation}
  \bar\Pi(-Q^2) \approx R_M^N(Q^2) = 
	\frac{\sum_{j=0}^{M} a_j Q^{2j} }
	{1 + \sum_{k=1}^{N}b_kQ^{2k}}.
\end{equation} 
Here, numerator and denominator are polynomials of degree $M$ and $N$,
respectively. Given that the subtracted HVP function is required to
vanish at zero momentum transfer $Q^2=0$, we impose the condition
$a_0=0$ during the fit to capture this behaviour. We find that
polynomials of degree $M=2$ and $N=3$ provide an accurate description
of the data across the accessible range of $Q^2$. Attempts to include higher-order terms result in poorly determined coefficients, reflecting insufficient sensitivity in the data to resolve additional parameters.  

The parameters of the Padé ansatz are fitted to a range of $Q^2$ covering all
windows of virtualities, specifically we take the values $Q^2\in \{0.25,\ldots,12\} \,
\text{GeV}^2$.
For the estimate of the covariance of the fit parameters of the Padé ansatz we
use \cite{Bruno:2022mfy}, which defines a  goodness-of-fit criterion for correlated data that
avoids inverting the covariance matrix during minimization while still
incorporating it into the error estimate.

The fitted parametrization provides a convenient analytic representation of $\bar\Pi^{(\gamma,\gamma)}$ (and analogously $\bar\Pi^{(Z,\gamma)}$) in the low-energy Euclidean region accessible non-perturbatively. This representation is intended for interpolation and phenomenological applications, and is not used in the determination of our final results. Isospin-breaking
effects, found to be by far subleading at $Q^2=9\ \mathrm{GeV}^2$ and
approximately constant in the range $5-9 \ \mathrm{GeV}^2$ (see section~\ref{sec:ib_effects}), are omitted from the Padé ansatz reported below. The rational approximation for $\bar\Pi^{(\gamma,\gamma)}$ we extract from the fit is 
\begin{equation}\label{eq:pade_pigg}
	\bar\Pi^{(\gamma,\gamma)}(-Q^2) \approx 
	\frac{
		0.102226x + 0.040299x^2
	}
	{
		1 + 2.074048x + 0.342168x^2 -0.002788x^3
	}, \qquad x = \frac{Q^2}{\mathrm{GeV}^2},
\end{equation}
together with the covariance matrix
\begin{equation}\label{eq:pade_pigg_corr_matrix}
	\mathrm{cov}^{(\gamma,\gamma)} \begin{bmatrix}
		a_1 \\ a_2 \\ b_1 \\ b_2 \\ b_3
	\end{bmatrix}
 =
  \begin{bmatrix}
 2.14338  &  1.76686   &  50.4751  &  15.7648  & -0.208102\\
 1.76686  &  3.54353   &  77.2778   & 35.8345  & -0.524144\\
 50.4751  &  77.2778   & 1907.63    & 748.194   & -9.87013\\
 15.7648  &  35.8345   &  748.194  &  370.938   & -5.67152\\
 -0.208102 & -0.524144 &   -9.87013 &  -5.67152 &  0.0988209
 \end{bmatrix} \times 10^{-6}.
\end{equation}
For $\bar\Pi^{(Z,\gamma)}$ we find
\begin{equation}
	\bar\Pi^{(Z,\gamma)}(-Q^2) \approx 
	\frac{
		0.024748x + 0.010791x^2
	}
	{
		1 + 2.106921x + 0.393233x^2 -0.002901x^3
	}, \qquad x = \frac{Q^2}{\mathrm{GeV}^2},
\end{equation}
together with the covariance matrix
\begin{equation}
	\mathrm{cov}^{(Z,\gamma)} \begin{bmatrix}
		a_1 \\ a_2 \\ b_1 \\ b_2 \\ b_3
	\end{bmatrix}
	=
	\begin{bmatrix}
	0.137083  &  0.136607   & 14.816    &  5.19524  & -0.0627519\\
	0.136607   & 0.295515  &  25.7522  &  12.2846  &  -0.156386\\
	14.816   &   25.7522  &  2467.08  &  1040.29   &  -12.2305\\
	5.19524  &  12.2846  &  1040.29  &   517.894   &  -6.82578\\
	-0.0627519 & -0.156386  & -12.2305  &  -6.82578  &  0.103273
	\end{bmatrix} \times 10^{-6}.
\end{equation}
We observe that the approximants are in very good agreement with the measured data. The deviation remains at the level of about $0.5$ per~mil across the entire $Q^2$ range, well below the overall uncertainty of our results.

\begin{figure}
	\centering
	\includegraphics[scale=0.3]{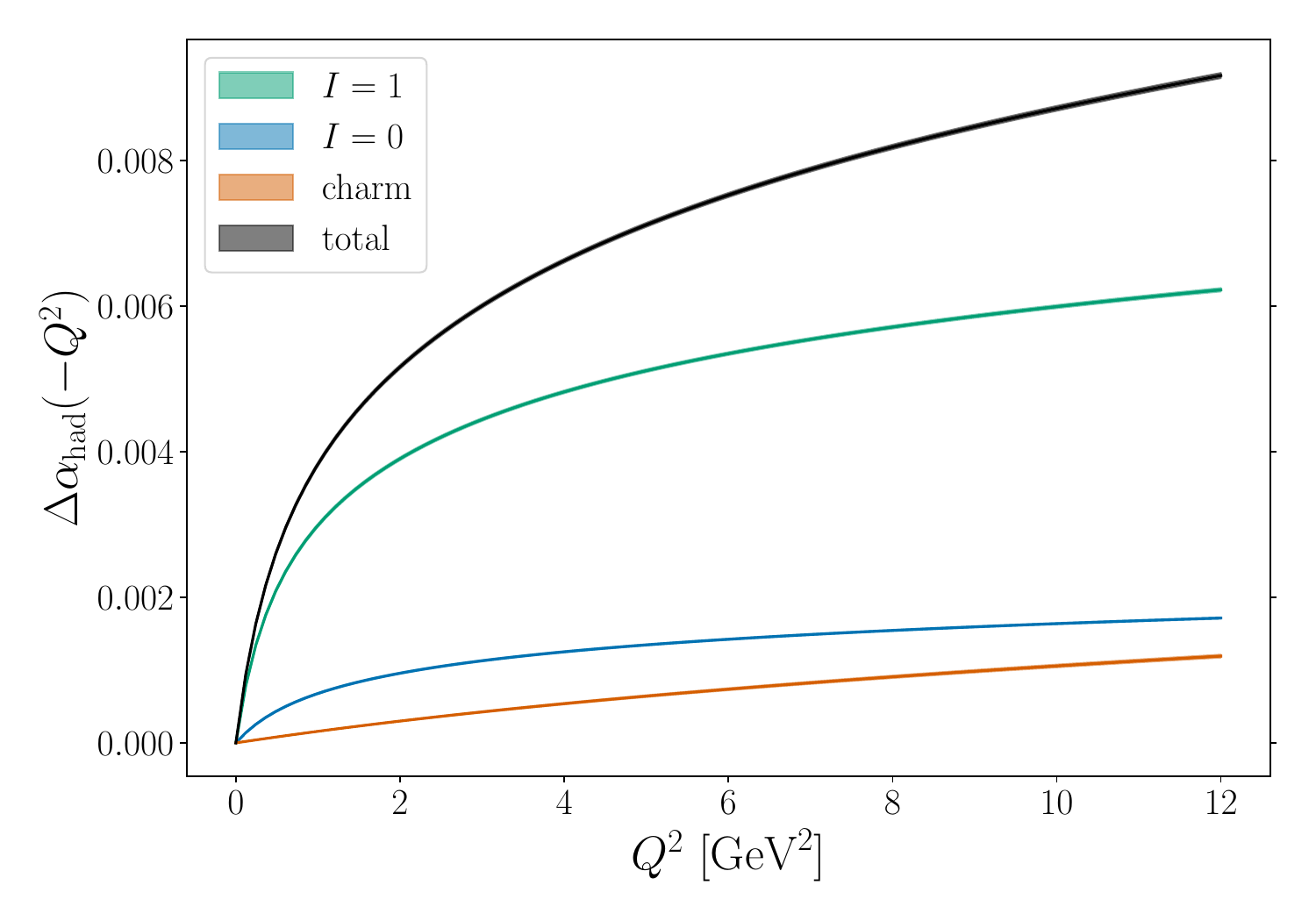}
	\includegraphics[scale=0.3]{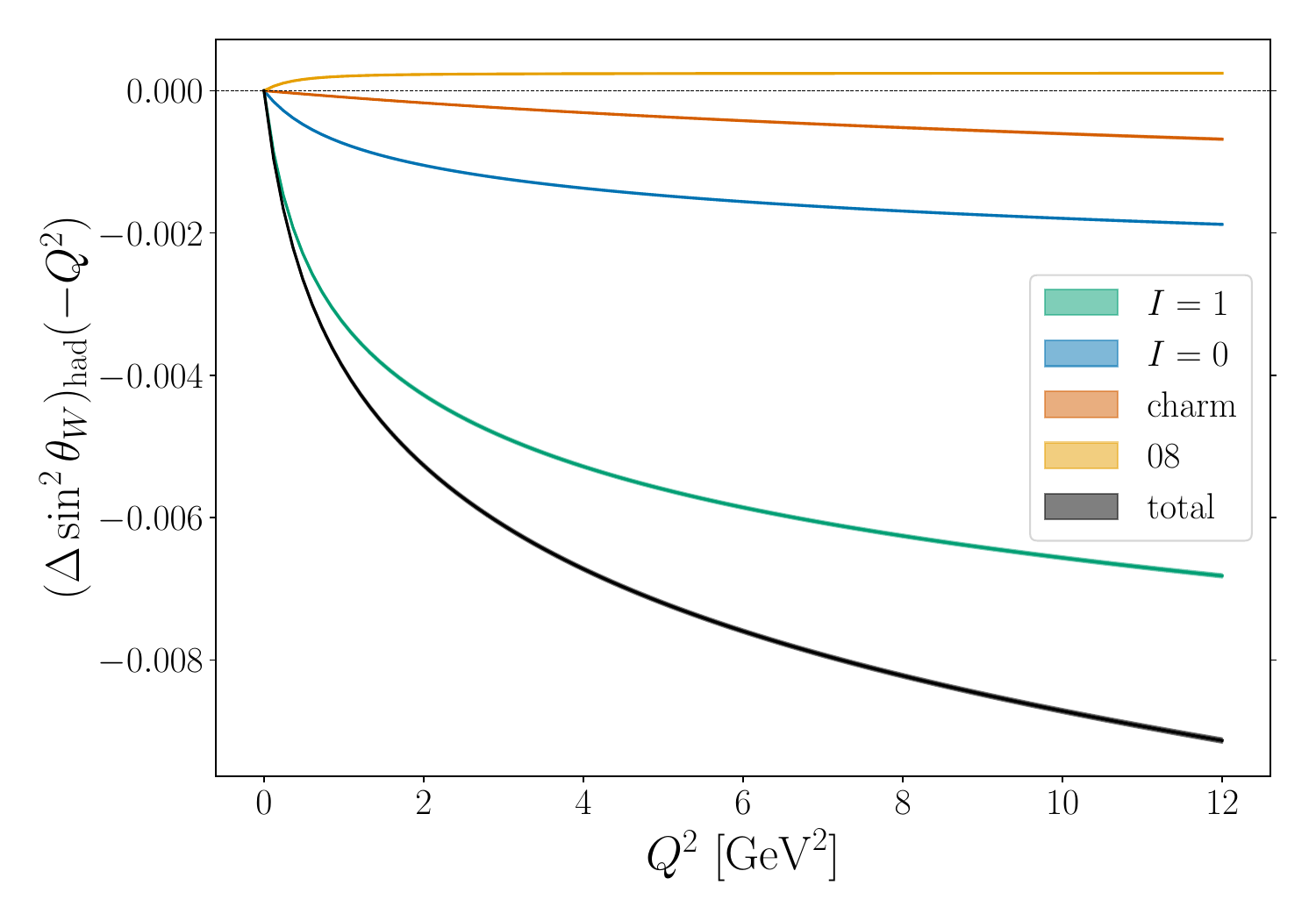}
	\caption{Total HVP contribution to the running of $\alpha$ (\textit{left}) and $\sin^2\theta_W$ (\textit{right}) as a function of the  space-like momentum transfer $Q^2$. Coloured bands correspond to the $I=1, \ I=0$, charm and, on the right panel, $Z\gamma$-mixing contributions.}
	\label{fig:space_like_running_alpha_sin}
\end{figure}

The  hadronic running of $\Delta\alpha_{\mathrm{had}}(-Q^2)$ and $(\Delta\sin^2\theta_W)_\mathrm{had}(-Q^2)$  are shown in figure~\ref{fig:space_like_running_alpha_sin} as a function of the space-like momentum  $Q^2$ using the fit results.

\subsection{Running of $\alpha$ to the $Z$-pole}


In this subsection, we describe the conversion of our results obtained
at space-like virtualities into an estimate for the hadronic
contributions to the running of $\alpha$ at the $Z$ boson mass,
$\Delta\alpha_{\mathrm{had}}^{(5)}(M_Z^2)$, which plays a critical
role in determining the accuracy of global electroweak (EW) precision
fits \cite{Jeger_yellow_rep, Jegerlehner:2011mw, Baak:2014ora}. An
established approach that provides the connection between space-like
virtualities and the $Z$~pole is the Euclidean split
technique \cite{Eidelman:1998vc, Jegerlehner:2008rs}.
This method evaluates the running entirely in the space-like region, combining the lattice determination at low energies with the perturbative Adler function at high energies. The connection to the time-like point at the $Z$-pole enters only through a small analytic continuation term, computed reliably in perturbation theory.
Altogether, this decomposition enables us to express the hadronic contribution to the running at the $Z$-pole as
\begin{equation}\label{eq:euclidean_splitting}
	\begin{split}
		\Delta\alpha_{\mathrm{had}}^{(5)}(M_Z^2) &= 		\Delta\alpha_{\mathrm{had}}^{(5)}(-Q_0^2) 
		\\
		&
		+ \big[
				\Delta\alpha_{\mathrm{had}}^{(5)}(-M_Z^2) 	-	\Delta\alpha_{\mathrm{had}}^{(5)}(-Q_0^2)
		\big]
		+
		\big[
				\Delta\alpha_{\mathrm{had}}^{(5)}(M_Z^2) - 		\Delta\alpha_{\mathrm{had}}^{(5)}(-M_Z^2)
		\big].
	\end{split}
\end{equation}
Here the first term denotes the space-like HVP as defined in Eq.~\eqref{eq:space_like_dalpha}, evaluated using our lattice results listed in table~\ref{tab:hvp_various_channel} at some threshold energy $Q_0^2$.

To compute the second term $\Delta\alpha_{\mathrm{had}}^{(5)}(-M_Z^2)
-	\Delta\alpha_{\mathrm{had}}^{(5)}(-Q_0^2)$, representing the
high-energy contribution to the hadronic running, we adopt the
approach developed in \cite{Jegerlehner:1999hg}: at sufficiently large
$Q_0^2$ values, the Adler function can be computed reliably in
perturbative QCD (pQCD), with only small corrections arising from
non-perturbative effects \cite{Eidelman:1998vc, Chetyrkin:1996cf}. We
evaluate this term from the public software\footnote{The code is
available at the GitHub repository
\url{https://github.com/rodofer2020/adlerpy}}  \texttt{AdlerPy}
\cite{Hernandez:2023ipz}. The code fully considers three-loop massive
perturbation theory with charm and bottom quark effects included,
together with the massless four- and five-loop  terms for a better
estimation of the high-energy tails. In addition, we include small
non-perturbative corrections arising from the quark and gluon
condensate as estimated in \cite{Davier:2023hhn}. The  heavy quark
masses in the $\overline{\mathrm{MS}}$ scheme and the strong coupling
at the $Z$-pole, used as input quantities, are taken from  FLAG24
\cite{FlavourLatticeAveragingGroupFLAG:2024oxs}. In particular, for
the charm and bottom quark masses, we use the FLAG24 average for
$N_f=2+1$, obtained by averaging the results from
Refs.~\cite{McNeile:2010ji, Petreczky:2019ozv, Bussone:2023kag,
  Yang:2014sea, Nakayama:2016atf, Heitger:2021apz} and
\cite{McNeile:2010ji, Petreczky:2019ozv}, respectively. The FLAG24
average for the strong coupling $\alpha_{\overline{\mathrm{MS}}}^{(5)}(M_Z)$ is obtained from Refs.~\cite{McNeile:2010ji, Chakraborty:2014aca, DallaBrida:2022eua, Petreczky:2020tky, Ayala:2020odx, Bazavov:2019qoo, Cali:2020hrj, Bruno:2017gxd, PACS-CS:2009zxm, Maltman:2008bx}.   

In addition, we crossed-checked our results using the publicly available \texttt{pQCDAdler} software \cite{Jegerlehner:2019pQCDAdler} to evaluate the Adler function, which had already been employed in the Mainz 2022 analysis \cite{Ce:2022eix} to estimate the perturbative running. In its original form, however, \texttt{pQCDAdler} does not allow for straightforward modifications of the input parameters. To enable a direct comparison, we updated the code so that the inputs could be set consistently with the FLAG24 values. With this modification, results from \texttt{pQCDAdler} move significantly closer to those obtained with \texttt{AdlerPy}, although perfect agreement is not expected due to the different renormalization schemes implemented in the two frameworks. 

In particular, we find a noticeable upward shift in the
\texttt{pQCDAdler} results once the input parameters are updated to
reflect the FLAG24 specifications. Concerning uncertainties,
\texttt{pQCDAdler} estimates the error by simultaneously varying input parameters within maximum bounds. On the other hand, in \texttt{AdlerPy} we implemented a bootstrap procedure to propagate the input uncertainties, which provides more direct control over the error budget. In addition, \texttt{AdlerPy} accounts for truncation errors in the light-and charm-quark contributions to the Adler function. 

After computing the Adler function $D(Q^2)$ within this framework, the high-energy contribution to the running can be determined through the following integral
\begin{equation}
	 \big[
	\Delta\alpha_{\mathrm{had}}^{(5)}(-M_Z^2) 	-	\Delta\alpha_{\mathrm{had}}^{(5)}(-Q_0^2)
	\big]
	= 
	\frac{\alpha}{3\pi} \int_{Q_0^2}^{M_Z^2} \frac{\dd Q^2}{Q^2} D(Q^2),
\end{equation}
with $\alpha$ the QED coupling in the Thomson limit. A summary of our results for several threshold energies $Q_0^2$ is given in table~\ref{tab:perturbative_running}. The quoted errors for \texttt{AdlerPy}, of the order of $0.2\%$ at $Q_0^2=9\ \mathrm{GeV}^2$, are dominated by the uncertainties of the strong coupling at the $Z$-pole  and the heavy-quark masses used as input quantities. At values of $Q_0^2\lesssim 1  \ \mathrm{GeV}^2$ the uncertainty from non-perturbative corrections to the Adler function increases substantially, and very small $Q_0^2$ cannot be probed due to the breakdown of perturbation theory associated with the infrared behaviour  of the strong coupling.

\begin{table}
	\centering
		\begin{tabular}{c  c  c  c}
			\toprule
			$Q_0^2 \ [\mathrm{GeV}^2]$ & $\mathrm{pQCD'}$ & $\mathrm{pQCD' \ updated}$ & $\mathrm{AdlerPy}$ \\
			\noalign{\smallskip}\hline\noalign{\smallskip}
			
			$1.0$ & $0.023\, 926(223)$ & $0.024\, 357(200)$ & $0.024\, 042(130)$\\
			
			$2.0$ & $0.022\, 489(148)$ & $0.022\, 671(122)$  & $0.022\, 617(71)$ \\
			
			$3.0$ & $ 0.021\, 638(128)$ & $0.021\, 758(104)$ & $0.021\, 769(57)$ \\
			
			$4.0$ & $0.021\, 018(116)$  & $0.021\, 114(94)$  & $0.021\, 115(50)$ \\
			
			$5.0$ & $0.020\, 525(106)$ & $0.020\, 610(87)$ & $0.020\, 660(44)$\\
			
			$6.0$ & $0.020\, 114(99)$    & $0.020\, 193(81)$ & $0.020\, 249(41)$ \\
			
			$7.0$ & $0.019\, 760(92)$   & $0.019\, 836(77)$ & $0.019\, 895(39)$ \\
			
			$8.0$ & $0.019\, 449(87)$  &  $0.019\, 523(72)$ & $0.019\, 583(37)$ \\
			
			$9.0$ & $0.019\, 170(82)$  & $0.019\, 244(69)$  & $0.019\, 304(35)$ \\

			$12.0$ & $0.018\, 475(71)$  & $0.018\, 549(59)$  & $0.018\, 605(31)$ \\			
			\bottomrule			
		\end{tabular}
		\caption{The perturbative running $[\Delta\alpha_{\mathrm{had}}^{5}(-M_Z^2)
			- \Delta\alpha_{\mathrm{had}}^{5}(-Q_0^2)
			]$ for various threshold energies $Q_0^2$. The
                  second column is a reproduction of the results
                  quoted in \cite{Ce:2022eix} using the pQCD code,
                  while the third column is obtained using a modified
                  version of the same code that consistently
                  incorporates updates in the input parameters. The
                  last column is obtained with the \texttt{AdlerPy}
                  package as explained in the main text. Input
                  parameters in the third and last columns are identical.}
		\label{tab:perturbative_running}
\end{table}

Finally, the last piece in Eq.~\ref{eq:euclidean_splitting}, provides the bridge between the space- and time-like regions at the $Z$-pole, and we quote the pQCD estimate from Jegerlehner \cite{Jeger_yellow_rep},
\begin{equation}
		\big[
	\Delta\alpha_{\mathrm{had}}^{(5)}(M_Z^2) - 		\Delta\alpha_{\mathrm{had}}^{(5)}(-M_Z^2)
	\big] = 0.000 \, 045(2).
\end{equation}

\begin{figure} 
	\centering
	\includegraphics[scale=0.4]{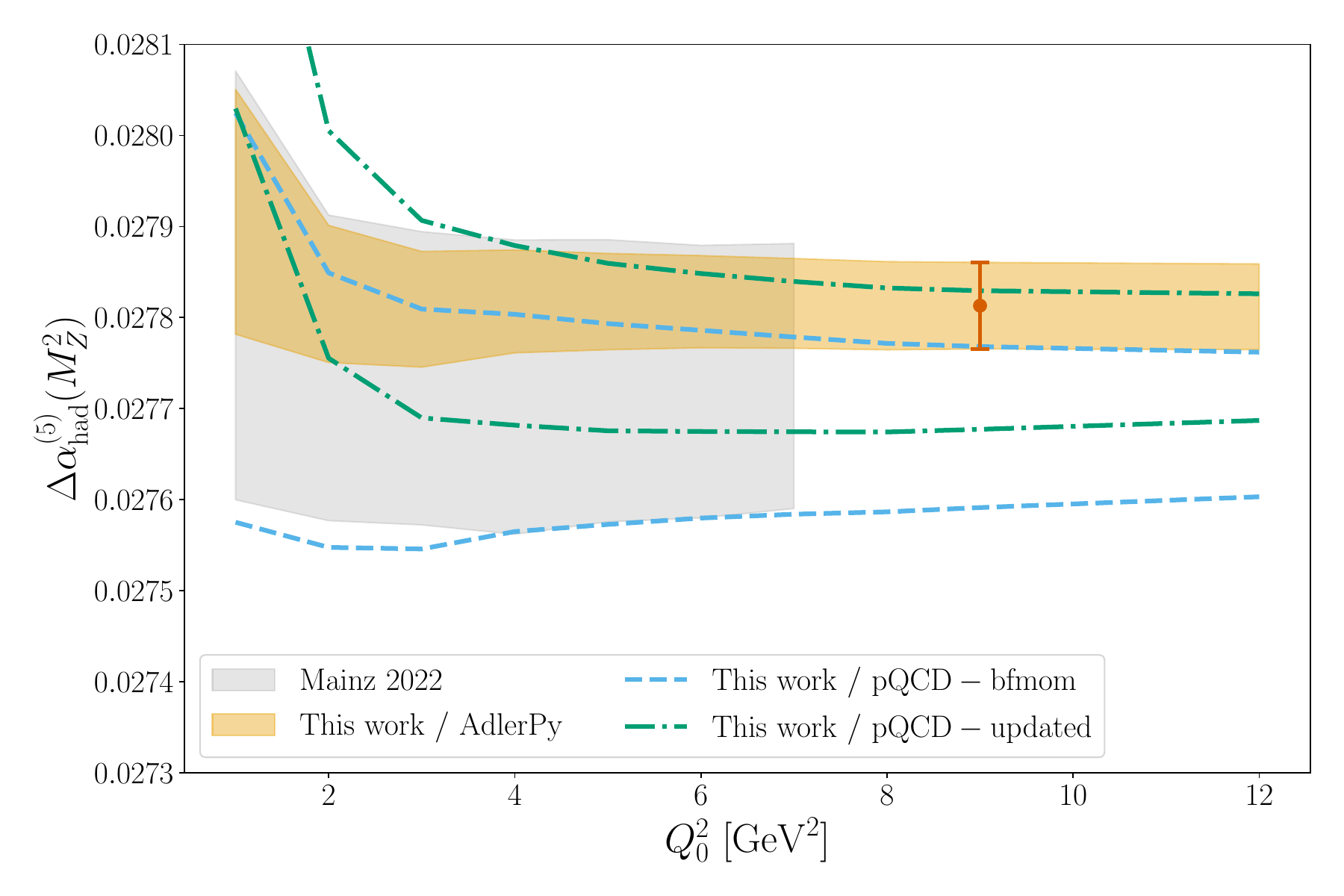}
	\caption{The five-flavour hadronic contribution to the running
          coupling at the $Z$-pole,
          $\Delta\alpha_{\mathrm{had}}^{(5)}(M_Z^2)$, evaluated from
          Eq.~\eqref{eq:euclidean_splitting} using our lattice
          determination of
          $\Delta\alpha_{\mathrm{had}}^{(5)}(-Q_0^2)$, shown as a
          function of the matching scale $Q_0^2$ in the Euclidean
          split technique. The curves connect direct lattice determinations evaluated at discrete momenta and do not involve any Padé fit procedure.}
	\label{fig:z_pole_different_q2}
\end{figure}

By combining these three pieces, we can now estimate the hadronic contribution to the running of $\alpha$ at the $Z$-pole, using our lattice determination $\Delta\alpha_{\mathrm{had}}^{(5)}(-Q_0^2)$ as input to Eq.~\ref{eq:euclidean_splitting}. Figure~\ref{fig:z_pole_different_q2}  shows our  results for $\Delta\alpha_{\mathrm{had}}^{(5)}(M_Z^2)$ as a function of the momentum threshold $Q_0^2$, where the lattice calculation is matched to the perturbative running. We display results obtained with the three perturbative strategies described above, along with the previous Mainz 2022 \cite{Ce:2022eix} determination. 
The orange band corresponds to the \texttt{AdlerPy} calculation, while the dashed lines represent the two versions of the \texttt{pQCDAdler} code. In all cases, the total errors is obtained by adding the lattice and perturbative errors in quadrature. We find excellent stability of the results for $Q_0^2> 3\ \mathrm{GeV}^2$, while for $Q_0^2< 2 \ \mathrm{GeV}^2$ an upward trend and loss of precision signal the breakdown of perturbation theory at low energies. 

\begin{figure} 
	\centering
	\includegraphics[scale=0.42]{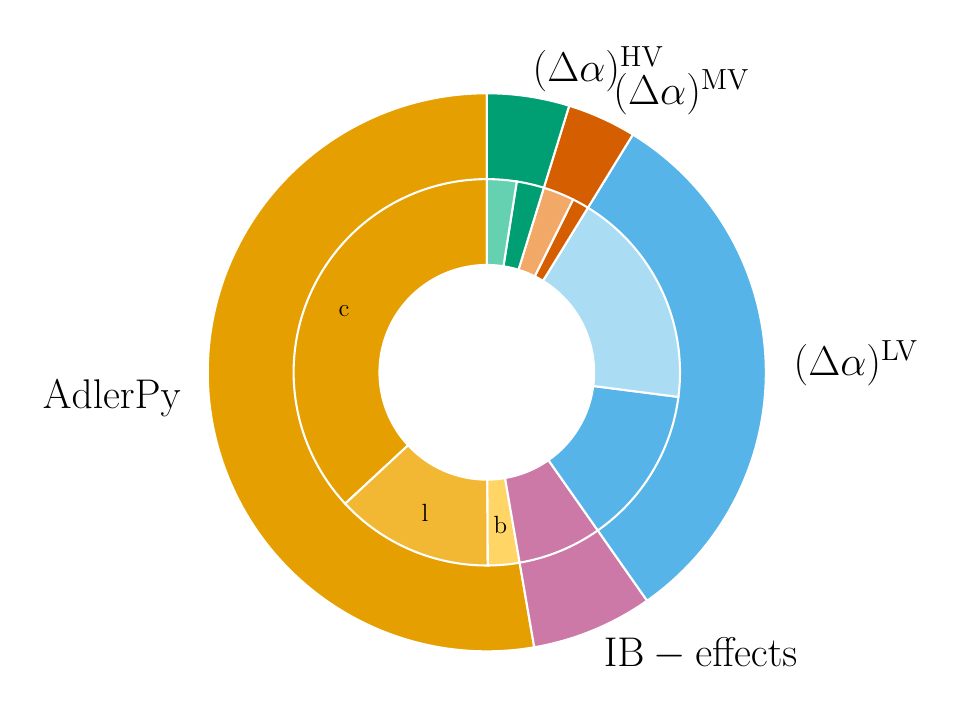}
	\caption{Breakdown of the squared uncertainty of our final estimate for $\Delta\alpha_{\mathrm{had}}^{(5)}(M_Z^2)$ from Eq~\eqref{eq:dalpha_z_final_result}. Each of the five contributions is decomposed into its main sources of uncertainty, shown in the inner circles. For the perturbative running, labelled AdlerPy, the uncertainty is further separated into the charm, light and bottom contributions to the Adler function, indicated by progressively lighter colours.  For the three momentum windows, statistical uncertainties are shown in lighter colours, while systematic components are displayed in darker tones.}
	\label{fig:variance_contrib}
\end{figure}

For our final determination, we adopt the \texttt{AdlerPy} result at
$Q_0^2= 9\ \mathrm{GeV}^2$, where the chiral-continuum extrapolation
of our lattice calculation is under good control. Choosing such a high
matching scale also reduces the size and error of the perturbative contribution to the total result. Our final quoted value for the hadronic running of the QED coupling at the $Z$-pole, based on the \texttt{AdlerPy} approach and including isospin-breaking corrections as estimated in Eq.~\eqref{eq:ib_effects}, is:
\begin{equation}\label{eq:dalpha_z_final_result}
	\Delta\alpha_{\mathrm{had}}^{(5)}(M_Z^2) = 0.027\, 821(34)_{\mathrm{lat}}(35)_\mathrm{pQCD}\ [49].
\end{equation} 
The first error arises from the total uncertainty of our lattice
estimate of $\Delta\alpha_{\mathrm{had}}^{(5)}(-9 \ \mathrm{GeV}^2)$
as listed in table~\ref{tab:dalpha_sin2_space_like_resl} and including
isospin-breaking effects, while the second error accounts for the
perturbative evaluation of the high-energy running
$[\Delta\alpha_{\mathrm{had}}^{5}(-M_Z^2)-
  \Delta\alpha_{\mathrm{had}}^{5}(-Q_0^2)]$ entering
Eq.~\ref{eq:euclidean_splitting}, and listed in the fourth column of table~\ref{tab:perturbative_running}.
Overall, we find a well-balanced distribution of uncertainties  between the perturbative running and the lattice contributions. A detailed decomposition of the squared uncertainty for our final estimate at the $Z$-pole is presented in figure~\ref{fig:variance_contrib}. On the lattice side, the dominant source of error arises from the LV region, while the HV and MV regions, as well as isospin-breaking effects provide smaller contributions.

\begin{figure}[h!] 
	\centering
	\includegraphics[scale=0.42]{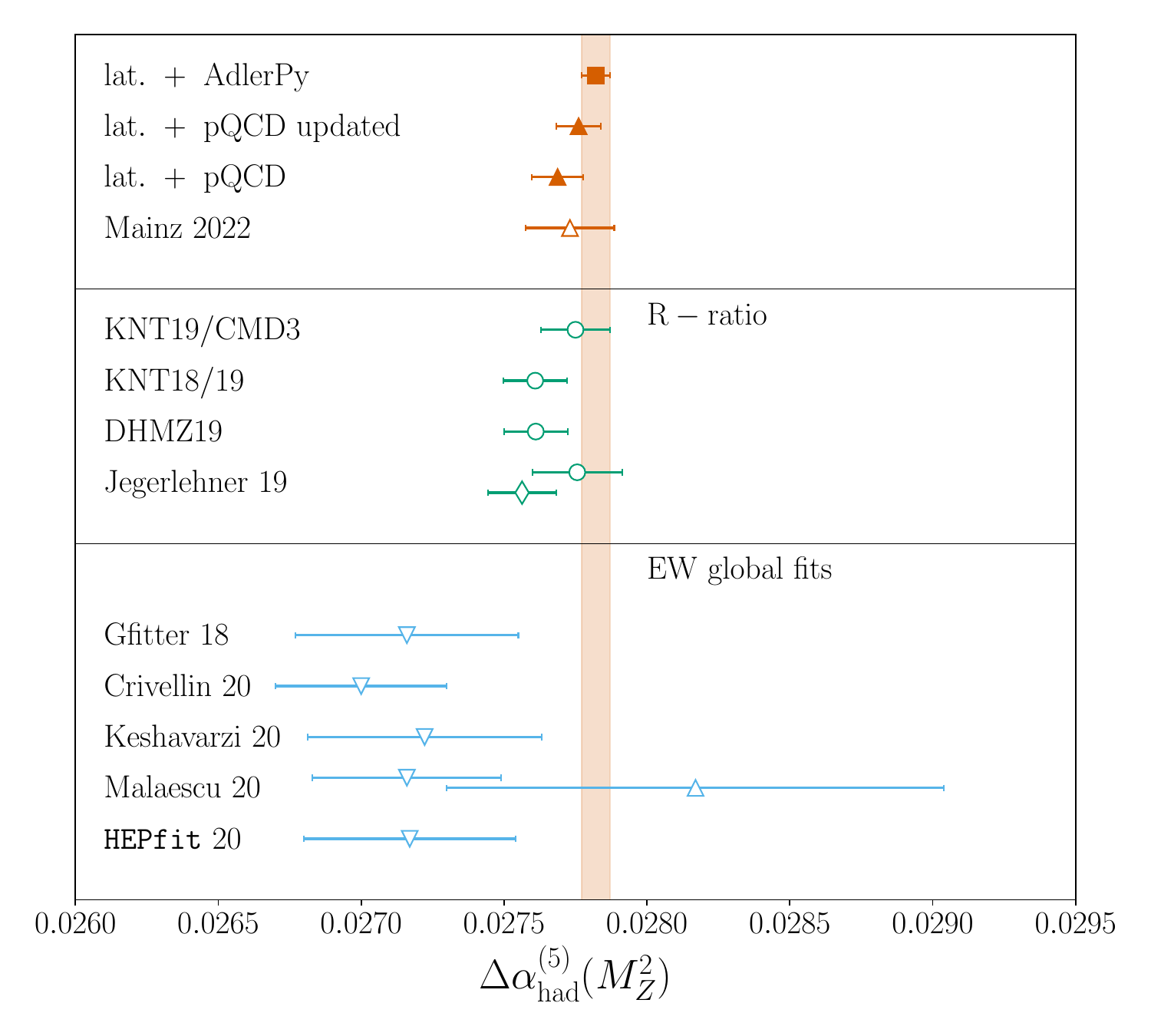}
	\caption{Summary of results for $\Delta\alpha_{\mathrm{had}}^{(5)}(M_Z^2)$. The first point corresponds to our determination using the perturbative results from \texttt{AdlerPy} listed in table~\ref{tab:perturbative_running} and including isospin-breaking corrections, with the orange band highlighting our final result. The second and third points are obtained using the \texttt{pQCDAdler} software package, followed by the previous Mainz 2022 result \cite{Ce:2022eix}. Results based on the standard dispersive approach, where the $R$-ratio is integrated over the full energy range, are shown as green circles. Blue symbols represent values extracted from global EW fits. We refer to the main text for additional details. }
	\label{fig:z_pole_comparison}
\end{figure}
In figure~\ref{fig:z_pole_comparison} we compare results for $\Delta\alpha_{\mathrm{had}}^{(5)}(M_Z^2)$ obtained from our lattice determination of the HVP, the standard dispersive approach and global EW fits. The first symbol (red filled square) represents our main result as given in Eq.~\ref{eq:dalpha_z_final_result} and obtained using the \texttt{AdlerPy} determination of the Adler function. The following two symbols (red filled triangle) represent our results from the two version of \texttt{pQCDAdler} code, while the empty red triangle shows our previous determination \cite{Ce:2022eix}. We find tensions at the level of $1-2\ \sigma$ with the dispersive evaluations (green points), while the determination of \cite{DiLuzio:2024sps} based on CMD-3 data \cite{CMD-3:2023alj, CMD-3:2023rfe} is fully compatible with our estimate.. The large tension with the dispersive approach observed in figure~\ref{fig:comparison_space_like_region} is substantially reduced at the $Z$-pole, due to the additional contribution to the uncertainty arising from the inclusion of the perturbative running.

Turning to global EW fits, we consider results from the Gfitter group
\cite{Haller:2018nnx}, from \cite{Crivellin:2020zul} (obtained using
the \texttt{HEPfit} code \cite{DeBlas:2019ehy}), from
\cite{Keshavarzi:2020bfy, Malaescu:2020zuc} (obtained from the Gfitter
library), and from \cite{deBlas:2021wap}. In
figure~\ref{fig:z_pole_comparison} these are shown as blue open lower
triangles, obtained by fitting EW precision data with
$\Delta\alpha_{\mathrm{had}}^{(5)}(M_Z^2)$ treated as a free
parameter. Such fits typically favour smaller values than both lattice
and dispersive determinations, but at a reduced precision. In
particular, we find a $2.7\sigma$ discrepancy with
\cite{Crivellin:2020zul}, while the other determinations remain
compatible with our estimate within $2 \sigma$. Overall, this marks
the first hint of a possible tension between lattice evaluations of the HVP contribution to the running of $\alpha$ and results from global EW fits. The point from \cite{Malaescu:2020zuc} (blue open upper triangle), where both the Higgs mass $M_H$ and $\Delta\alpha_{\mathrm{had}}^{(5)}(M_Z^2)$ are fitted without priors, yields a larger central value with significantly increased uncertainty, highlighting the stabilizing role of precise $M_H$ input. 

\subsection{Running of the electroweak mixing angle}
\begin{figure} 
	\centering
	\includegraphics[scale=0.44]{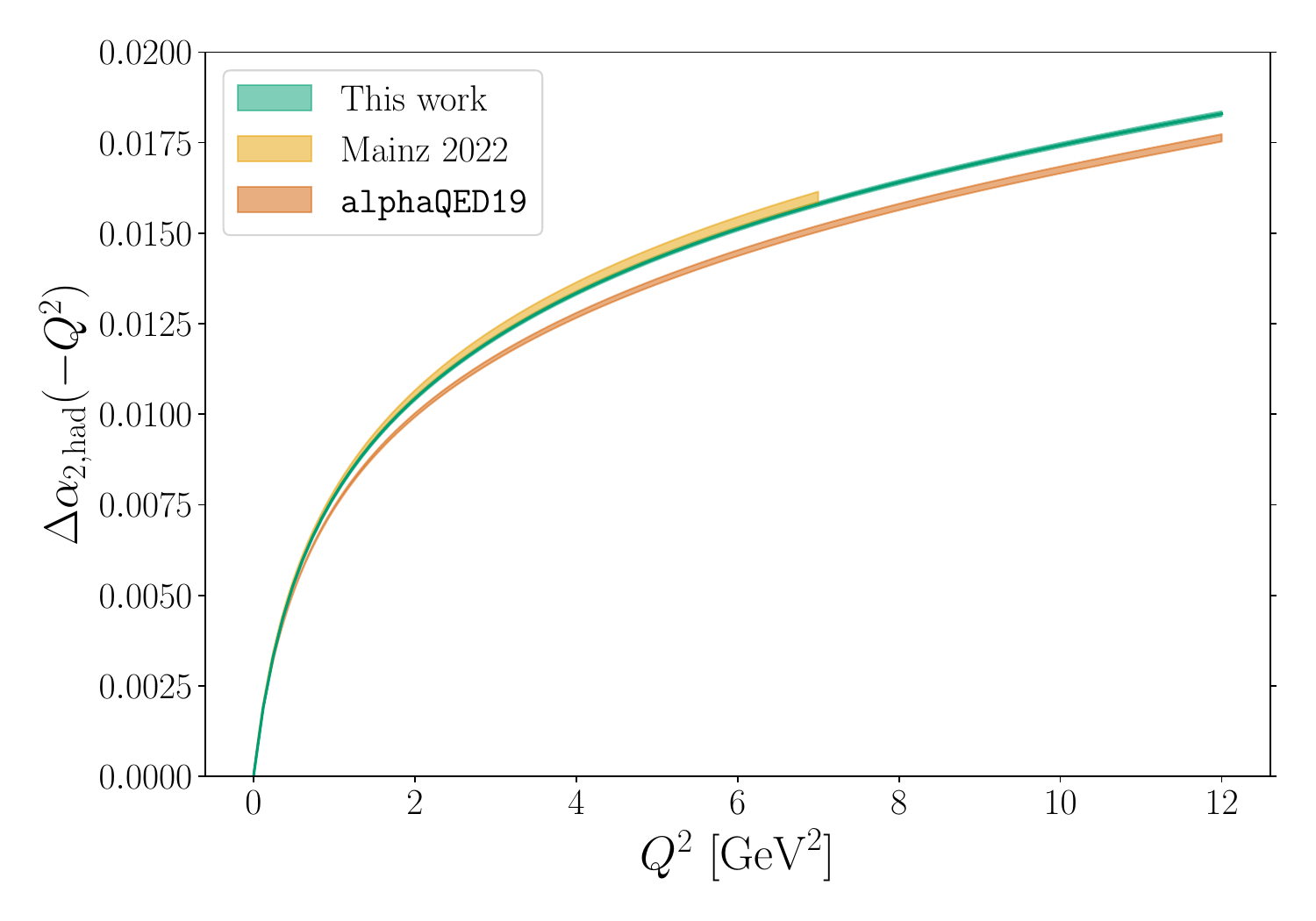}
	\caption{Hadronic contribution to the running of the weak coupling $\alpha_2$ from our analysis, shown as a function of the space-like momentum transfer $Q^2$, and compared with lattice results from \cite{Ce:2022eix} and with the phenomenological estimate obtained using \texttt{alphaQCDc19} \cite{Jegerlehner:2019alphaQEDc19}.}
	\label{fig:dalpha2_running}
\end{figure}
The lattice formulation provides exact control over the quark flavour charges entering the vector currents, which allows us to separate the HVP function $\bar\Pi^{(Z,\gamma)}$ into individual valence-quark contributions. This feature removes the need for channel reweighting in hadronic cross-section data, therefore avoiding a key source of systematic uncertainty in phenomenological determinations of the hadronic running of $\sin^2\theta_W$.

The results presented in table~\ref{tab:dalpha_sin2_space_like_resl}
at space-like momenta up to $Q^2 = 12 \ \mathrm{GeV}^2$ can be used
directly to replace data-driven estimates for studies of the running
of the electroweak mixing angle. For comparison, phenomenological
estimates are available from Jegerlehner's \texttt{alphaQEDc19}
package \cite{Jegerlehner:2019alphaQEDc19}, which provides both
$\Delta\alpha_{\mathrm{had}}$ and $\Delta\alpha_{2,\mathrm{had}}$. The
latter can be directly compared with our lattice determination for the
running of the $\rm SU(2)$ gauge coupling $\alpha_2$.  The package
employs a modified $\rm SU(3)$-symmetric flavour separation scheme \cite{Jegerlehner:2008rs, Jegerlehner:2019alphaQEDc19}, designed to better match earlier lattice results.  A direct comparison between this estimate, our lattice data and the previous Mainz 2022 result \cite{Ce:2022eix} is given in figure~\ref{fig:dalpha2_running}. We observe a good agreement between our result and Mainz 2022, while the phenomenological estimate still lies systematically below our determinations, by about $3.5\%$ at $Q^2=12\ \mathrm{GeV}^2$.

For completeness, we provide a Pad\'e parametrization of the running of $\alpha_2$, without the inclusion of the subleading isospin-breaking effects. The rational approximant we extract from the fit is
\begin{equation}
	\Delta\alpha_{2,\mathrm{had}}(-Q^2) \approx 
	\frac{
		0.018885x + 0.007822x^2
	}
	{
		1 + 2.088887x + 0.3667x^2 -0.002853x^3 
	}, \qquad x = \frac{Q^2}{\mathrm{GeV}^2},
\end{equation}
together with the covariance matrix
\begin{equation}
	\mathrm{cov}^{\Delta\alpha_{2,\mathrm{had}}} \begin{bmatrix}
		a_1 \\ a_2 \\ b_1 \\ b_2  \\ b_3
	\end{bmatrix}
	=
	\begin{bmatrix}
		0.0768698   & 0.0684375 &   10.1641  &  3.35298  & -0.042378\\
		0.0684375 &  0.142874  & 16.6493 &   7.78842  & -0.105787\\
		10.1641   &  16.6493   & 2163.42  &  878.042   & -10.8963\\
		3.35298   &  7.78842   &  878.042  & 432.15    &  -6.10704\\
		-0.042378  & -0.105787   & -10.8963 &  -6.10704  &  0.0986334
	\end{bmatrix} \times 10^{-6}.
\end{equation}
Also here we observe that the approximant is in good agreement with the measured data and reproduces the error band very accurately.

An alternative way to estimate the hadronic contribution to the running of the electroweak mixing angle at low energies is to  combine phenomenological evaluations of $\Delta\alpha_{\mathrm{had}}(-Q^2)$ from $R$-ratio data with lattice input that provides exact flavour separation.  From Eqs.~\eqref{eq:corr_building_blocks_gg} and \eqref{eq:corr_building_blocks_zg}, the difference between $\bar\Pi^{(\gamma,\gamma)}$ and $\bar\Pi^{(Z,\gamma)}$ is proportional to the isoscalar mixing function $\bar\Pi^{(0,8)}$. 

\begin{figure}[h!]
	\centering
	\includegraphics[scale=0.44]{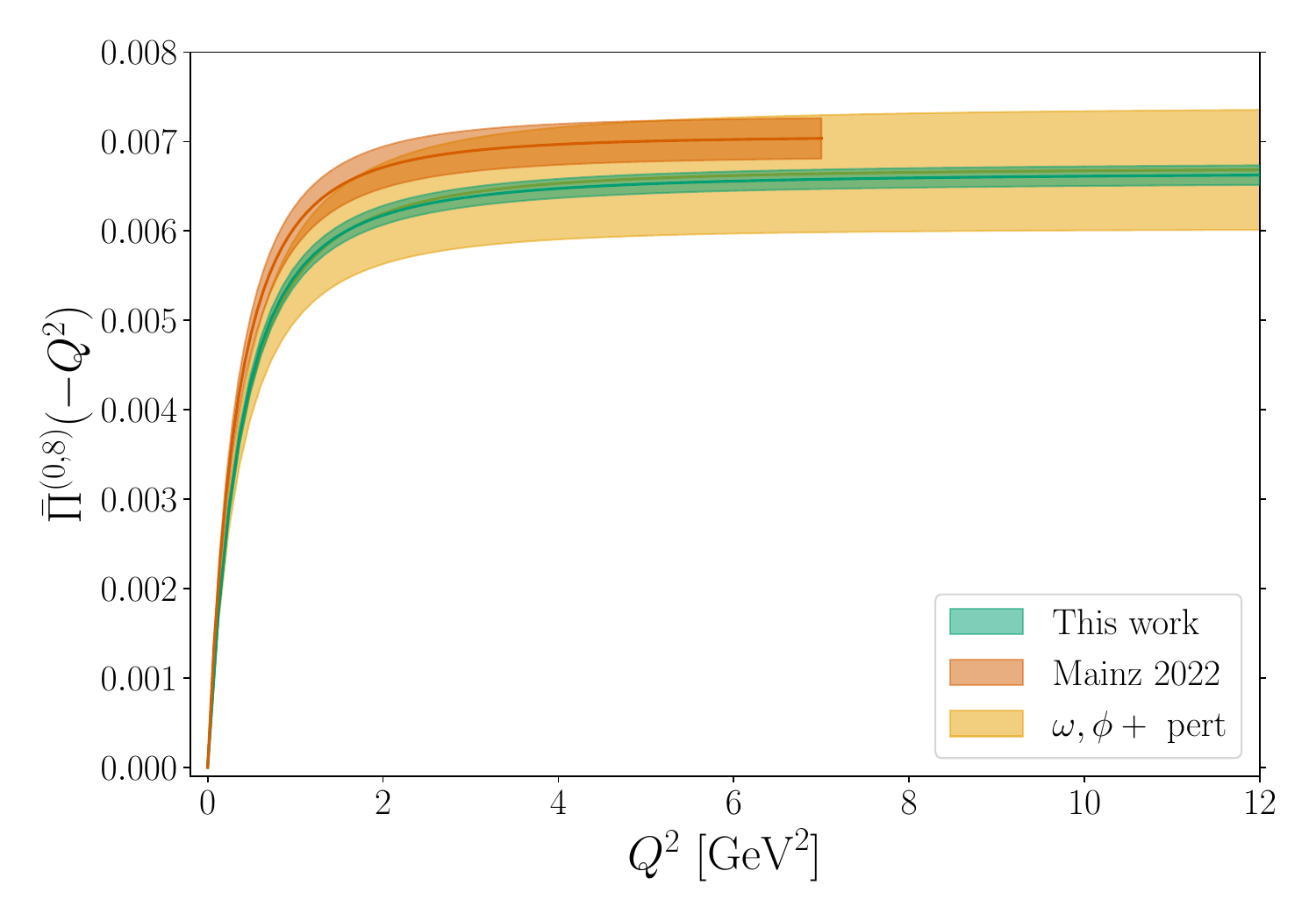}
	\caption{Lattice results for the HVP mixing function $\bar\Pi^{(0,8)}(Q^2)$ as a function of $Q^2$, compared with the Mainz 2022 results \cite{Ce:2022eix} and the phenomenological model obtained by combining the perturbative QCD contribution  and the resonance one from the $\omega$ and $\phi$ mesons.}
	\label{fig:pi08_running}
\end{figure}

Our results for the running of this channel are presented in
figure~\ref{fig:pi08_running}. For comparison, we also display the
results from our earlier Mainz 2022 analysis \cite{Ce:2022eix}, where
we observe a tension of about $2\sigma$ with respect to the updated
analysis. The difference can be largely attributed to the inclusions
of significantly more ensembles with increased  statistics, and  an
additional ensembles at the finest lattice spacing in the present
work. In particular, the telescopic window decomposition allows for a
clean identification of cutoff effects in the HV and MV regions that
were previously obscured by statistical noise, leading to a more
robust determination across all momentum regions. Finally,
figure~\ref{fig:pi08_running} also shows a phenomenological model
estimate (detailed in appendix D of \cite{Ce:2022eix}), whose
uncertainty is dominated by the experimental errors on the $\omega$
and $\phi$ leptonic widths. The model assumes that the disconnected
$(s,s)$ and $(l,s)$ diagrams can be neglected, an approximation known
to work well in analogous models used to estimate the strange and light isoscalar
contributions to $a_\mu^{\mathrm{HVP, LO}}$. Within uncertainties, the model agrees with our results, but the lattice data are significantly more precise across the entire $Q^2$ range.

Using a rational approximant of order $[2/2]$ for the running of $\bar\Pi^{(0,8)}$ over the whole $Q^2$ range we obtain
\begin{equation}
	\bar\Pi^{(0,8)}(-Q^2) \approx 
	\frac{
		0.017001x + 0.021094x^2
	}
	{
		1 + 2.763262x + 3.160956x^2 
	}, \qquad x = \frac{Q^2}{\mathrm{GeV}^2},
\end{equation}
together with the covariance matrix
\begin{equation}
	\mathrm{cov}^{(0,8)} \begin{bmatrix}
		a_1 \\ a_2 \\ b_1 \\ b_2 
	\end{bmatrix}
	=
	\begin{bmatrix}
		0.0375666  &  0.0216062  &   3.45356    &  0.620977\\
		0.0216062  &  1.16285   &   33.1381  &   166.475\\
		3.45356   &  33.1381   &  1287.57   &   4612.1\\
		0.620977  & 166.475  &    4612.1   &   24048.8
	\end{bmatrix} \times 10^{-5}.
\end{equation}
In the limit of $Q^2\to\infty$, the approximant tends to the ratio
$\frac{1}{6\sqrt{3}}a_2/b_2 = 64.2(1.1) \times 10^{-5}$, which agrees very well with our value at the largest $Q^2=12\ \mathrm{GeV}^2$
\begin{equation}
	\frac{1}{6\sqrt(3)}\bar{\Pi}^{(0,8)} = 64.0(1.1) \times 10^{-5},
\end{equation}
that we quote as our main result for the $I=0$ $Z\gamma$-mixing HVP contribution, and is only $3\%$ larger than the result at $Q^2=3\ \mathrm{GeV}^2$.

\section{Conclusions and outlook}
\label{sec:conc}
In this work, we have presented an updated lattice QCD determination of the HVP contribution to the running of the electromagnetic coupling $\Delta\alpha_{\mathrm{had}}^{(5)}(-Q^2)$, and of the electroweak mixing angle in the space-like region up to $Q^2=12 \ \mathrm{GeV}^2$.  Our analysis benefits from a  substantial increase in statistical precision and improved control of systematics, achieved through an expanded ensemble set, including additional fine lattice spacings, and a refined strategy for the chiral-continuum extrapolation. 

A key innovation of this study is the implementation of a telescopic window strategy, which enables a clean separation of the HVP into low-, mid- and high- virtuality regions. This decomposition disentangles discretization effects at short distances from the strong chiral dependence at large Euclidean times. As a result, the chiral-continuum extrapolation is much better controlled, allowing for high-precision determinations across all considered momentum regions. Complementary to this, we have employed an extensive set of noise-reduction techniques, including low-mode averaging, bounding method and explicit reconstruction of the LD correlators from spectroscopy data, further enhancing the signal quality at large Euclidean separations.

Our determination of $\Delta\alpha_{\mathrm{had}}^{(5)}(-Q^2)$  at space-like values of the momenta exhibits a clear and persistent tension with phenomenological estimates based on the $R$-ratio, reaching up to $7\sigma$ at $Q^2=1 \ \mathrm{GeV}^2$ and remaining above $4\sigma$ up to $9 \ \mathrm{GeV}^2$. This confirms, with improved precision, the discrepancy previously observed in our 2022 analysis \cite{Ce:2022eix}.

When combining our lattice results with the perturbative determination of the Adler function through the Euclidean split technique, we obtain a high-precision determination of the five-flavour hadronic running up to the $Z$-pole. Thanks to several improvements in our calculation, the precision of our final value for $\Delta\alpha_{\mathrm{had}}^{(5)}(M_Z^2)$ exceeds that of state-of-the-art phenomenological determinations by a factor two. The inclusion of the perturbative running, which contributes  more than half of the squared total uncertainty, substantially reduces the tension observed in the space-like region relative to dispersive results, leaving a mild $1-2\sigma$ deviation. Estimates from EW global fits, which have a significantly lower precision compared to our lattice determination, are systematically lower than our final result, but the latter generally remains within their $2\sigma$ bands.

For the electroweak mixing angle, we have computed the hadronic contribution to its running in the space-like region up to $Q^2=12 \ \mathrm{GeV}^2$. Our lattice framework allows for an exact flavour separation of the mixed HVP function $\bar\Pi^{(Z,\gamma)}$, eliminating the need for a model-dependent flavour reweighting present in phenomenological analyses.
We find overall good agreement with the \texttt{alphaQEDc19} estimate, though our results  are  consistently larger by about $4\%$. In addition, we provide a rational representation of the octet-singlet mixing contribution $\bar\Pi^{(0,8)}(-Q^2)$, which asymptotically approaches  a constant at large $Q^2$ values. We observe a tension of about $ 2\sigma$ compared to our previous result \cite{Ce:2022eix}, primarily driven by the inclusion of an additional, fine lattice spacing and a far more advanced treatment of lattice artefacts.

This improved analysis demonstrates that lattice QCD provides a fully first-principles determination of the hadronic running of the electroweak couplings with competitive precision and well-controlled systematics. Future work will focus on extending the momentum range of the lattice calculation in order to reduce the uncertainty of the perturbative running, incorporating full strong isospin-breaking and QED corrections, and further reducing uncertainties associated with large Euclidean times. 
One particular direction to be explored is the use of the covariant coordinate-space method~\cite{Meyer:2017hjv}, which might reduce discretization errors on $\widehat{\Pi}(-Q^2)$ at high virtualities~\cite{ToelleMasterArbeit}.

\section*{Acknowledgements}
We thank Andrew Hanlon, Nolan Miller, Ben H\"{o}rz, Daniel Mohler, Colin Morningstar and Srijit Paul for the collaboration on the data generation and analysis for the spectral reconstruction.
We thank Volodymyr Biloshytskyi for providing auxiliary data used in the 
estimate of isospin-breaking effects.  
A.C. is grateful to Arnau Beltran for valuable discussions and crosschecks performed throughout the analysis.
We are grateful to Marco Cè for sharing data used to produce comparison plots.
We are grateful  to Bogdan Malaescu and the authors of refs. \cite{Davier:2019can} for sharing their data for the running of $\alpha$ at $Q^2>0$.
We thank Alex Keshavarzi and the authors of refs. \cite{Keshavarzi:2018mgv} for sharing the $R$-ratio data with covariance matrix, which were used to calculate tabulated data for the running of $\alpha$ at $Q^2>0$.
We are grateful to our colleagues in the CLS initiative for sharing ensembles.
Calculations for this project were performed on the HPC
clusters Clover and HIMster-II at the Helmholtz Institute Mainz and
Mogon-II and Mogon-NHR at Johannes Gutenberg-Universität (JGU)
Mainz, as well as on the GCS Supercomputers JUQUEEN and JUWELS at 
the Jülich Supercomputing Centre (JSC), HAZELHEN and HAWK at the 
Höchstleistungsrechenzentrum Stuttgart (HLRS), and SuperMUC at the 
Leibniz Supercomputing Centre (LRZ).
The authors gratefully acknowledge the support of the Gauss Centre
for Supercomputing (GCS) and the John von Neumann-Institut für
Computing (NIC) by providing computing time via the projects HMZ21, HMZ23
and HINTSPEC at JSC, as well as projects GCS-HQCD and GCS-MCF300 
at HLRS and LRZ. We also gratefully acknowledge the scientific 
support and HPC resources provided by NHR-SW of Johannes 
Gutenberg-Universität Mainz (project NHR-Gitter).
This work has been supported by Deutsche Forschungsgemeinschaft
(German Research Foundation, DFG) through the Collaborative Research 
Center 1660 “Hadrons and Nuclei as Discovery Tools”, under grant HI~2048/1-2
(Project No.\ 399400745), and through the Cluster of Excellence
``Precision Physics, Fundamental Interactions and Structure of
Matter'' (PRISMA+ EXC 2118/1), funded within the German Excellence
strategy (Project No.\ 390831469). This project has received funding
from the European Union's Horizon Europe research and innovation
programme under the Marie Sk\l{}odowska-Curie grant agreement
No.\ 101106243. 
A.R. was supported by the programme Netzwerke 2021, an initiative of the Ministry of Culture and Science of the State of Northrhine Westphalia, in the NRW-FAIR network, funding code NW21-024-A.

\begin{appendix}
\section{Tables}
This appendix contains tables with detailed results for individual gauge ensembles. Table~\ref{tab:masses_and_t0} lists the pseudoscalar masses and the gluonic observable $t_0/a^2$.  Tables~\ref{tab:hvp_results} and~\ref{tab:hvp_results_deltas} report the HVP contributions for the various channels entering the analysis, while Tables~\ref{tab:fvc_lref_HV},~\ref{tab:fvc_lref_IV} and~\ref{tab:fvc_lref_LV} present the finite-volume correction to the reference volume $L_{\mathrm{ref}}$ in the HV, IV, and LV regions, respectively.

\begin{table}[h!]
	\centering
		\renewcommand{\arraystretch}{1.2}
		\begin{tabular}{l   c c  c c c}
			\toprule
			\text{id} & $am_\pi$ & $am_K$ & $t_0/a^2$ & $\phi_2$ & $\phi_4$ \\
			\hline 
			H101 & 0.18311(53) & 0.18311(53) & 2.8442(45) & 0.7629(34) & 1.1444(51) \\ 
			H102 & 0.15439(69) & 0.19161(66) & 2.8779(52) & 0.5488(38) & 1.1197(62) \\ 
			N101 & 0.12157(62) & 0.20161(33) & 2.8929(48) & 0.3420(29) & 1.1117(38) \\ 
			C101 & 0.09646(52) & 0.20564(32) & 2.9108(40) & 0.2167(22) & 1.0931(33) \\ 
			C102 & 0.09660(77) & 0.21792(40) & 2.8649(49) & 0.2139(32) & 1.1953(47) \\ 
			D150 & 0.05645(76) & 0.20874(47) & 2.9462(40) & 0.0751(20) & 1.0645(45) \\ 
			\hline 
            B450 & 0.16102(59) & 0.16102(59) & 3.680(18) & 0.7633(52) & 1.1449(78) \\ 
			N451 & 0.11079(50) & 0.17827(28) & 3.6890(74) & 0.3622(30) & 1.1189(39) \\ 
			N452 & 0.13548(33) & 0.17031(29) & 3.6748(91) & 0.5396(20) & 1.1225(30) \\ 
			D450 & 0.08331(70) & 0.18399(36) & 3.6992(27) & 0.2054(33) & 1.1046(51) \\ 
			D451 & 0.08338(29) & 0.19388(20) & 3.6640(33) & 0.2038(13) & 1.2038(20) \\ 
			D452 & 0.05963(59) & 0.18663(17) & 3.7242(59) & 0.1059(20) & 1.0907(22) \\ 
			\hline 
			N202 & 0.13433(38) & 0.13433(38) & 5.166(14) & 0.7458(39) & 1.1187(58) \\ 
			N203 & 0.11217(28) & 0.14378(23) & 5.1407(53) & 0.5175(24) & 1.1088(37) \\ 
			N200 & 0.09246(30) & 0.15062(23) & 5.1622(53) & 0.3530(21) & 1.1134(33) \\ 
			D251 & 0.09198(24) & 0.15046(14) & 5.1661(55) & 0.3496(17) & 1.1104(21) \\ 
			D200 & 0.06526(22) & 0.15649(13) & 5.1769(60) & 0.1764(11) & 1.1024(19) \\ 
			D201 & 0.06541(37) & 0.16313(21) & 5.1366(66) & 0.1758(19) & 1.1814(31) \\ 
			E250 & 0.04240(23) & 0.159364(83) & 5.2012(25) & 0.07481(74) & 1.0942(12) \\
			\hline 
			J307 & 0.10519(24) & 0.10519(24) & 8.605(24) & 0.7616(37) & 1.1425(56) \\ 
			J306 & 0.08703(33) & 0.11333(34) & 8.591(14) & 0.5206(42) & 1.1430(76) \\ 
			J303 & 0.06472(25) & 0.11965(19) & 8.614(13) & 0.2887(21) & 1.1309(37) \\ 
			J304 & 0.06532(19) & 0.13172(18) & 8.493(11) & 0.2899(15) & 1.3239(33) \\ 
			E300 & 0.04408(10) & 0.12399(12) & 8.6178(53) & 0.13399(60) & 1.1269(20) \\ 
			F300 & 0.03370(24) & 0.12332(16) & 8.6566(39) & 0.0786(11) & 1.0924(30) \\ 
			\hline 
			J500 & 0.08152(22) & 0.08152(22) & 13.965(22) & 0.7424(34) & 1.1135(51) \\ 
			J501 & 0.06591(18) & 0.08787(18) & 13.972(43) & 0.4855(20) & 1.1058(34) \\ 
			\bottomrule
		\end{tabular}
	\caption{Pseudoscalar masses in lattice units. Estimates of the gluonic observable $t_0/a^2$ and the two dimensionless hadronic quantities $\phi_2$ and $\phi_4$ are provided in the last three columns. }
	\label{tab:masses_and_t0}
\end{table}


\begin{table}[h!]
	\centering
	\renewcommand{\arraystretch}{1.2}
	\begin{tabular}{l | c c | c c |  c | c c}
		\toprule
		& \multicolumn{2}{c|}{$\bar\Pi^{(3,3)}_{\mathrm{sub}}$} & 	\multicolumn{2}{c|}{$\frac{1}{3}\bar\Pi^{(8,8)}$}  &  	$\frac{1}{6\sqrt{3}}\bar\Pi^{(0,8)}$  &   \multicolumn{2}{c}{$\frac{4}{9}\bar\Pi^{(c,c)}_\mathrm{sub}$}  \\
		\hline
		\text{id} & $(LL)$ & $(LC)$ & $(LL)$ & $(LC)$ &  $(LC)$ & $(LL)$ & $(LC)$ \\
		\hline
		H101 & 4503(9) & 4933(8) & 1256.9(2.6) & 1338.7(2.3) &  - & 90.3(1.4) & 581.7(2.5)\\ 
		H102 & 4661(11) & 5085(10) & 1217.7(3.0) & 1299.5(2.7) & 20.38(72) & 98.1(1.3) & 588.5(2.5)\\ 
		N101 & 4913(12) & 5335(10) & 1184.1(2.4) & 1266.5(2.2) & 43.6(1.8) & 110.9(1.2) & 599.9(2.5)\\ 
		C101 & 5102(14) & 5520(14) & 1170.9(3.3) & 1254.4(3.2) & 56.3(1.3) & 115.6(1.2) & 604.0(2.3)\\ 
		C102 & 5089(15) & 5504(14) & 1142.8(4.1) & 1226.9(4.1) & 62.9(1.5) &  - & - \\ 
		D150 & 5402(14) & 5814(13) & 1161.0(7.4) & 1243.8(7.4) & 64.7(3.6) &  - & - \\ 
		\hline
		B450 & 4626(12) & 4921(11) & 1276.5(3.6) & 1334.3(3.5) &  - & 302.8(1.7) & 608.9(2.4)\\ 
		N451 & 5051(9) & 5341(8) & 1226.2(1.9) & 1284.5(1.6) & 42.00(93) &  - & - \\ 
		N452 & 4861(16) & 5158(15) & 1253.2(2.2) & 1311.3(2.1) & 21.7(1.0) &  - & - \\ 
		D450 & 5266(7) & 5556(5) & 1204.2(2.3) & 1264.0(2.1) & 57.6(1.1) & 334.6(1.9) & 640.1(2.6)\\ 
		D451 & 5278(8) & 5571(7) & 1178.2(2.5) & 1238.0(2.3) & 62.7(1.9) &  - & - \\ 
		D452 & 5455(9) & 5742(8) & 1191.0(3.8) & 1250.8(3.6) & 63.3(2.9) & 338.7(1.4) & 644.1(2.2)\\ 
		\hline
		N202 & 4857(15) & 5024(14) & 1335.7(4.6) & 1370.5(4.6) &  - & 484.7(2.6) & 639.1(2.9)\\ 
		N203 & 5016(13) & 5184(12) & 1290.6(3.1) & 1325.8(3.1) & 21.96(75) & 496.2(2.1) & 649.2(2.4)\\ 
		N200 & 5169(12) & 5333(11) & 1255.9(2.6) & 1291.4(2.5) & 41.1(1.1) & 509.2(2.2) & 662.4(2.6)\\ 
		D251 & 5202(6) & 5368(5) & 1257.9(1.3) & 1293.4(1.1) & 40.41(80) &  - & - \\ 
		D200 & 5420(11) & 5585(11) & 1236.5(3.4) & 1272.6(3.3) & 58.4(1.5) & 518.6(2.7) & 671.9(2.9)\\ 
		D201 & 5408(9) & 5574(8) & 1212.5(3.9) & 1249.3(3.9) & 62.1(3.4) &  - & - \\ 
		E250 & 5711(7) & 5874(6) & 1223.8(4.4) & 1260.2(4.2) & 69.7(2.1) & 525.7(2.3) & 679.6(2.6)\\ 
		\hline
		J307 & 4914(10) & 4978(10) & 1348.8(3.2) & 1364.6(3.1) &  - &  - & - \\ 
		J306 & 5100(11) & 5163(10) &  - & - & - &  - & - \\ 
		J303 & 5294(15) & 5362(21) & 1265.4(3.1) & 1283.6(4.6) & 46.2(1.4) & 617.0(2.7) & 670.1(2.7)\\ 
		J304 & 5290(10) & 5354(9) & 1219.3(3.2) & 1236.0(3.2) & 55.1(1.3) &  - & - \\ 
		E300 & 5591(10) & 5684(10) & 1253.2(4.9) & 1273.0(4.6) & 63.7(2.2) & 638.3(1.9) & 686.3(2.1)\\ 
		F300 & 5778(12) & 5840(12) &  - & - & - &  - & - \\ 
		\hline
		J500 & 4909(13) & 4934(14) & 1354.1(5.7) & 1362.3(5.8) &  - & 614.3(3.6) & 628.1(3.2)\\ 
		J501 & 5067(13) & 5084(15) & 1300.1(2.7) & 1308.0(3.0) & 23.8(1.2) &  - & - \\
		\bottomrule
	\end{tabular}
	\caption{Values of the HVP for the isovector, isoscalar, mixed $Z\gamma$ and charm-connected contributions at $Q^2=9\ \mathrm{GeV}^2$, in units of $10^{-5}$. Results are shown for both local-local $(LL)$ and local-conserved $(LC)$ discretizations of the vector currents. For the isovector and charm-connected contributions, values obtained using the subtracted kernel are reported. In the isoscalar case, we show results combining the MV and LV momenta regions; in the HV region,  the isoscalar channel is extracted  from  $\Delta_{ls}$, as described in section~\ref{subsec:isoscalar_contrib}.   All values correspond to  the improvement coefficient set 2.   }
	\label{tab:hvp_results}
\end{table}

\begin{table}[h!]
	\centering
	\renewcommand{\arraystretch}{1.2}
	\begin{tabular}{l | c c | c c }
		\toprule
		& \multicolumn{2}{c|}{$-\frac{1}{3}\Delta_{ls}$} & 	\multicolumn{2}{c|}{$-\frac{4}{9}\Delta_{lc}b$}    \\
		\hline
		\text{id} & $(LL)$ & $(LC)$ & $(LL)$ & $(LC)$ \\
		\hline
		H101 &  - & - & 946.7(1.7) & 571.4(1.9)\\ 
		H102 & 9.48(26) & 8.80(26) & 949.5(1.8) & 578.4(2.5)\\ 
		N101 & 21.08(54) & 19.61(50) & 945.5(1.7) & 579.9(2.0)\\ 
		C101 & 27.55(38) & 25.69(35) & 948.9(1.6) & 584.4(1.4)\\ 
		C102 & 30.87(23) & 28.75(20) &  - & - \\ 
		D150 & 33.86(31) & 31.67(29) &  - & - \\ 
		\hline
		B450 &  - & - & 767.9(2.2) & 566.6(2.5)\\ 
		N451 & 16.92(13) & 16.34(12) &  - & - \\ 
		N452 & 8.94(30) & 8.60(30) &  - & - \\ 
		D450 & 24.11(13) & 23.26(10) & 758.6(1.7) & 565.0(1.6)\\ 
		D451 & 27.64(14) & 26.65(12) &  - & - \\ 
		D452 & 27.96(14) & 26.94(11) & 760.4(1.3) & 567.2(1.2)\\ 
		\hline
		N202 &  - & - & 605.4(2.9) & 544.8(2.7)\\ 
		N203 & 6.73(46) & 6.77(44) & 597.8(2.1) & 539.7(1.7)\\ 
		N200 & 13.69(46) & 13.59(44) & 593.2(2.3) & 535.7(2.1)\\ 
		D251 & 14.404(99) & 14.300(90) &  - & - \\ 
		D200 & 19.87(39) & 19.68(38) & 591.3(2.8) & 535.8(2.4)\\ 
		D201 & 22.75(15) & 22.52(13) &  - & - \\ 
		E250 & 24.32(48) & 24.16(46) & 590.9(1.8) & 534.6(1.4)\\ 
		\hline
		J303 & 12.78(38) & 13.08(42) & 492.0(3.0) & 507.6(2.7)\\ 
		J304 & 17.36(16) & 17.46(15) &  - & - \\ 
		E300 & 18.54(17) & 18.74(16) & 469.2(1.4) & 492.33(89)\\ 
		\hline
		J500 &  - & - & 456.6(3.5) & 489.9(2.6)\\ 
		J501 & 5.05(20) & 4.94(24) &  - & - \\ 
		\bottomrule
	\end{tabular}
	\caption{ Values of $\Delta_{ls}$  at $Q^2=9 \ \mathrm{GeV}^2$ and of $\Delta_{lc}b$  at $Q^2=4Q_m^2=36 \ \mathrm{GeV}^2$, for both  local-local $(LL)$ and local-conserved $(LC)$ discretizations of the vector currents. For  ensembles at the $SU(3)$-symmetric point, $\Delta_{ls}=0$ by construction. Results are given in units of $10^{-5}$ and correspond to  improvement coefficient set 2. }
	\label{tab:hvp_results_deltas}
\end{table}

\begin{table}
	\centering
	\renewcommand{\arraystretch}{1.2}
	\begin{tabular}{l  | c c c | c}
		\toprule
		\text{id} & \text{HP\&MLL} & HP & Kaon & Total \\
		\hline 
		H101 & -7.580(77) & -7.554(99) & - & -7.580(77) \\ 
		H102 & -2.337(19) & -2.345(24) & -1.151(15) & -3.488(33) \\ 
		N101 & -2.257(34) & -2.251(36) & -0.2606(40) & -2.518(38) \\ 
		C101 & -0.535(11) & -0.538(13) & -0.015\ 27(26) & -0.550(11) \\ 
		C102 & -0.580(13) & -0.583(15) & -0.011\ 10(24) & -0.591(13) \\ 
		D150 & 0.805(21) & 0.813(25) & 0.000\ 3850(62) & 0.806(21) \\ 
		\hline
		B450 & -5.847(52) & -5.833(67) & - & -5.847(52) \\ 
		N451 & -2.034(18) & -2.037(22) & -0.2959(37) & -2.330(21) \\ 
		N452 & -3.962(32) & -3.960(33) & -1.723(20) & -5.685(52) \\ 
		D450 & -1.187(16) & -1.186(17) & -0.02194(27) & -1.209(16) \\ 
		D451 & -1.170(16) & -1.169(17) & -0.01405(16) & -1.184(16) \\ 
		D452 & 0.722(14) & 0.727(17) & 0.001\ 455(18) & 0.723(14) \\ 
		\hline
		N202 & -8.124(64) & -8.114(74) & - & -8.124(64) \\ 
		N203 & -3.040(22) & -3.040(30) & -1.292(16) & -4.332(38) \\ 
		N200 & -0.4504(40) & -0.4544(54) & -0.072\ 83(96) & -0.5232(49) \\ 
		D251 & -2.373(18) & -2.372(19) & -0.2966(32) & -2.670(21) \\ 
		D200 & 0.2371(27) & 0.2375(35) & 0.003\ 674(46) & 0.2408(27) \\ 
		D201 & 0.2407(37) & 0.2409(46) & 0.002\ 627(36) & 0.2433(38) \\ 
		E250 & 0.1992(24) & 0.1989(27) & 0.000\ 033\ 14(29) & 0.1992(24) \\ 
		\hline
		J307 & -8.545(89) & -8.539(94) & - & -8.545(89) \\ 
		J306 & -3.208(24) & -3.208(33) & -1.247(16) & -4.455(39) \\ 
		J303 & 0.4546(32) & 0.4532(43) & 0.04117(52) & 0.4958(36) \\ 
		J304 & 0.2634(18) & 0.2618(25) & 0.01407(17) & 0.2774(20) \\ 
		E300 & 0.089\ 27(67) & 0.089\ 09(84) & 0.000\ 3658(45) & 0.089\ 64(68) \\ 
		F300 & -0.030\ 05(55) & -0.030\ 11(61) & -0.000\ 005\ 23(16) & -0.030\ 05(55) \\ 
		\hline
		J500 & -5.957(37) & -5.941(48) & - & -5.957(37) \\ 
		J501 & 0.3040(26) & 0.2972(32) & 0.1217(19) & 0.4256(43) \\ 
		\bottomrule
	\end{tabular}
	\caption{Overview of finite-volume correction to $(m_\pi
		L)_{\mathrm{ref}}$  in the HV region. The column "HP\&MLL"
		illustrates the results obtained using the Hansen-Patella
		method for $t<t^\star$ and the MLL formalism beyond that
		threshold, while the column denoted by "HP" shows results
		from the Hansen-Patella method only. The column "Kaon" shows
		the correction from the kaon, computed with HP, and already
		included in the pion correction for $SU(3)$-symmetric
		ensembles. Finally, in the last column we list the sum of "HP\&MLL" and "Kaon", which represents the total finite-volume correction entering the isovector contribution. Results are shown for the subtracted kernel with $Q^2= Q_m^2 = 9\ \mathrm{GeV}^2$ and in units of $ 10^{-5}$. }
	\label{tab:fvc_lref_HV}
\end{table}

\begin{table}
	\centering
	\renewcommand{\arraystretch}{1.2}
	\begin{tabular}{l  | c c c | c}
		\toprule
		\text{id} & \text{HP\&MLL} & HP & Kaon & Total \\
		\hline
		H101 & -23.83(32) & -23.75(38) & - & -23.83(32) \\ 
		H102 & -7.659(84) & -7.720(90) & -3.891(58) & -11.55(13) \\ 
		N101 & -7.90(12) & -7.88(12) & -0.922(15) & -8.82(13) \\ 
		C101 & -1.972(41) & -1.974(45) & -0.055\ 56(87) & -2.027(42) \\ 
		C102 & -2.133(49) & -2.135(53) & -0.040\ 24(90) & -2.174(50) \\ 
		D150 & 3.092(80) & 3.087(89) & 0.001\ 481(23) & 3.094(81) \\ 
		\hline
		B450 & -18.03(16) & -17.90(25) & - & -18.03(16) \\ 
		N451 & -7.087(71) & -7.087(78) & -1.038(13) & -8.125(84) \\ 
		N452 & -13.19(13) & -13.19(13) & -5.901(84) & -19.09(21) \\ 
		D450 & -4.395(57) & -4.390(61) & -0.080\ 62(89) & -4.476(58) \\ 
		D451 & -4.329(57) & -4.323(60) & -0.051\ 43(57) & -4.380(57) \\ 
		D452 & 2.738(55) & 2.737(61) & 0.005\ 500(59) & 2.743(55) \\ 
		\hline
		N202 & -25.42(28) & -25.41(30) & - & -25.42(28) \\ 
		N203 & -10.057(98) & -10.06(12) & -4.408(66) & -14.47(16) \\ 
		N200 & -1.527(13) & -1.559(19) & -0.2530(36) & -1.780(17) \\ 
		D251 & -8.344(71) & -8.338(75) & -1.047(12) & -9.391(83) \\ 
		D200 & 0.879(11) & 0.879(13) & 0.013\ 48(16) & 0.892(11) \\ 
		D201 & 0.885(14) & 0.885(16) & 0.009\ 61(13) & 0.895(14) \\ 
		E250 & 0.7662(89) & 0.765(10) & 0.000\ 128 \ 89(96) & 0.7664(89) \\ 
		\hline
		J307 & -26.86(39) & -26.85(40) & - & -26.86(39) \\ 
		J306 & -10.57(12) & -10.57(14) & -4.253(62) & -14.82(19) \\ 
		J303 & 1.573(11) & 1.574(16) & 0.1444(18) & 1.717(13) \\ 
		J304 & 0.9158(65) & 0.9109(91) & 0.048\ 86(65) & 0.9647(70) \\ 
		E300 & 0.3346(28) & 0.3337(31) & 0.001\ 374(13) & 0.3360(28) \\ 
		F300 & -0.1158(20) & -0.1159(23) & -0.000\ 020 \ 38(60) & -0.1158(20) \\ 
		\hline
		J500 & -18.53(15) & -18.37(20) & - & -18.53(15) \\ 
		J501 & 1.063(12) & 0.956(12) & 0.4096(70) & 1.473(16) \\ 
		\bottomrule
	\end{tabular}
	\caption{ Same as table~\ref{tab:fvc_lref_HV} for the MV region. Results are shown for $Q^2/4=2.25\ \mathrm{GeV}^2$ and in units of $ 10^{-5}$. }
	\label{tab:fvc_lref_IV}
\end{table}

\begin{table}
	\centering
	\renewcommand{\arraystretch}{1.2}
	\begin{tabular}{l  | c c c | c}
		\toprule
		\text{id} & \text{HP\&MLL} & HP & Kaon & Total \\
		\hline 
		H101 & -54.8(1.0) & -53.4(1.7) & - & -54.8(1.0) \\ 
		H102 & -19.16(65) & -19.56(55) & -10.53(48) & -29.68(82) \\ 
		N101 & -26.25(61) & -24.91(77) & -2.77(11) & -29.02(69) \\ 
		C101 & -7.94(19) & -7.99(26) & -0.1806(64) & -8.12(20) \\ 
		C102 & -8.60(27) & -8.60(30) & -0.1280(54) & -8.73(27) \\ 
		D150 & 22.14(83) & 21.64(88) & 0.006\ 19(19) & 22.14(83) \\ 
		\hline
		B450 & -39.04(63) & -38.0(1.2) & - & -39.04(63) \\ 
		N451 & -21.81(35) & -21.68(52) & -3.04(12) & -24.86(41) \\ 
		N452 & -34.85(57) & -34.47(89) & -16.50(73) & -51.4(1.2) \\ 
		D450 & -19.30(31) & -19.08(43) & -0.2721(88) & -19.58(31) \\ 
		D451 & -19.16(28) & -18.83(44) & -0.1698(56) & -19.33(28) \\ 
		D452 & 15.90(38) & 15.64(48) & 0.020\ 88(60) & 15.92(38) \\ 
		\hline
		N202 & -57.34(88) & -56.4(1.5) & - & -57.34(88) \\ 
		N203 & -25.96(37) & -25.77(71) & -12.19(54) & -38.15(76) \\ 
		N200 & -4.367(96) & -4.55(13) & -0.722(30) & -5.09(10) \\ 
		D251 & -26.96(34) & -26.67(59) & -3.13(12) & -30.09(44) \\ 
		D200 & 3.95(13) & 3.84(13) & 0.0451(15) & 3.99(13) \\ 
		D201 & 3.90(11) & 3.78(11) & 0.0316(10) & 3.93(11) \\ 
		E250 & 6.138(89) & 6.04(12) & 0.000\ 575(14) & 6.139(89) \\ 
		\hline
		J307 & -61.2(1.5) & -60.3(2.1) & - & -61.2(1.5) \\ 
		J306 & -26.99(59) & -26.75(91) & -11.72(50) & -38.71(90) \\ 
		J303 & 5.032(95) & 4.87(12) & 0.421(16) & 5.453(97) \\ 
		J304 & 2.968(55) & 2.830(68) & 0.1366(52) & 3.105(56) \\ 
		E300 & 1.784(18) & 1.725(31) & 0.005\ 06(13) & 1.789(18) \\ 
		F300 & -0.911(20) & -0.916(26) & -000\ 0914(35) & -0.911(20) \\ 
		\hline
		J500 & -40.87(89) & -39.70(96) & - & -40.87(89) \\ 
		J501 & 2.725(63) & 2.280(60) & 1.094(46) & 3.819(79) \\ 
		
		\bottomrule
	\end{tabular}
	\caption{ Same as table~\ref{tab:fvc_lref_HV} for the LV region. Results are shown for $Q^2/16=0.5625\ \mathrm{GeV}^2$ and in units of $ 10^{-5}$. }
	\label{tab:fvc_lref_LV}
\end{table}

\section{The strange contribution}\label{app:strange_contribution}
In addition to the other channels discussed in the main text, we also
determine the strange quark contribution to the running of $\alpha$ in
the space-like momentum range $0.025 \ \mathrm{GeV}^2 \leq Q^2 \leq 12
\ \mathrm{GeV}^2$. This contribution is phenomenologically relevant
since, together with the isovector and isoscalar results, it enables
the transition from the strong isospin to the full flavour basis, as
advocated in \cite{Erler:2024lds}. The strange contribution is
evaluated following the same decomposition introduced in
Eq.~\ref{eq:hvp_splitting}, with each term computed using the
non-subtracted kernels in Eq.~\ref{eq:non_sub_kernel_QSD_QID} and
Eq.~\ref{eq:non_sub_kernel_QLD} for the HV, MV and LV virtuality regions, respectively.

\begin{figure}
	\centering
	\includegraphics[scale=0.3]{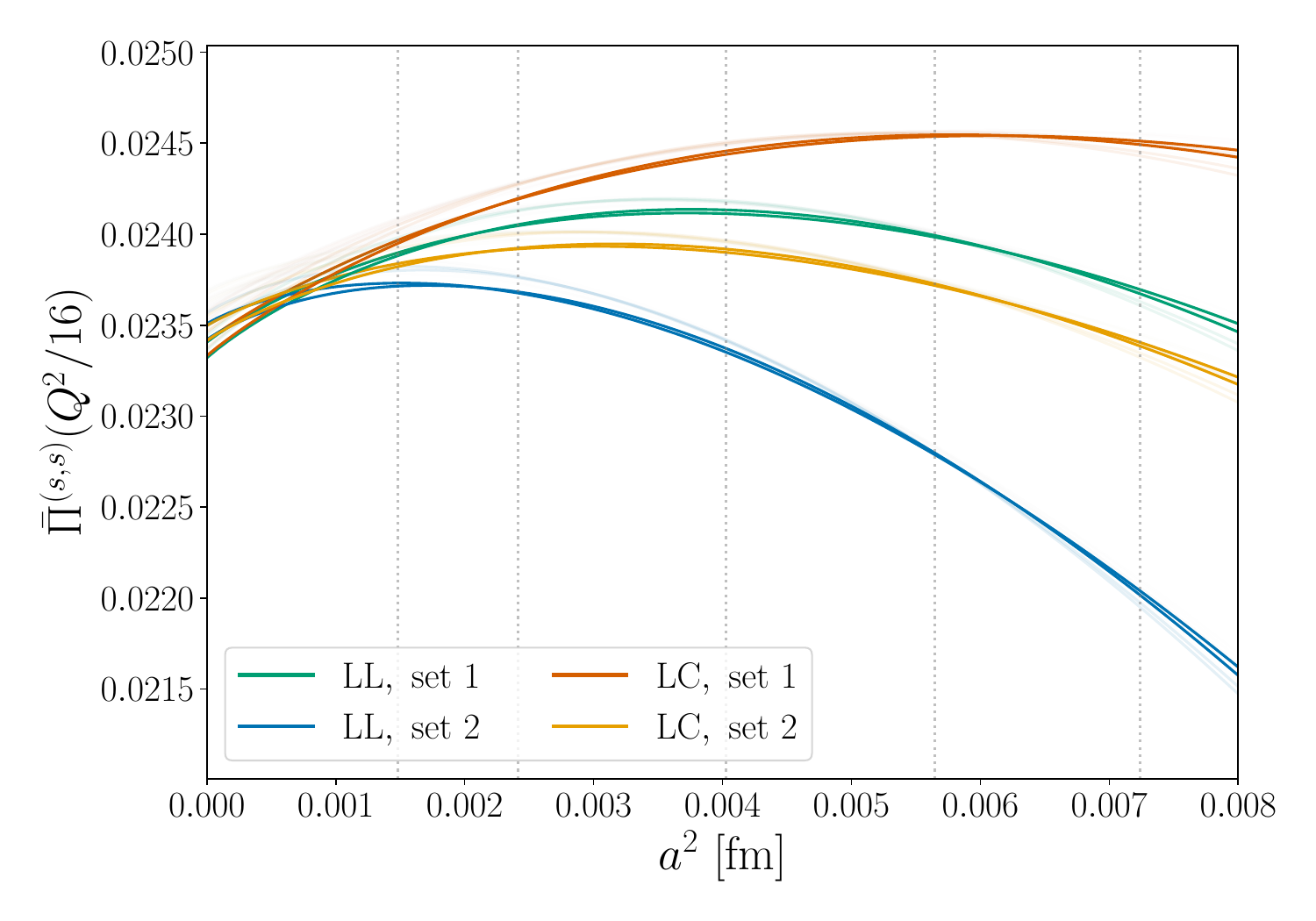}
	\includegraphics[scale=0.31]{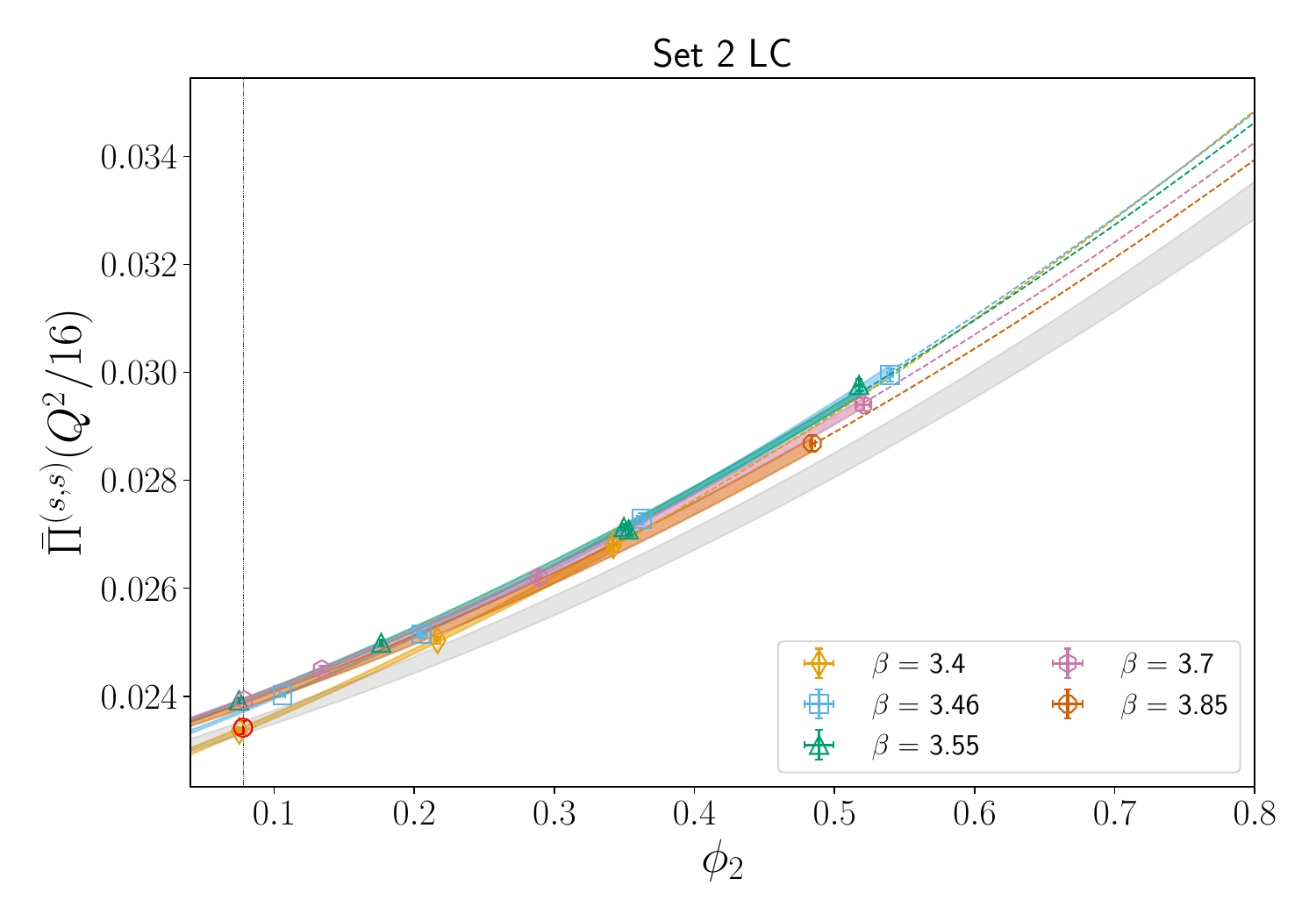}
	\caption{Illustration of fits to the strange contribution
          $\widehat{\Pi}^{(s,s)}(Q^2/16)$ at $Q^2=9\ \mathrm{GeV}^2$
          in the LV region. \textit{Left:} continuum limit behaviour
          for the four sets of data based on different improvement schemes and discretizations of the vector current. Each line corresponds to a single fit, with the opacity associated to the weights as given by our model average prescription. \textit{Right}: chiral approach to the physical pion mass for one of the fits with the highest weight. Data points are projected to $\phi_4^{\mathrm{phys}}$. Coloured lines denote the chiral trajectories at finite lattice spacings, while the grey band shows the dependence on $\phi_2$ in the continuum.
	}
	\label{fig:piss_cl}
\end{figure}

For the chiral-continuum extrapolation, we employ fit ans\"{a}tze
analogous to those described in section~\ref{sec:phys_point_extrap},
with the inclusion of $O(a^4)$ lattice artefacts in the HV region
only. In the LV region instead, we observe enhanced sensitivity to
chiral effects, though fits with chirally divergent terms are not
included in our final analysis. In the same region, higher-order
discretization effects of $O(a^3)$ are required to adequately describe the approach to the continuum. A summary of our fits for the LV region is shown in figure~\ref{fig:piss_cl}. The left panel illustrates the continuum dependence for the four lattice datasets, while the right panel displays the light-quark mass dependence to the physical point.

\begin{table}	
		\centering
		\begin{tabular}{c c }
			\toprule
			$Q^2 \ [\mathrm{GeV}^2]$ & $\bar\Pi^{(s,s)}$   \\
			\noalign{\smallskip}\hline\noalign{\smallskip}
			
			$0.5$ & 2329(12)(7)(30)[33]    \\
			
			$1.0$ & 3729(16)(7)(36)[40]  \\ 
			
			$2.0$ & 5451(18)(11)(40)[45]  \\ 
			
			$3.0$ & 6547(19)(15)(42)[49] \\ 
			
			$4.0$ &  7245(27)(36)(38)[59] \\
			
			$5.0$ & 7858(26)(29)(39)[55]    \\
			
			$6.0$ &  8365(25)(24)(39)[52]  \\ 
			
			$7.0$ &  8797(24)(22)(40)[51] \\ 
			
			$8.0$ & 9173(24)(20)(40)[51]  \\ 
			
			$9.0$ & 9507(23)(19)(41)[50] \\ 
			
			$12.0$ & 10327(23)(18)(42)[51]  \\ 
			\bottomrule			
		\end{tabular}
	\caption{Contribution to the running for the strange channel at the physical point, in units of $ 10^{-5}$. The first quoted uncertainty corresponds to the statistical error, the second to systematics from model exploration, and the third to the scale-setting error. The final uncertainty in squared brackets is the sum in quadrature of the previous ones. }
	\label{tab:strange_res}
\end{table}

\begin{figure} [h!]
	\centering
	\includegraphics[scale=0.42]{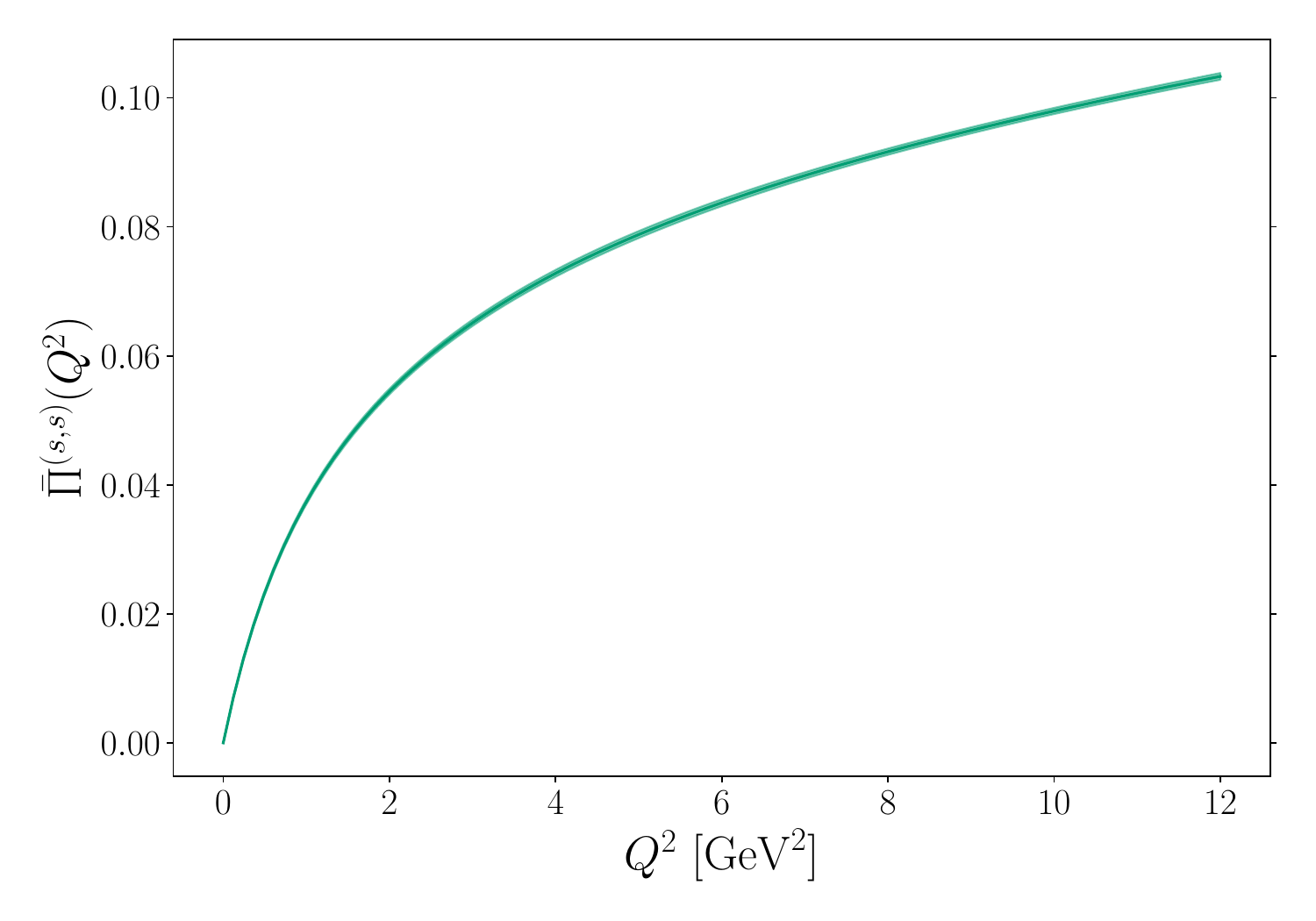}
	\caption{Strange contribution $\bar\Pi^{(s,s)}$ to the running at the physical point as a function of the squared momentum transfer $Q^2$.}
	\label{fig:piss_running}
\end{figure}

A summary of our results for several values of $Q^2$ is given in
table~\ref{tab:strange_res}. In addition, we provide the analytic solution to the running of the strange component. The rational approximant we extract from the fit is 
\begin{equation}
	\bar\Pi^{(s,s)}(-Q^2) \approx 
	\frac{
		0.062164x + 0.004783x^2
	}
	{
		1 + 0.764044x + 0.025822x^2  
	}, \qquad x = \frac{Q^2}{\mathrm{GeV}^2},
\end{equation}
together with the covariance matrix
\begin{equation}
	\mathrm{cov}^{(s,s)} \begin{bmatrix}
		a_1 \\ a_2 \\ b_1 \\ b_2  
	\end{bmatrix}
	=
	\begin{bmatrix}
		 0.141398  & 0.0223985  & 1.93356 &  0.119207\\
		 0.0223985 & 0.0661053  & 1.62774 &  0.512815\\
		 1.93356  &  1.62774   & 55.9891   & 11.9994\\
		 0.119207  & 0.512815 &  11.9994  &  4.0041
	\end{bmatrix} \times 10^{-5}.
\end{equation}
Also here we observe that the approximant is in good agreement with
the directly computed data and reproduces the error band very accurately. Our results for the running of the strange component are displayed in figure~\ref{fig:piss_running}

For completeness, we also provide the statistical correlations between the different HVP components, $\bar\Pi^{(3,3)}$, $\bar\Pi^{(8,8)}$, $\bar\Pi^{(0,8)}$, $\bar\Pi^{(c,c)}$, and $\bar\Pi^{(s,s)}$, at representative momenta $Q^2 = 1$ and $9~\mathrm{GeV}^2$. These correlations can be useful for phenomenological applications where combined flavour contributions enter global analyses of the hadronic running of the electroweak couplings. At $Q^2=1 \ \mathrm{GeV}^2$, the correlation matrix reads
\begin{equation}
	\mathrm{corr} \begin{bmatrix}
		\bar\Pi^{(3,3)} \\ \bar\Pi^{(8,8)} \\ \bar\Pi^{(0,8)} \\ \bar\Pi^{(c,c)}  \\ \bar\Pi^{(s,s)}
	\end{bmatrix}
	=
	\begin{bmatrix}
		1.0     &   0.377044  & 0.260078 & 0.0615657 & 0.546812\\
		0.377044  & 1.0   &     0.248847 & 0.0553464 & 0.36846\\
		0.260078 &  0.248847 &  1.0    &   0.102147 &  0.38444\\
		0.0615657 & 0.0553464 & 0.102147 & 1.0    &    0.136224\\
		0.546812 &  0.36846  &  0.38444 &  0.136224  & 1.0
	\end{bmatrix},
\end{equation}
while at $Q^2= 9 \ \mathrm{GeV}^2$ we find
\begin{equation}
	\mathrm{corr} \begin{bmatrix}
		\bar\Pi^{(3,3)} \\ \bar\Pi^{(8,8)} \\ \bar\Pi^{(0,8)} \\ \bar\Pi^{(c,c)}  \\ \bar\Pi^{(s,s)}
	\end{bmatrix}
	=
	\begin{bmatrix}
		1.0     &   0.342392  & 0.200783 & 0.0333197 & 0.390514\\
		0.342392  & 1.0   &     0.187271 & 0.0542744  & 0.310952\\
		0.200783 &  0.187271 &  1.0    &   0.120811  & 0.306019\\
		0.0333197 & 0.0542744  & 0.120811 & 1.0    &    0.156293\\
		0.390514 &  0.310952  & 0.306019 & 0.156293  & 1.0
	\end{bmatrix}.
\end{equation}

\section{Lattice results for $\Delta\alpha_{\mathrm{had}}(-Q^2)$ versus data-driven estimates using a modified $R$-ratio}\label{app:modified_Rratio}

As discussed in section~\ref{sec:had_run}, our lattice results for $\Delta\alpha_{\mathrm{had}}(-Q^2)$ show tensions to those obtained
via the data driven-approach, 
\begin{align}
\Delta\alpha_{\mathrm{had}}(-Q^2)|_{R} = \frac{\alpha Q^2}{3\pi}\int_0^{\infty} ds\frac{R(s)}{s(s + Q^2)}\ ,
\end{align}
where $R(s)$ denotes the so called $R$-ratio obtained by $e^+e^-\to \text{hadron}$ cross-section data:
\begin{align}
R(s) = \frac{\sigma(e^+e^- \to \gamma^{*}\to \text{had})}{4\pi\alpha^2(s)/(3s)}\ .
\end{align}
In this appendix, we investigate the origin of the tension.

\begin{figure}[!h]
\centering{
\includegraphics[width=0.48\textwidth]{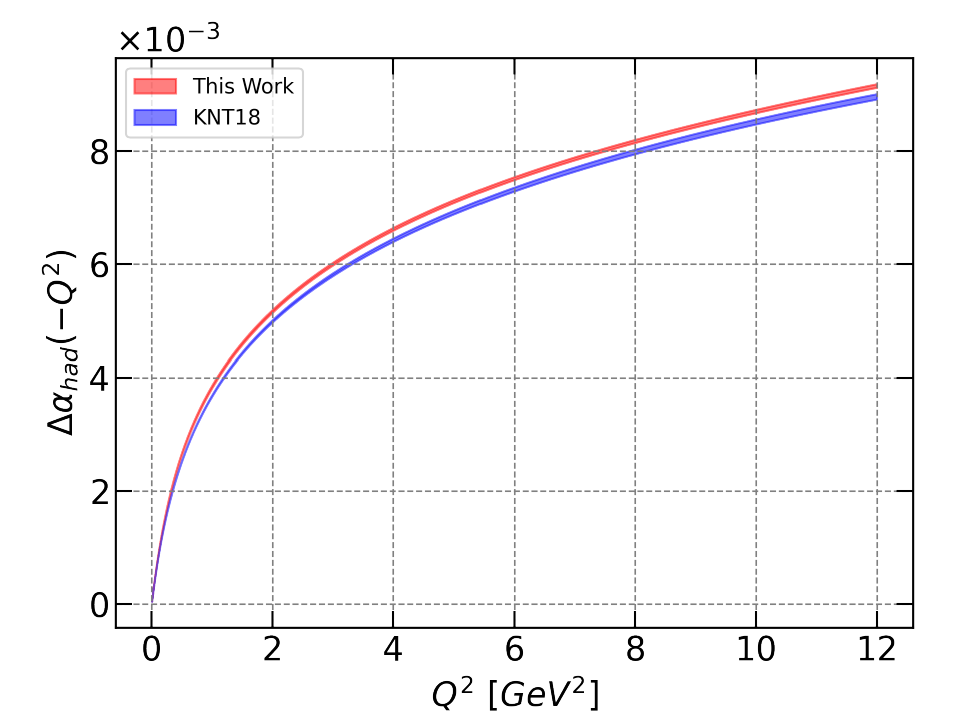}
\includegraphics[width=0.48\textwidth]{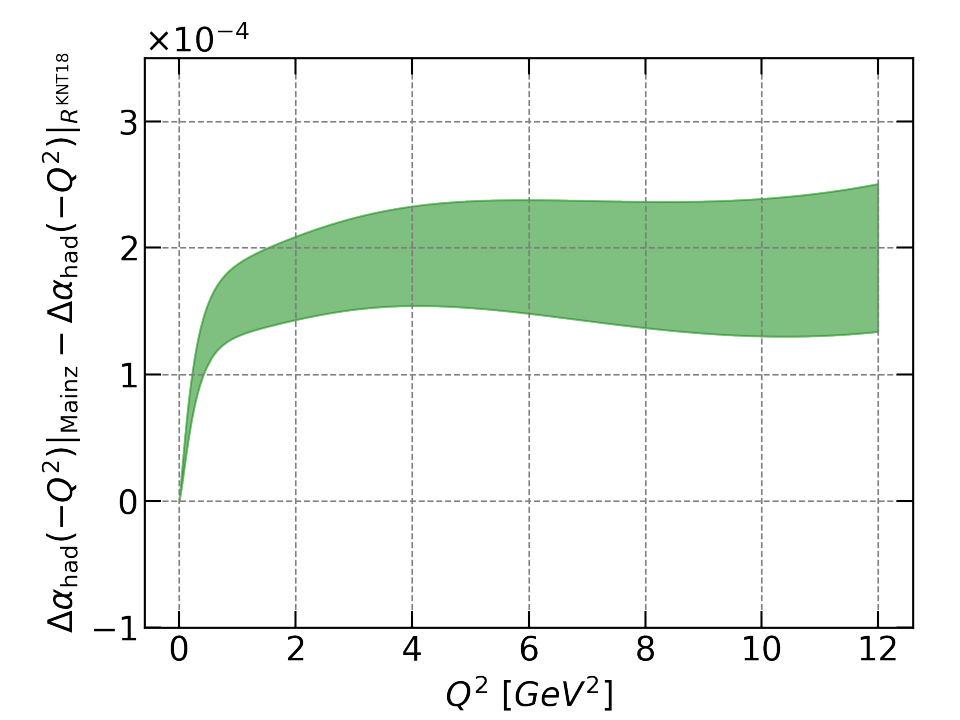}
\caption{\textit{Left}: Our lattice result for $\Delta\alpha_{\mathrm{had}}(-Q^2)$ (red band) compared with the one
by the data-driven approach (blue band)~\protect\cite{Keshavarzi:2018mgv,Keshavarzi:2019abf}. \textit{Right}: The difference between two methods shown in the left panel.
}
\label{fig:alp_lat_vs_r}
}
\end{figure}

In the left panel of figure~\ref{fig:alp_lat_vs_r},
the red band shows our $\Delta\alpha_{\mathrm{had}}(-Q^2)$ obtained
from the Pad\'{e} fits in Eqs.~\eqref{eq:pade_pigg} and~\eqref{eq:pade_pigg_corr_matrix}.
This is compared with the one obtained via the data-driven approach (blue band)
using the $R$-ratio (KNT18)~\protect\cite{Keshavarzi:2018mgv,Keshavarzi:2019abf}.
Across the entire $Q^2$ range, our results lie systematically above the phenomenological estimates. The right panel displays the difference between the two determinations, showing that the tension exceed $5\sigma$ in the low-energy region $Q^2 \lesssim 1$~GeV${}^2$. Here, we investigate whether such a discrepancy could be accounted for by modifying the $R$-ratio in some specific intervals of the centre-of-mass energy $\sqrt{s}$.

\begin{figure}[!h]
\centering{
\includegraphics[width=0.6\textwidth]{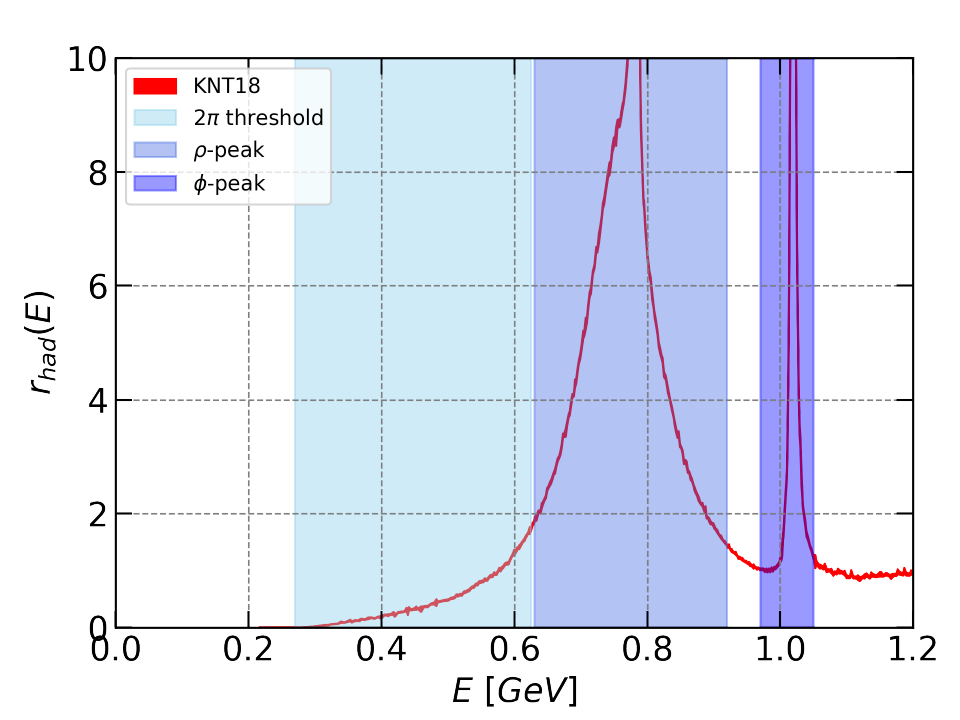}
\caption{ The $R$-ratio provided by Keshavarzi-Nomura-Teubner (KNT18)
and used in Refs.~\protect\cite{Keshavarzi:2018mgv,Keshavarzi:2019abf}. The coloured region will be modified by 6\%, respectively.
}\label{fig:r_mod}
}
\end{figure}

We consider a modification for three regions in the $R$-ratio:
two-pion threshold ($E = \sqrt{s} \leq 0.63$~GeV),
$\rho$ meson peak ($E \in [0.63,\ 0.92]$~GeV),
and $\phi$ meson peak regions ($E \in [0.97,\ 1.05]$~GeV).
The selection scheme for the range is illustrated in figure~\ref{fig:r_mod}.

\begin{figure}[!h]
\centering{
\includegraphics[width=0.32\textwidth]{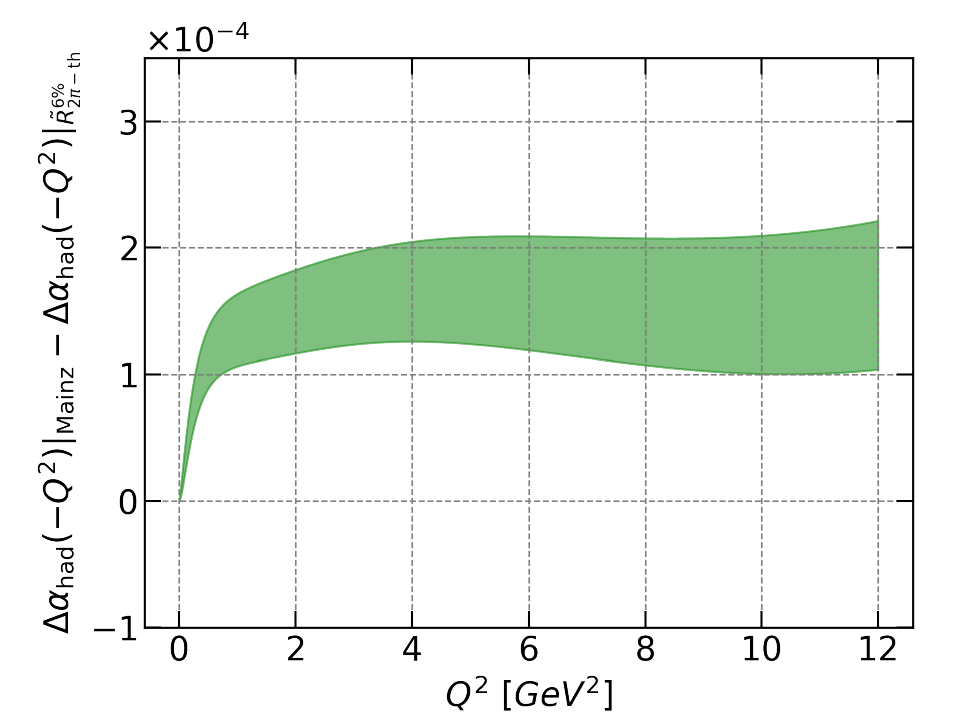}
\includegraphics[width=0.32\textwidth]{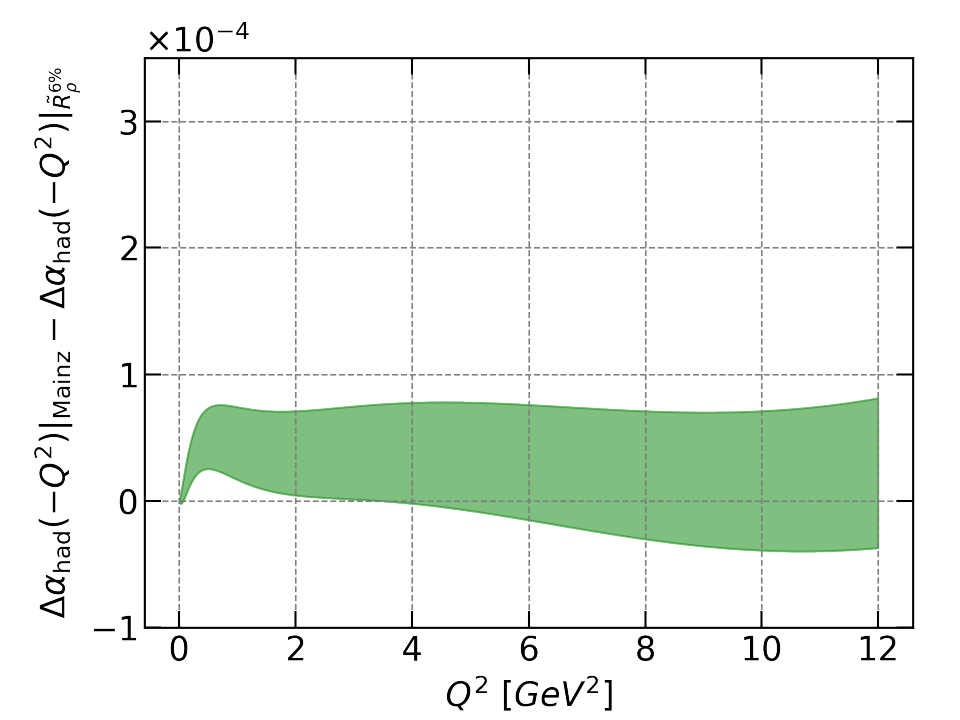}
\includegraphics[width=0.32\textwidth]{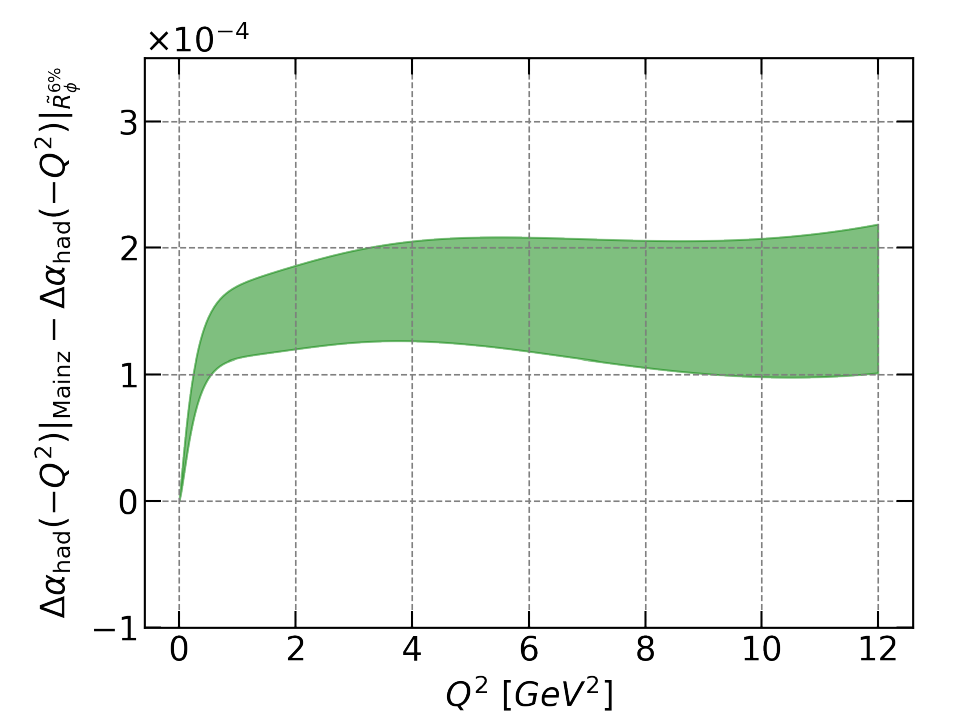}
\caption{The difference between our $\Delta\alpha_{\mathrm{had}}(-Q^2)$ and the data-driven results with 6\% modifications
of two pion threshold (\textit{left}),
$\rho$ meson peak (\textit{middle}),
$\phi$ meson peak (\textit{right}) regions in the $R$-ratio.
}
\label{fig:diff_alp_lat_vs_r}
}
\end{figure}

In figure~\ref{fig:diff_alp_lat_vs_r},
we show the difference between our $\Delta\alpha_{\mathrm{had}}(-Q^2)$ and the data-driven results
with a 6\% modification in each range of the $R$-ratio.
In the left panel, the two-pion threshold region is modified as indicated by a suffix ($\tilde{R}^{6\%}_{2\pi\text{-th}}$).
The impact of the modification is very limited; the modified data-driven results are not reconciled with our lattice data.

The middle panel of figure~\ref{fig:diff_alp_lat_vs_r} shows the
result after modifying the region around the $\rho$ meson peak
(suffix: $\tilde{R}^{6\%}_{\rho}$), which shows that the tension has
disappeared for most of the $Q^2$ region. A small remaining tension at
low $Q^2$ may come from the fact that
$\Delta\alpha_{\mathrm{had}}(-Q^2)|_{\text{Mainz}}$ represents the
isoQCD result.  Since the vector current correlator in isoQCD is
dominated by neutral pions at large distances, it tends to become
larger than the real-world counterpart with charged pions (slightly
heavier than the neutral pions). If we take into account
isospin-breaking corrections, our estimate for
$\Delta\alpha_{\mathrm{had}}(-Q^2)|_{\text{Mainz}}$ will be slightly
suppressed in the low $Q^2$ region via the pion mass shift and become
more consistent with the data-driven results with the $\rho$ meson
peak modification.

Finally, in the right panel figure~\ref{fig:diff_alp_lat_vs_r}, we show the result with a $\phi$ meson peak modification.
Similarly to the two-pion threshold modification (left panel),
we find a tiny effect which cannot explain the existing tension.
In summary, the tension shown in figure~\ref{fig:alp_lat_vs_r}
can be explained by modifying solely the region around the $\rho$ peak in the $R$-ratio.
Our results are consistent with the recent work~\cite{Davier:2023cyp} based on the similar analyses.

\end{appendix}

\bibliographystyle{JHEPg6l10}
\bibliography{biblio}

@misc{SuppMat,
	howpublished = {See Supplemental Material at \url{https://link.aps.org/supplemental/10.1103/bk78-wtvd} for chiral-continuum extrapolation plots and fit quality indicators for all channels entering the analysis.}
}

@article{Conigli:2026xxx,
	author = "Conigli, Alessandro and Djukanovic, Dalibor and von Hippel, Georg and Kuberski, Simon and Meyer, Harvey B. and Miura, Kohtaroh and Ottnad, Konstantin and Risch, Andreas and Wittig, Hartmut",
	title = "{Running of the Electroweak Gauge Couplings from First Principles}",
	eprint = "2607.03370",
	archivePrefix = "arXiv",
	primaryClass = "hep-lat",
	reportNumber = "MITP-26-001, CERN-TH-2026-004, KEK-TH-2795",
	doi = "10.1103/r2cx-tl7s",
	journal = "Phys. Rev. Lett.",
	volume = "137",
	number = "7",
	pages = "071901",
	year = "2026"
}

@article{Morel:2020dww,
    author = {Morel, L{\'e}o and Yao, Zhibin and Clad{\'e}, Pierre and Guellati-Kh{\'e}lifa, Sa{\"\i}da},
    title = "{Determination of the fine-structure constant with an accuracy of 81 parts per trillion}",
    doi = "10.1038/s41586-020-2964-7",
    journal = "Nature",
    volume = "588",
    number = "7836",
    pages = "61--65",
    year = "2020"
}

@article{Erler:2004in,
	author = "Erler, Jens and Ramsey-Musolf, Michael J.",
	title = "{The Weak mixing angle at low energies}",
	eprint = "hep-ph/0409169",
	archivePrefix = "arXiv",
	reportNumber = "FT-2004-03, CALTECH-MAP-300",
	doi = "10.1103/PhysRevD.72.073003",
	journal = "Phys. Rev. D",
	volume = "72",
	pages = "073003",
	year = "2005"
}

@article{Fan:2022eto,
    author = "Fan, X. and Myers, T. G. and Sukra, B. A. D. and Gabrielse, G.",
    title = "{Measurement of the Electron Magnetic Moment}",
    eprint = "2209.13084",
    archivePrefix = "arXiv",
    primaryClass = "physics.atom-ph",
    doi = "10.1103/PhysRevLett.130.071801",
    journal = "Phys. Rev. Lett.",
    volume = "130",
    number = "7",
    pages = "071801",
    year = "2023"
}

@article{Parker:2018sci,
    author = {Parker, Richard H. and Yu, Chenghui and Zhong, Weicheng and Estey, Brian and M{\"u}ller, Holger},
    title = "{Measurement of the fine-structure constant as a test of the Standard Model}",
    eprint = "1812.04130",
    archivePrefix = "arXiv",
    primaryClass = "physics.atom-ph",
    doi = "10.1126/science.aap7706",
    journal = "Science",
    volume = "360",
    pages = "191",
    year = "2018"
}

@article{DiLuzio:2024sps,
	author = "Di Luzio, Luca and Keshavarzi, Alexander and Masiero, Antonio and Paradisi, Paride",
	title = "{Model-Independent Tests of the Hadronic Vacuum Polarization Contribution to the Muon g-2}",
	eprint = "2408.01123",
	archivePrefix = "arXiv",
	primaryClass = "hep-ph",
	doi = "10.1103/PhysRevLett.134.011902",
	journal = "Phys. Rev. Lett.",
	volume = "134",
	number = "1",
	pages = "011902",
	year = "2025"
}

@article{CMD-3:2023alj,
	author = "Ignatov, F. V. and others",
	collaboration = "CMD-3",
	title = "{Measurement of the e+e-{\textrightarrow}{\ensuremath{\pi}}+{\ensuremath{\pi}}- cross section from threshold to 1.2~GeV with the CMD-3 detector}",
	eprint = "2302.08834",
	archivePrefix = "arXiv",
	primaryClass = "hep-ex",
	doi = "10.1103/PhysRevD.109.112002",
	journal = "Phys. Rev. D",
	volume = "109",
	number = "11",
	pages = "112002",
	year = "2024"
}

@article{CMD-3:2023rfe,
	author = "Ignatov, F. V. and others",
	collaboration = "CMD-3",
	title = "{Measurement of the Pion Form Factor with CMD-3 Detector and its Implication to the Hadronic Contribution to Muon (g-2)}",
	eprint = "2309.12910",
	archivePrefix = "arXiv",
	primaryClass = "hep-ex",
	doi = "10.1103/PhysRevLett.132.231903",
	journal = "Phys. Rev. Lett.",
	volume = "132",
	number = "23",
	pages = "231903",
	year = "2024"
}

@article{Ramos:2018vgu,
	author = "Ramos, Alberto",
	title = "{Automatic differentiation for error analysis of Monte Carlo data}",
	eprint = "1809.01289",
	archivePrefix = "arXiv",
	primaryClass = "hep-lat",
	doi = "10.1016/j.cpc.2018.12.020",
	journal = "Comput. Phys. Commun.",
	volume = "238",
	pages = "19--35",
	year = "2019"
}

@article{Meyer:2017hjv,
    author = "Meyer, Harvey B.",
    title = "{Lorentz-covariant coordinate-space representation of the leading hadronic contribution to the anomalous magnetic moment of the muon}",
    eprint = "1706.01139",
    archivePrefix = "arXiv",
    primaryClass = "hep-lat",
    doi = "10.1140/epjc/s10052-017-5200-3",
    journal = "Eur. Phys. J. C",
    volume = "77",
    number = "9",
    pages = "616",
    year = "2017"
}

@mastersthesis{ToelleMasterArbeit,
    author = "T{\"o}lle, C.",
    title = "{Coordinate space methods for vacuum polarization calculations in lattice regularization}",
    school = "{Johannes Gutenberg Universit\"at Mainz}",
    year = "2025"
}

@article{Harris:2025xvk,
	author = "Harris, Tim and Meyer, Harvey B.",
	title = "{Nonsinglet vector current in lattice QCD: O(a)-improvement from large volumes}",
	eprint = "2510.06869",
	archivePrefix = "arXiv",
	primaryClass = "hep-lat",
	reportNumber = "CERN-TH-2025-189, MITP-25-060",
	doi = "10.1103/hpqh-tstb",
	journal = "Phys. Rev. D",
	volume = "113",
	number = "11",
	pages = "114510",
	year = "2026"
}

@article{Gerardin:2018kpy,
        author = "Gerardin, Antoine and Harris, Tim and Meyer, Harvey B.",
	title = "{Nonperturbative renormalization and $O(a)$-improvement of the nonsinglet vector current with $N_f=2+1$ Wilson fermions and tree-level Symanzik improved gauge action}",
	eprint = "1811.08209",
	archivePrefix = "arXiv",
	primaryClass = "hep-lat",
	reportNumber = "MITP/18-106",
	doi = "10.1103/PhysRevD.99.014519",
	journal = "Phys. Rev. D",
	volume = "99",
	number = "1",
	pages = "014519",
	year = "2019"
}

@article{Hansen:2019rbh,
	author = "Hansen, Maxwell T. and Patella, Agostino",
	title = "{Finite-volume effects in $(g-2)^{\text{HVP,LO}}_\mu$}",
	eprint = "1904.10010",
	archivePrefix = "arXiv",
	primaryClass = "hep-lat",
	reportNumber = "CERN-TH-2019-051",
	doi = "10.1103/PhysRevLett.123.172001",
	journal = "Phys. Rev. Lett.",
	volume = "123",
	pages = "172001",
	year = "2019"
}

@article{Hansen:2020whp,
	author = "Hansen, Maxwell T. and Patella, Agostino",
	title = "{Finite-volume and thermal effects in the leading-HVP contribution to muonic ($g - 2$)}",
	eprint = "2004.03935",
	archivePrefix = "arXiv",
	primaryClass = "hep-lat",
	reportNumber = "CERN-TH-2020-053, HU-EP-20/08",
	doi = "10.1007/JHEP10(2020)029",
	journal = "JHEP",
	volume = "10",
	pages = "029",
	year = "2020"
}

@article{Davier:2023hhn,
	author = "Davier, M. and D\'\i{}az-Calder\'on, D. and Malaescu, B. and Pich, A. and Rodr\'\i{}guez-S\'anchez, A. and Zhang, Z.",
	title = "{The Euclidean Adler function and its interplay with $ \Delta {\alpha}_{\textrm{QED}}^{\textrm{had}} $ and \ensuremath{\alpha}$_{s}$}",
	eprint = "2302.01359",
	archivePrefix = "arXiv",
	primaryClass = "hep-ph",
	doi = "10.1007/JHEP04(2023)067",
	journal = "JHEP",
	volume = "04",
	pages = "067",
	year = "2023"
}

@article{ParticleDataGroup:2024cfk,
	author = "Navas, S. and others",
	collaboration = "Particle Data Group",
	title = "{Review of particle physics}",
	doi = "10.1103/PhysRevD.110.030001",
	journal = "Phys. Rev. D",
	volume = "110",
	number = "3",
	pages = "030001",
	year = "2024"
}

@article{Keshavarzi:2018mgv,
	author = "Keshavarzi, Alexander and Nomura, Daisuke and Teubner, Thomas",
	title = "{Muon $g-2$ and $\alpha(M_Z^2)$: a new data-based analysis}",
	eprint = "1802.02995",
	archivePrefix = "arXiv",
	primaryClass = "hep-ph",
	reportNumber = "LTH 1153, KEK-TH-2035, LTH-1153, YITP-18-09, LTH 1153; KEK-TH-2035; YITP-18-09",
	doi = "10.1103/PhysRevD.97.114025",
	journal = "Phys. Rev. D",
	volume = "97",
	number = "11",
	pages = "114025",
	year = "2018"
}

@article{Jegerlehner:2017zsb,
	author = "Jegerlehner, Fred",
	editor = "Denig, A. and Redmer, C. -F.",
	title = "{Variations on Photon Vacuum Polarization}",
	eprint = "1711.06089",
	archivePrefix = "arXiv",
	primaryClass = "hep-ph",
	reportNumber = "DESY-17-194, HU-EP-17-24",
	doi = "10.1051/epjconf/201921801003",
	journal = "EPJ Web Conf.",
	volume = "218",
	pages = "01003",
	year = "2019"
}

@article{Jeger1986,
    author = "Jegerlehner, F.",
    title = "{Hadronic Contributions to Electroweak Parameter Shifts: A Detailed Analysis}",
    reportNumber = "BI-TP 85/28",
    doi = "10.1007/BF01552495",
    journal = "Z. Phys. C",
    volume = "32",
    pages = "195",
    year = "1986"
}

@article{Czarnecki:2000ic,
	author = "Czarnecki, Andrzej and Marciano, William J.",
	editor = "Heusch, C. A.",
	title = "{Polarized M{\o}ller scattering asymmetries}",
	eprint = "hep-ph/0003049",
	archivePrefix = "arXiv",
	reportNumber = "BNL-HET-00-2",
	doi = "10.1016/S0217-751X(00)00243-0",
	journal = "Int. J. Mod. Phys. A",
	volume = "15",
	pages = "2365--2376",
	year = "2000"
}

@article{Burger:2015lqa,
	author = "Burger, Florian and Jansen, Karl and Petschlies, Marcus and Pientka, Grit",
	title = "{Leading hadronic contributions to the running of the electroweak coupling constants from lattice QCD}",
	eprint = "1505.03283",
	archivePrefix = "arXiv",
	primaryClass = "hep-lat",
	reportNumber = "DESY-15-056, HU-EP-15-23, SFB-CPP-14-126",
	doi = "10.1007/JHEP11(2015)215",
	journal = "JHEP",
	volume = "11",
	pages = "215",
	year = "2015"
}

@misc{Jegerlehner:2019alphaQEDc19,
	author       = "F. Jegerlehner",
	title        = "{alphaQEDc19}",
	year         = {2019},
  howpublished = {\url{http://www-com.physik.hu-berlin.de/~fjeger/software.html}}
}

@misc{Jegerlehner:2019pQCDAdler,
	author       = "F. Jegerlehner",
	title        = "{pQCDAdler}",
	year         = {2012},
	howpublished = {\url{ http://www-com.physik.hu-berlin.de/~fjeger/ software.html}}
}

@article{Francis:2015grz,
	author = {Francis, Anthony and G{\"u}lpers, Vera and Herdo{\'\i}za, Gregorio and Horch, Hanno and J{\"a}ger, Benjamin and Meyer, Harvey B. and Wittig, Hartmut},
	title = "{Study of the hadronic contributions to the running of the QED coupling and the weak mixing angle}",
	eprint = "1511.04751",
	archivePrefix = "arXiv",
	primaryClass = "hep-lat",
	journal = "PoS",
	volume = "LATTICE2015",
	pages = "110",
	year = "2015"
}

@article{Budapest-Marseille-Wuppertal:2017okr,
	author = "Borsanyi, Sz. and others",
	collaboration = "Budapest-Marseille-Wuppertal",
	title = "{Hadronic vacuum polarization contribution to the anomalous magnetic moments of leptons from first principles}",
	eprint = "1711.04980",
	archivePrefix = "arXiv",
	primaryClass = "hep-lat",
	doi = "10.1103/PhysRevLett.121.022002",
	journal = "Phys. Rev. Lett.",
	volume = "121",
	number = "2",
	pages = "022002",
	year = "2018"
}

@article{Haller:2018nnx,
	author = {Haller, Johannes and Hoecker, Andreas and Kogler, Roman and M{\"o}nig, Klaus and Peiffer, Thomas and Stelzer, J{\"o}rg},
	title = "{Update of the global electroweak fit and constraints on two-Higgs-doublet models}",
	eprint = "1803.01853",
	archivePrefix = "arXiv",
	primaryClass = "hep-ph",
	doi = "10.1140/epjc/s10052-018-6131-3",
	journal = "Eur. Phys. J. C",
	volume = "78",
	number = "8",
	pages = "675",
	year = "2018"
}

@article{Crivellin:2020zul,
	author = "Crivellin, Andreas and Hoferichter, Martin and Manzari, Claudio Andrea and Montull, Marc",
	title = "{Hadronic Vacuum Polarization: $(g-2)_\mu$ versus Global Electroweak Fits}",
	eprint = "2003.04886",
	archivePrefix = "arXiv",
	primaryClass = "hep-ph",
	reportNumber = "PSI-PR-20-04, UZ-TH 06/20",
	doi = "10.1103/PhysRevLett.125.091801",
	journal = "Phys. Rev. Lett.",
	volume = "125",
	number = "9",
	pages = "091801",
	year = "2020"
}

@article{DeBlas:2019ehy,
	author = "De Blas, J. and others",
	title = "{$\texttt{HEPfit}$: a code for the combination of indirect and direct constraints on high energy physics models}",
	eprint = "1910.14012",
	archivePrefix = "arXiv",
	primaryClass = "hep-ph",
	reportNumber = "CERN-TH-2019-178, CPHT-RR060.102019, DESY-19-184, DESY 19-184, FTUV/19-1031, FTUV/19-1031,
	IFIC/19-44, KEK-TH-2163, LPT-Orsay-19-36, PSI-PR-19-22, UCI-TR-2019-26, IFIC/19-44",
	doi = "10.1140/epjc/s10052-020-7904-z",
	journal = "Eur. Phys. J. C",
	volume = "80",
	number = "5",
	pages = "456",
	year = "2020"
}

@article{Keshavarzi:2020bfy,
	author = "Keshavarzi, Alexander and Marciano, William J. and Passera, Massimo and Sirlin, Alberto",
	title = "{Muon $g-2$ and $\Delta \alpha$ connection}",
	eprint = "2006.12666",
	archivePrefix = "arXiv",
	primaryClass = "hep-ph",
	reportNumber = "MAN/HEP/2020/006",
	doi = "10.1103/PhysRevD.102.033002",
	journal = "Phys. Rev. D",
	volume = "102",
	number = "3",
	pages = "033002",
	year = "2020"
}

@article{Malaescu:2020zuc,
	author = "Malaescu, Bogdan and Schott, Matthias",
	title = "{Impact of correlations between $a_{\mu }$ and $\alpha _\text {QED}$ on the EW fit}",
	eprint = "2008.08107",
	archivePrefix = "arXiv",
	primaryClass = "hep-ph",
	doi = "10.1140/epjc/s10052-021-08848-9",
	journal = "Eur. Phys. J. C",
	volume = "81",
	number = "1",
	pages = "46",
	year = "2021"
}

@article{deBlas:2021wap,
	author = "de Blas, J. and Ciuchini, M. and Franco, E. and Goncalves, A. and Mishima, S. and Pierini, M. and Reina, L. and Silvestrini, L.",
	title = "{Global analysis of electroweak data in the Standard Model}",
	eprint = "2112.07274",
	archivePrefix = "arXiv",
	primaryClass = "hep-ph",
	reportNumber = "KEK-TH-2378",
	doi = "10.1103/PhysRevD.106.033003",
	journal = "Phys. Rev. D",
	volume = "106",
	number = "3",
	pages = "033003",
	year = "2022"
}

@article{Adler:1974gd,
	author = "Adler, Stephen L.",
	title = "{Some Simple Vacuum Polarization Phenomenology: $e^+ e^- \to$ hadrons; The Muonic-Atom x-Ray Discrepancy and $g_{\mu}-2$}",
	reportNumber = "FNAL-PUB-74-063-THY, FERMILAB-PUB-74-063-T",
	doi = "10.1103/PhysRevD.10.3714",
	journal = "Phys. Rev. D",
	volume = "10",
	pages = "3714",
	year = "1974"
}

@article{Erler:2017knj,
	author = "Erler, Jens and Ferro-Hern{\'a}ndez, Rodolfo",
	title = "{Weak Mixing Angle in the Thomson Limit}",
	eprint = "1712.09146",
	archivePrefix = "arXiv",
	primaryClass = "hep-ph",
	doi = "10.1007/JHEP03(2018)196",
	journal = "JHEP",
	volume = "03",
	pages = "196",
	year = "2018"
}

@proceedings{Proceedings:2019vxr,
	author = "Abreu, S. and others",
	editor = "Blondel, A. and Gluza, J. and Jadach, S. and Janot, P. and Riemann, T.",
	title = "{Theory for the FCC-ee}: {Report on the 11th FCC-ee Workshop Theory and Experiments}",
	eprint = "1905.05078",
	archivePrefix = "arXiv",
	primaryClass = "hep-ph",
	reportNumber = "CERN-2020-003, BU-HEPP-19-03, CERN-TH-2019-061, CP3-19-22, DESY-19-072, FR-PHENO-2019-005, IFIC/19-23, IFT-UAM-CSIC-19-058, IPhT-19-050, IPPP/19/32, KW 19-003, MPP-2019-84, LTH 1203, ZU-TH-22-19, TUM-HEP-1200-19, TTP19-008, TTK-19-19",
	doi = "10.23731/CYRM-2020-003",
	publisher = "CERN",
	address = "Geneva",
	series = "CERN Yellow Reports: Monographs",
	volume = "3/2020",
	month = "5",
	year = "2019"
}

@article{deDivitiis:2013xla,
	author = "de Divitiis, G. M. and Frezzotti, R. and Lubicz, V. and Martinelli, G. and Petronzio, R. and Rossi, G. C. and Sanfilippo, F. and Simula, S. and Tantalo, N.",
	collaboration = "RM123",
	title = "{Leading isospin breaking effects on the lattice}",
	eprint = "1303.4896",
	archivePrefix = "arXiv",
	primaryClass = "hep-lat",
	doi = "10.1103/PhysRevD.87.114505",
	journal = "Phys. Rev. D",
	volume = "87",
	number = "11",
	pages = "114505",
	year = "2013"
}

@article{Risch:2021hty,
	author = "Risch, Andreas and Wittig, Hartmut",
	title = "{Leading isospin breaking effects in the HVP contribution to $a_{\mu}$ and to the running of $\alpha$}",
	eprint = "2112.00878",
	archivePrefix = "arXiv",
	primaryClass = "hep-lat",
	reportNumber = "DESY-21-207, MITP-21-067",
	doi = "10.22323/1.396.0106",
	journal = "PoS",
	volume = "LATTICE2021",
	pages = "106",
	year = "2022"
}

@article{Risch:2019xio,
	author = "Risch, Andreas and Wittig, Hartmut",
	title = "{Leading isospin breaking effects in the hadronic vacuum polarisation with open boundaries}",
	eprint = "1911.04230",
	archivePrefix = "arXiv",
	primaryClass = "hep-lat",
	reportNumber = "MITP/19-065",
	doi = "10.22323/1.363.0296",
	journal = "PoS",
	volume = "LATTICE2019",
	pages = "296",
	year = "2019"
}

@article{Risch:2018ozp,
	author = "Risch, Andreas and Wittig, Hartmut",
	title = "{Towards leading isospin breaking effects in mesonic masses with open boundaries}",
	eprint = "1811.00895",
	archivePrefix = "arXiv",
	primaryClass = "hep-lat",
	reportNumber = "MITP/18-101",
	doi = "10.22323/1.334.0059",
	journal = "PoS",
	volume = "LATTICE2018",
	pages = "059",
	year = "2018"
}

@article{Risch:2017xxe,
	author = "Risch, Andreas and Wittig, Hartmut",
	editor = "Della Morte, M. and Fritzsch, P. and G{\'a}miz S{\'a}nchez, E. and Pena Ruano, C.",
	title = "{Towards leading isospin breaking effects in mesonic masses with $O(a)$ improved Wilson fermions}",
	eprint = "1710.06801",
	archivePrefix = "arXiv",
	primaryClass = "hep-lat",
	reportNumber = "MITP-17-067, HIM-2017-05",
	doi = "10.1051/epjconf/201817514019",
	journal = "EPJ Web Conf.",
	volume = "175",
	pages = "14019",
	year = "2018"
}

@article{Hayakawa:2008an,
	author = "Hayakawa, Masashi and Uno, Shunpei",
	title = "{QED in finite volume and finite size scaling effect on electromagnetic properties of hadrons}",
	eprint = "0804.2044",
	archivePrefix = "arXiv",
	primaryClass = "hep-ph",
	doi = "10.1143/PTP.120.413",
	journal = "Prog. Theor. Phys.",
	volume = "120",
	pages = "413--441",
	year = "2008"
}

@article{Glashow:1961tr,
	author = "Glashow, S. L.",
	title = "{Partial Symmetries of Weak Interactions}",
	doi = "10.1016/0029-5582(61)90469-2",
	journal = "Nucl. Phys.",
	volume = "22",
	pages = "579--588",
	year = "1961"
}

@article{Kumar:2013yoa,
	author = "Kumar, K. S. and Mantry, Sonny and Marciano, W. J. and Souder, P. A.",
	title = "{Low Energy Measurements of the Weak Mixing Angle}",
	eprint = "1302.6263",
	archivePrefix = "arXiv",
	primaryClass = "hep-ex",
	doi = "10.1146/annurev-nucl-102212-170556",
	journal = "Ann. Rev. Nucl. Part. Sci.",
	volume = "63",
	pages = "237--267",
	year = "2013"
}

@article{Dzuba:2012kx,
	author = "Dzuba, V. A. and Berengut, J. C. and Flambaum, V. V. and Roberts, B.",
	title = "{Revisiting parity non-conservation in cesium}",
	eprint = "1207.5864",
	archivePrefix = "arXiv",
	primaryClass = "hep-ph",
	doi = "10.1103/PhysRevLett.109.203003",
	journal = "Phys. Rev. Lett.",
	volume = "109",
	pages = "203003",
	year = "2012"
}

@article{SLACE158:2005uay,
	author = "Anthony, P. L. and others",
	collaboration = "SLAC E158",
	title = "{Precision measurement of the weak mixing angle in M{\o}ller scattering}",
	eprint = "hep-ex/0504049",
	archivePrefix = "arXiv",
	reportNumber = "SLAC-PUB-11149",
	doi = "10.1103/PhysRevLett.95.081601",
	journal = "Phys. Rev. Lett.",
	volume = "95",
	pages = "081601",
	year = "2005"
}

@article{NuTeV:2001whx,
	author = "Zeller, G. P. and others",
	collaboration = "NuTeV",
	title = "{A Precise Determination of Electroweak Parameters in Neutrino Nucleon Scattering}",
	eprint = "hep-ex/0110059",
	archivePrefix = "arXiv",
	reportNumber = "FERMILAB-PUB-01-341-E",
	doi = "10.1103/PhysRevLett.88.091802",
	journal = "Phys. Rev. Lett.",
	volume = "88",
	pages = "091802",
	year = "2002",
	note = "[Erratum: Phys.Rev.Lett. 90, 239902 (2003)]"
}

@article{Ramsey-Musolf:1999qyv,
	author = "Ramsey-Musolf, M. J.",
	title = "{Low-energy parity violation and new physics}",
	eprint = "hep-ph/9903264",
	archivePrefix = "arXiv",
	reportNumber = "DOE-ER-40561-49",
	doi = "10.1103/PhysRevC.60.015501",
	journal = "Phys. Rev. C",
	volume = "60",
	pages = "015501",
	year = "1999"
}

@article{Erler:2013xha,
	author = "Erler, Jens and Su, Shufang",
	title = "{The Weak Neutral Current}",
	eprint = "1303.5522",
	archivePrefix = "arXiv",
	primaryClass = "hep-ph",
	doi = "10.1016/j.ppnp.2013.03.004",
	journal = "Prog. Part. Nucl. Phys.",
	volume = "71",
	pages = "119--149",
	year = "2013"
}

@article{Chang:2009yw,
	author = "Chang, We-Fu and Ng, John N. and Wu, Jackson M. S.",
	title = "{Non-Supersymmetric New Physics and Polarized Moller Scattering}",
	eprint = "0901.0613",
	archivePrefix = "arXiv",
	primaryClass = "hep-ph",
	doi = "10.1103/PhysRevD.79.055016",
	journal = "Phys. Rev. D",
	volume = "79",
	pages = "055016",
	year = "2009"
}

@article{Husung:2019ytz,
	author = "Husung, Nikolai and Marquard, Peter and Sommer, Rainer",
	title = "{Asymptotic behavior of cutoff effects in Yang{\textendash}Mills theory and in Wilson{\textquoteright}s lattice QCD}",
	eprint = "1912.08498",
	archivePrefix = "arXiv",
	primaryClass = "hep-lat",
	reportNumber = "DESY-19-188, DESY 19-188",
	doi = "10.1140/epjc/s10052-020-7685-4",
	journal = "Eur. Phys. J. C",
	volume = "80",
	number = "3",
	pages = "200",
	year = "2020"
}

@article{Husung:2024cgc,
	author = "Husung, Nikolai",
	title = "{Lattice artifacts of local fermion bilinears up to $\text {O}(\text {a}^2)$}",
	eprint = "2409.00776",
	archivePrefix = "arXiv",
	primaryClass = "hep-lat",
	reportNumber = "IFT-UAM/CSIC-24-125",
	doi = "10.1140/epjc/s10052-025-13825-7",
	journal = "Eur. Phys. J. C",
	volume = "85",
	number = "4",
	pages = "427",
	year = "2025"
}

@article{McNeile:2010ji,
	author = "McNeile, C. and Davies, C. T. H. and Follana, E. and Hornbostel, K. and Lepage, G. P.",
	title = "{High-Precision c and b Masses, and QCD Coupling from Current-Current Correlators in Lattice and Continuum QCD}",
	eprint = "1004.4285",
	archivePrefix = "arXiv",
	primaryClass = "hep-lat",
	doi = "10.1103/PhysRevD.82.034512",
	journal = "Phys. Rev. D",
	volume = "82",
	pages = "034512",
	year = "2010"
}

@article{Petreczky:2019ozv,
	author = "Petreczky, P. and Weber, J. H.",
	title = "{Strong coupling constant and heavy quark masses in (2+1)-flavor QCD}",
	eprint = "1901.06424",
	archivePrefix = "arXiv",
	primaryClass = "hep-lat",
	doi = "10.1103/PhysRevD.100.034519",
	journal = "Phys. Rev. D",
	volume = "100",
	number = "3",
	pages = "034519",
	year = "2019"
}

@article{Yang:2014sea,
	author = "Yang, Yi-Bo and others",
	title = "{Charm and strange quark masses and $f_{D_s}$ from overlap fermions}",
	eprint = "1410.3343",
	archivePrefix = "arXiv",
	primaryClass = "hep-lat",
	doi = "10.1103/PhysRevD.92.034517",
	journal = "Phys. Rev. D",
	volume = "92",
	number = "3",
	pages = "034517",
	year = "2015"
}

@article{Nakayama:2016atf,
	author = "Nakayama, Katsumasa and Fahy, Brendan and Hashimoto, Shoji",
	title = {{Short-distance charmonium correlator on the lattice with M{\"o}bius domain-wall fermion and a determination of charm quark mass}},
	eprint = "1606.01002",
	archivePrefix = "arXiv",
	primaryClass = "hep-lat",
	reportNumber = "KEK-CP-345",
	doi = "10.1103/PhysRevD.94.054507",
	journal = "Phys. Rev. D",
	volume = "94",
	number = "5",
	pages = "054507",
	year = "2016"
}

@article{DallaBrida:2022eua,
	author = {Dalla Brida, Mattia and H{\"o}llwieser, Roman and Knechtli, Francesco and Korzec, Tomasz and Nada, Alessandro and Ramos, Alberto and Sint, Stefan and Sommer, Rainer},
	collaboration = "ALPHA",
	title = "{Determination of $\alpha _s(m_Z)$ by the non-perturbative decoupling method}",
	eprint = "2209.14204",
	archivePrefix = "arXiv",
	primaryClass = "hep-lat",
	reportNumber = "IFIC/22-25, WUB/22-00, DESY-22-051, CERN-TH-2022-015, HU-EP-22/04",
	doi = "10.1140/epjc/s10052-022-10998-3",
	journal = "Eur. Phys. J. C",
	volume = "82",
	number = "12",
	pages = "1092",
	year = "2022"
}

@article{Petreczky:2020tky,
	author = "Petreczky, Peter and Weber, Johannes Heinrich",
	title = "{Strong coupling constant from moments of quarkonium correlators revisited}",
	eprint = "2012.06193",
	archivePrefix = "arXiv",
	primaryClass = "hep-lat",
	reportNumber = "HU-EP-20/39-RTG",
	doi = "10.1140/epjc/s10052-022-09998-0",
	journal = "Eur. Phys. J. C",
	volume = "82",
	number = "1",
	pages = "64",
	year = "2022"
}

@article{Ayala:2020odx,
	author = "Ayala, Cesar and Lobregat, Xabier and Pineda, Antonio",
	title = "{Determination of $\alpha(M_z)$ from an hyperasymptotic approximation to the energy of a static quark-antiquark pair}",
	eprint = "2005.12301",
	archivePrefix = "arXiv",
	primaryClass = "hep-ph",
	doi = "10.1007/JHEP09(2020)016",
	journal = "JHEP",
	volume = "09",
	pages = "016",
	year = "2020"
}

@article{Bazavov:2019qoo,
	author = "Bazavov, Alexei and Brambilla, Nora and Garcia i Tormo, Xavier and Petreczky, P{\'e}ter and Soto, Joan and Vairo, Antonio and Weber, Johannes Heinrich",
	collaboration = "TUMQCD",
	title = "{Determination of the QCD coupling from the static energy and the free energy}",
	eprint = "1907.11747",
	archivePrefix = "arXiv",
	primaryClass = "hep-lat",
	reportNumber = "TUM-EFT 111/18; INT-PUB-19-028;",
	doi = "10.1103/PhysRevD.100.114511",
	journal = "Phys. Rev. D",
	volume = "100",
	number = "11",
	pages = "114511",
	year = "2019"
}

@article{Cali:2020hrj,
	author = "Cali, Salvatore and Cichy, Krzysztof and Korcyl, Piotr and Simeth, Jakob",
	title = "{Running coupling constant from position-space current-current correlation functions in three-flavor lattice QCD}",
	eprint = "2003.05781",
	archivePrefix = "arXiv",
	primaryClass = "hep-lat",
	doi = "10.1103/PhysRevLett.125.242002",
	journal = "Phys. Rev. Lett.",
	volume = "125",
	pages = "242002",
	year = "2020"
}

@article{PACS-CS:2009zxm,
	author = "Aoki, S. and others",
	collaboration = "PACS-CS",
	title = "{Precise determination of the strong coupling constant in $N_f$ = 2+1 lattice QCD with the Schr{\"o}dinger functional scheme}",
	eprint = "0906.3906",
	archivePrefix = "arXiv",
	primaryClass = "hep-lat",
	reportNumber = "UTHEP-584, UTCCS-P-54",
	doi = "10.1088/1126-6708/2009/10/053",
	journal = "JHEP",
	volume = "10",
	pages = "053",
	year = "2009"
}

@article{Maltman:2008bx,
	author = "Maltman, K. and Leinweber, D. and Moran, P. and Sternbeck, A.",
	title = "{The Realistic Lattice Determination of $\alpha_s(M_Z)$ Revisited}",
	eprint = "0807.2020",
	archivePrefix = "arXiv",
	primaryClass = "hep-lat",
	doi = "10.1103/PhysRevD.78.114504",
	journal = "Phys. Rev. D",
	volume = "78",
	pages = "114504",
	year = "2008"
}

@article{Bussone:2025wlf,
	author = "Bussone, Andrea and Conigli, Alessandro and Frison, Julien and Herdo{\'\i}za, Gregorio and Pena, Carlos and Preti, David and Romero, Jos{\'e} {\'A}ngel and S{\'a}ez, Alejandro and Ugarrio, Javier",
	collaboration = "ALPHA",
	title = "{Hadronic physics from a Wilson fermion mixed-action approach: setup and scale setting}",
	eprint = "2510.20450",
	archivePrefix = "arXiv",
	primaryClass = "hep-lat",
	reportNumber = "IFT-UAM/CSIC-25-108",
	doi = "10.1140/epjc/s10052-026-15518-1",
	journal = "Eur. Phys. J. C",
	volume = "86",
	number = "5",
	pages = "577",
	year = "2026"
}

@article{Davier:2017zfy,
      author         = "Davier, Michel and Hoecker, Andreas and Malaescu, Bogdan and Zhang, Zhiqing",
      title          = "{Reevaluation of the hadronic vacuum polarisation contributions to the Standard Model predictions of the muon $g-2$ and $\alpha(m_Z^2)$ using newest hadronic cross-section data}",
      journal        = "Eur. Phys. J. C",
      volume         = "77",
      year           = "2017",
      number         = "12",
      pages          = "827",
      doi            = "10.1140/epjc/s10052-017-5161-6",
      eprint         = "1706.09436",
      archivePrefix  = "arXiv",
      primaryClass   = "hep-ph",
      SLACcitation   = "%%CITATION = ARXIV:1706.09436;%%"
}

@article{Davier:2019can,
	author = "Davier, M. and Hoecker, A. and Malaescu, B. and Zhang, Z.",
	title = "{A new evaluation of the hadronic vacuum polarisation contributions to the muon anomalous magnetic moment and to $\mathbf{\boldsymbol\alpha(m_Z^2)}$}",
	eprint = "1908.00921",
	archivePrefix = "arXiv",
	primaryClass = "hep-ph",
	doi = "10.1140/epjc/s10052-020-7792-2",
	journal = "Eur. Phys. J. C",
	volume = "80",
	number = "3",
	pages = "241",
	year = "2020",
	note = "[Erratum: Eur.Phys.J.C 80, 410 (2020)]"
}

@article{Keshavarzi:2019abf,
	author = "Keshavarzi, Alexander and Nomura, Daisuke and Teubner, Thomas",
	title = "{$g-2$ of charged leptons, $\alpha (M^2_Z)$ , and the hyperfine splitting of muonium}",
	eprint = "1911.00367",
	archivePrefix = "arXiv",
	primaryClass = "hep-ph",
	reportNumber = "MAN/HEP/2019/010, LTH 1216, KEK-TH-2165",
	doi = "10.1103/PhysRevD.101.014029",
	journal = "Phys. Rev. D",
	volume = "101",
	number = "1",
	pages = "014029",
	year = "2020"
}

@article{Bali:2023sdi,
	author = {Bali, Gunnar S. and Collins, Sara and Heybrock, Simon and L{\"o}ffler, Marius and R{\"o}dl, Rudolf and S{\"o}ldner, Wolfgang and Weish{\"a}upl, Simon},
	collaboration = "RQCD",
	title = "{Octet baryon isovector charges from Nf=2+1 lattice QCD}",
	eprint = "2305.04717",
	archivePrefix = "arXiv",
	primaryClass = "hep-lat",
	doi = "10.1103/PhysRevD.108.034512",
	journal = "Phys. Rev. D",
	volume = "108",
	number = "3",
	pages = "034512",
	year = "2023"
}

@article{Meyer:2011um,
	author = "Meyer, Harvey B.",
	title = "{Lattice QCD and the Timelike Pion Form Factor}",
	eprint = "1105.1892",
	archivePrefix = "arXiv",
	primaryClass = "hep-lat",
	doi = "10.1103/PhysRevLett.107.072002",
	journal = "Phys. Rev. Lett.",
	volume = "107",
	pages = "072002",
	year = "2011"
}

@inbook{Akaike:1998zah,
	author = "Akaike, Hirotogu",
	booktitle = "Selected Papers of Hirotugu Akaike",
	title = "{Information Theory and an Extension of the Maximum Likelihood Principle}",
	publisher = "Springer Science+Business Media",
	address = "New York",
	year = "1998"
}

@article{Giusti:2004yp,
	author = "Giusti, Leonardo and Hernandez, P. and Laine, M. and Weisz, P. and Wittig, H.",
	title = "{Low-energy couplings of QCD from current correlators near the chiral limit}",
	eprint = "hep-lat/0402002",
	archivePrefix = "arXiv",
	reportNumber = "BI-TP-2004-03, CPT-2003-P-4622, DESY-04-009, FTUV-04-0203, IFIC-04-04, MPP-2004-6",
	doi = "10.1088/1126-6708/2004/04/013",
	journal = "JHEP",
	volume = "04",
	pages = "013",
	year = "2004"
}

@article{DeGrand:2004qw,
	author = "DeGrand, Thomas A. and Schaefer, Stefan",
	title = "{Improving meson two point functions in lattice QCD}",
	eprint = "hep-lat/0401011",
	archivePrefix = "arXiv",
	reportNumber = "COLO-HEP-497",
	doi = "10.1016/j.cpc.2004.02.006",
	journal = "Comput. Phys. Commun.",
	volume = "159",
	pages = "185--191",
	year = "2004"
}

@article{RBC:2018dos,
	author = {Blum, T. and Boyle, P. A. and G\"ulpers, V. and Izubuchi, T. and Jin, L. and Jung, C. and J\"uttner, A. and Lehner, C. and Portelli, A. and Tsang, J. T.},
	collaboration = "RBC, UKQCD",
	title = "{Calculation of the hadronic vacuum polarization contribution to the muon anomalous magnetic moment}",
	eprint = "1801.07224",
	archivePrefix = "arXiv",
	primaryClass = "hep-lat",
	doi = "10.1103/PhysRevLett.121.022003",
	journal = "Phys. Rev. Lett.",
	volume = "121",
	number = "2",
	pages = "022003",
	year = "2018"
}

@article{Borsanyi:2016lpl,
	author = "Borsanyi, Sz. and Fodor, Z. and Kawanai, T. and Krieg, S. and Lellouch, L. and Malak, R. and Miura, K. and Szabo, K. K. and Torrero, C. and Toth, B.",
	title = "{Slope and curvature of the hadronic vacuum polarization at vanishing virtuality from lattice QCD}",
	eprint = "1612.02364",
	archivePrefix = "arXiv",
	primaryClass = "hep-lat",
	doi = "10.1103/PhysRevD.96.074507",
	journal = "Phys. Rev. D",
	volume = "96",
	number = "7",
	pages = "074507",
	year = "2017"
}

@article{Hernandez:2023ipz,
	author = "Hern\'andez, Rodolfo Ferro",
	title = "{AdlerPy: A Python Package for the Perturbative Adler Function}",
	eprint = "2311.04849",
	archivePrefix = "arXiv",
	primaryClass = "hep-ph",
	month = "11",
	year = "2023"
}

@article{FlavourLatticeAveragingGroupFLAG:2024oxs,
	author = "Aoki, Y. and others",
	collaboration = "Flavour Lattice Averaging Group (FLAG)",
	title = "{FLAG review 2024}",
	eprint = "2411.04268",
	archivePrefix = "arXiv",
	primaryClass = "hep-lat",
	reportNumber = "CERN-TH-2024-192, FERMILAB-PUB-24-0785-T",
	doi = "10.1103/nfzp-p5dn",
	journal = "Phys. Rev. D",
	volume = "113",
	number = "1",
	pages = "014508",
	year = "2026"
}

@article{Aliberti:2025beg,
    author = "Aliberti, R. and others",
    title = "{The anomalous magnetic moment of the muon in the Standard Model: an update}",
    eprint = "2505.21476",
    archivePrefix = "arXiv",
    primaryClass = "hep-ph",
    reportNumber = "CERN-TH-2025-101, FERMILAB-PUB-25-0344-T, INT-PUB-25-015, IPARCOS-UCM-25-029, KEK Preprint 2025-22, LTH 1403, MITP-25-037, UWThPh 2025-15, UWThPh
  2025-15, ZU-TH 37/25, IPARCOS-UCM-25-029",
    doi = "10.1016/j.physrep.2025.08.002",
    journal = "Phys. Rept.",
    volume = "1143",
    pages = "1--158",
    year = "2025"
}

@article{Baikov:2010je,
	author = "Baikov, P. A. and Chetyrkin, K. G. and K{\"u}hn, J. H.",
	title = "{Adler Function, Bjorken Sum Rule, and the Crewther Relation to Order $\alpha^4_s$ in a General Gauge Theory}",
	eprint = "1001.3606",
	archivePrefix = "arXiv",
	primaryClass = "hep-ph",
	reportNumber = "TTP10-05",
	doi = "10.1103/PhysRevLett.104.132004",
	journal = "Phys. Rev. Lett.",
	volume = "104",
	pages = "132004",
	year = "2010"
}

@article{Baikov:2008jh,
	author = "Baikov, P. A. and Chetyrkin, K. G. and K{\"u}hn, Johann H.",
	title = "{Order $\alpha^4_{\mathrm{s}}$ QCD Corrections to $Z$ and $\tau$ Decays}",
	eprint = "0801.1821",
	archivePrefix = "arXiv",
	primaryClass = "hep-ph",
	reportNumber = "SFB-CPP-08-04, TTP08-01",
	doi = "10.1103/PhysRevLett.101.012002",
	journal = "Phys. Rev. Lett.",
	volume = "101",
	pages = "012002",
	year = "2008"
}

@article{Bulava:2013cta,
	author = "Bulava, John and Schaefer, Stefan",
	title = "{Improvement of $N_f$ = 3 lattice QCD with Wilson fermions and tree-level improved gauge action}",
	eprint = "1304.7093",
	archivePrefix = "arXiv",
	primaryClass = "hep-lat",
	reportNumber = "CERN-PH-TH-2013-082, TCD-MATH-13-06",
	doi = "10.1016/j.nuclphysb.2013.05.019",
	journal = "Nucl. Phys. B",
	volume = "874",
	pages = "188--197",
	year = "2013"
}

@article{Bernecker:2011gh,
	author = "Bernecker, David and Meyer, Harvey B.",
	title = "{Vector Correlators in Lattice QCD: Methods and applications}",
	eprint = "1107.4388",
	archivePrefix = "arXiv",
	primaryClass = "hep-lat",
	doi = "10.1140/epja/i2011-11148-6",
	journal = "Eur. Phys. J. A",
	volume = "47",
	pages = "148",
	year = "2011"
}

@article{Kuberski:2023zky,
	author = "Kuberski, Simon",
	title = "{Low-mode deflation for twisted-mass and RHMC reweighting in lattice QCD}",
	eprint = "2306.02385",
	archivePrefix = "arXiv",
	primaryClass = "hep-lat",
	reportNumber = "MITP-23-021",
	doi = "10.1016/j.cpc.2024.109173",
	journal = "Comput. Phys. Commun.",
	volume = "300",
	pages = "109173",
	year = "2024"
}

@article{Bali:2016umi,
	author = {Bali, Gunnar S. and Scholz, Enno E. and Simeth, Jakob and S\"oldner, Wolfgang},
	collaboration = "RQCD",
	title = "{Lattice simulations with $N_f=2+1$ improved Wilson fermions at a fixed strange quark mass}",
	eprint = "1606.09039",
	archivePrefix = "arXiv",
	primaryClass = "hep-lat",
	doi = "10.1103/PhysRevD.94.074501",
	journal = "Phys. Rev. D",
	volume = "94",
	number = "7",
	pages = "074501",
	year = "2016"
}

@article{Ce:2021xgd,
	author = {C\`e, Marco and Harris, Tim and Meyer, Harvey B. and Toniato, Arianna and T\"or\"ok, Csaba},
	title = "{Vacuum correlators at short distances from lattice QCD}",
	eprint = "2106.15293",
	archivePrefix = "arXiv",
	primaryClass = "hep-lat",
	reportNumber = "MITP/21-032, CERN-TH-2021-100",
	doi = "10.1007/JHEP12(2021)215",
	journal = "JHEP",
	volume = "12",
	pages = "215",
	year = "2021"
}

@article{Sommer:2022wac,
	author = "Sommer, Rainer and Chimirri, Leonardo and Husung, Nikolai",
	title = "{Log-enhanced discretization errors in integrated correlation functions}",
	eprint = "2211.15750",
	archivePrefix = "arXiv",
	primaryClass = "hep-lat",
	doi = "10.22323/1.430.0358",
	journal = "PoS",
	volume = "LATTICE2022",
	pages = "358",
	year = "2023"
}

@article{Borsanyi:2020mff,
	author = "Borsanyi, Sz. and others",
	title = "{Leading hadronic contribution to the muon magnetic moment from lattice QCD}",
	eprint = "2002.12347",
	archivePrefix = "arXiv",
	primaryClass = "hep-lat",
	doi = "10.1038/s41586-021-03418-1",
	journal = "Nature",
	volume = "593",
	number = "7857",
	pages = "51",
	year = "2021"
}

@article{Becker:2018ggl,
	author = "Becker, Dominik and others",
	title = "{The P2 experiment}",
	eprint = "1802.04759",
	archivePrefix = "arXiv",
	primaryClass = "nucl-ex",
	doi = "10.1140/epja/i2018-12611-6",
	journal = "Eur. Phys. J. A",
	volume = "54",
	number = "11",
	pages = "208",
	year = "2018"
}

@article{MOLLER:2014iki,
    author = "Benesch, J. and others",
    collaboration = "MOLLER",
    title = "{The MOLLER Experiment: An Ultra-Precise Measurement of the Weak Mixing Angle Using Møller Scattering}",
    eprint = "1411.4088",
    archivePrefix = "arXiv",
    primaryClass = "nucl-ex",
    reportNumber = "JLAB-PHY-14-1986",
    month = "11",
    year = "2014"
}

@article{Djukanovic:2024cmq,
    author = "Djukanovic, Dalibor and von Hippel, Georg and Kuberski, Simon and Meyer, Harvey B. and Miller, Nolan and Ottnad, Konstantin and Parrino, Julian and Risch, Andreas and Wittig, Hartmut",
    title = "{The hadronic vacuum polarization contribution to the muon g {\ensuremath{-}} 2 at long distances}",
    eprint = "2411.07969",
    archivePrefix = "arXiv",
    primaryClass = "hep-lat",
    reportNumber = "CERN-TH-2024-196, MITP-24-080",
    doi = "10.1007/JHEP04(2025)098",
    journal = "JHEP",
    volume = "04",
    pages = "098",
    year = "2025"
}

@article{Gerardin:2019rua,
	author = {G\'erardin, Antoine and C\`e, Marco and von Hippel, Georg and H\"orz, Ben and Meyer, Harvey B. and Mohler, Daniel and Ottnad, Konstantin and Wilhelm, Jonas and Wittig, Hartmut},
	title = "{The leading hadronic contribution to $(g-2)_\mu$ from lattice QCD with $N_{\rm f}=2+1$ flavours of O($a$) improved Wilson quarks}",
	eprint = "1904.03120",
	archivePrefix = "arXiv",
	primaryClass = "hep-lat",
	reportNumber = "MITP-19-021, MITP/19-021",
	doi = "10.1103/PhysRevD.100.014510",
	journal = "Phys. Rev. D",
	volume = "100",
	number = "1",
	pages = "014510",
	year = "2019"
}

@article{Aubin2012,
	title = {Model-independent parametrization of the hadronic vacuum polarization and $g$-2 for the muon on the lattice},
	author = {Aubin, Christopher and Blum, Thomas and Golterman, Maarten and Peris, Santiago},
	eprint="1205.3695",
	archivePrefix = "arXiv",
	journal = {Phys. Rev. D},
	volume = {86},
	issue = {5},
	pages = {054509},
	numpages = {10},
	year = {2012},
	publisher = {American Physical Society},
	doi = {10.1103/PhysRevD.86.054509},
	url = {https://link.aps.org/doi/10.1103/PhysRevD.86.054509}
}

@misc{Lehner2016,
	author       = {C. Lehner},
	title        = {The hadronic vacuum polarization contribution to the muon anomalous magnetic moment},
	howpublished = {Talk at RBRC Workshop on Lattice Gauge Theories},
	year         = {(2016)},
	doi        = {\url{https://indico.bnl.gov/event/1628/contributions/2819/}},
    url        = {https://indico.bnl.gov/event/1628/contributions/2819/}
}

@misc{Jeger_yellow_rep,
	author = {Jegerlehner, F},
	title = {{$\alpha_{\mathrm{QED,eff}}(s)$ for precision physics at the FCC-ee/ILC}},
	year = {(2020)},
	howpublished = {CERN Yellow Reports: Monographs \textbf{3}},
	doi = {https://doi.org/10.23731/CYRM-2020-003.9}
}

@article{Jegerlehner:2011mw,
	author = "Jegerlehner, Fred",
	title = "{Electroweak effective couplings for future precision experiments}",
	eprint = "1107.4683",
	archivePrefix = "arXiv",
	primaryClass = "hep-ph",
	reportNumber = "HU-EP-11-33, DESY-11-117, HU-EP-11/33, DESY 11-117",
	doi = "10.1393/ncc/i2011-11011-0",
	journal = "Nuovo Cim. C",
	volume = "034S1",
	pages = "31--40",
	year = "2011"
}

@article{Baak:2014ora,
	author = {Baak, M. and C\'uth, J. and Haller, J. and Hoecker, A. and Kogler, R. and M\"onig, K. and Schott, M. and Stelzer, J.},
	collaboration = "Gfitter Group",
	title = "{The global electroweak fit at NNLO and prospects for the LHC and ILC}",
	eprint = "1407.3792",
	archivePrefix = "arXiv",
	primaryClass = "hep-ph",
	reportNumber = "DESY-14-124",
	doi = "10.1140/epjc/s10052-014-3046-5",
	journal = "Eur. Phys. J. C",
	volume = "74",
	pages = "3046",
	year = "2014"
}

@article{Eidelman:1998vc,
	author = "Eidelman, S. and Jegerlehner, F. and Kataev, A. L. and Veretin, O.",
	title = "{Testing nonperturbative strong interaction effects via the Adler function}",
	eprint = "hep-ph/9812521",
	archivePrefix = "arXiv",
	reportNumber = "DESY-98-206",
	doi = "10.1016/S0370-2693(99)00389-5",
	journal = "Phys. Lett. B",
	volume = "454",
	pages = "369--380",
	year = "1999"
}

@article{Jegerlehner:2008rs,
	author = "Jegerlehner, F.",
	editor = "Bini, Cesare and Venanzoni, Graziano",
	title = "{The Running fine structure constant $\alpha(E)$ via the Adler function}",
	eprint = "0807.4206",
	archivePrefix = "arXiv",
	primaryClass = "hep-ph",
	reportNumber = "IFJPAN-IV-2008-3, HU-EP-08-22, DESY-08-078",
	doi = "10.1016/j.nuclphysbps.2008.09.010",
	journal = "Nucl. Phys. B Proc. Suppl.",
	volume = "181-182",
	pages = "135--140",
	year = "2008"
}

@inproceedings{Jegerlehner:1999hg,
	author = "Jegerlehner, F.",
	title = "{Hadronic effects in $(g - 2)_\mu$ and $\alpha_{\rm QED}(M_Z)$: Status and perspectives}",
	booktitle = "{4th International Symposium on Radiative Corrections: Applications of Quantum Field Theory to Phenomenology}",
	eprint = "hep-ph/9901386",
	archivePrefix = "arXiv",
	reportNumber = "DESY-99-007",
	pages = "75--89",
	month = "1",
	year = "1999"
}

@article{Chetyrkin:1996cf,
	author = "Chetyrkin, K. G. and K{\"u}hn, Johann H. and Steinhauser, M.",
	title = "{Three loop polarization function and O ($\alpha_{\mathrm{s}}^2$) corrections to the production of heavy quarks}",
	eprint = "hep-ph/9606230",
	archivePrefix = "arXiv",
	reportNumber = "TTP-96-13, MPI-PHT-96-27",
	doi = "10.1016/S0550-3213(96)00534-2",
	journal = "Nucl. Phys. B",
	volume = "482",
	pages = "213--240",
	year = "1996"
}

@article{Francis:2013fzp,
	author = "Francis, Anthony and J{\"a}ger, Benjamin and Meyer, Harvey B. and Wittig, Hartmut",
	title = "{A new representation of the Adler function for lattice QCD}",
	eprint = "1306.2532",
	archivePrefix = "arXiv",
	primaryClass = "hep-lat",
	reportNumber = "MITP-13-032, HIM-2013-05",
	doi = "10.1103/PhysRevD.88.054502",
	journal = "Phys. Rev. D",
	volume = "88",
	pages = "054502",
	year = "2013"
}

@article{Kuberski:2024bcj,
	author = "Kuberski, Simon and C\`e, Marco and von Hippel, Georg and Meyer, Harvey B. and Ottnad, Konstantin and Risch, Andreas and Wittig, Hartmut",
	title = "{Hadronic vacuum polarization in the muon g \ensuremath{-} 2: the short-distance contribution from lattice QCD}",
	eprint = "2401.11895",
	archivePrefix = "arXiv",
	primaryClass = "hep-lat",
	reportNumber = "MITP-24-011, CERN-TH-2024-011",
	doi = "10.1007/JHEP03(2024)172",
	journal = "JHEP",
	volume = "03",
	pages = "172",
	year = "2024"
}

@article{Urech:1994hd,
	author = "Urech, Res",
	title = "{Virtual photons in chiral perturbation theory}",
	eprint = "hep-ph/9405341",
	archivePrefix = "arXiv",
	reportNumber = "BUTP-94-9",
	doi = "10.1016/0550-3213(95)90707-N",
	journal = "Nucl. Phys. B",
	volume = "433",
	pages = "234",
	year = "1995"
}

@article{Neufeld:1995mu,
	author = "Neufeld, H. and Rupertsberger, H.",
	title = "{The Electromagnetic interaction in chiral perturbation theory}",
	eprint = "hep-ph/9506448",
	archivePrefix = "arXiv",
	reportNumber = "UWTHPH-1995-18",
	doi = "10.1007/s002880050156",
	journal = "Z. Phys. C",
	volume = "71",
	pages = "131",
	year = "1996"
}

@article{Heitger:2020zaq,
	author = "Heitger, Jochen and Joswig, Fabian",
	collaboration = "ALPHA",
	title = "{The renormalised $\mathrm{O}(a)$ improved vector current in three-flavour lattice QCD with Wilson quarks}",
	eprint = "2010.09539",
	archivePrefix = "arXiv",
	primaryClass = "hep-lat",
	reportNumber = "MS-TP-20-37",
	doi = "10.1140/epjc/s10052-021-09037-4",
	journal = "Eur. Phys. J. C",
	volume = "81",
	number = "3",
	pages = "254",
	year = "2021"
}

@article{Ce:2022kxy,
	author = "C\`e, Marco and others",
	title = "{Window observable for the hadronic vacuum polarization contribution to the muon g-2 from lattice QCD}",
	eprint = "2206.06582",
	archivePrefix = "arXiv",
	primaryClass = "hep-lat",
	reportNumber = "MITP-22-038, CERN-TH-2022-098, DESY-22-105",
	doi = "10.1103/PhysRevD.106.114502",
	journal = "Phys. Rev. D",
	volume = "106",
	number = "11",
	pages = "114502",
	year = "2022"
}

@article{Ce:2022eix,
	author = "C\`e, Marco and G\'erardin, Antoine and von Hippel, Georg and Meyer, Harvey B. and Miura, Kohtaroh and Ottnad, Konstantin and Risch, Andreas and San Jos\'e, Teseo and Wilhelm, Jonas and Wittig, Hartmut",
	title = "{The hadronic running of the electromagnetic coupling and the electroweak mixing angle from lattice QCD}",
	eprint = "2203.08676",
	archivePrefix = "arXiv",
	primaryClass = "hep-lat",
	reportNumber = "MITP-22-019, CERN-TH-2022-035, DESY-22-050",
	doi = "10.1007/JHEP08(2022)220",
	journal = "JHEP",
	volume = "08",
	pages = "220",
	year = "2022"
}

@article{Bruno:2022mfy,
	author = "Bruno, Mattia and Sommer, Rainer",
	title = "{On fits to correlated and auto-correlated data}",
	eprint = "2209.14188",
	archivePrefix = "arXiv",
	primaryClass = "hep-lat",
	doi = "10.1016/j.cpc.2022.108643",
	journal = "Comput. Phys. Commun.",
	volume = "285",
	pages = "108643",
	year = "2023"
}

@article{Jay:2020jkz,
	author = "Jay, William I. and Neil, Ethan T.",
	title = "{Bayesian model averaging for analysis of lattice field theory results}",
	eprint = "2008.01069",
	archivePrefix = "arXiv",
	primaryClass = "stat.ME",
	reportNumber = "FERMILAB-PUB-20-374-T",
	doi = "10.1103/PhysRevD.103.114502",
	journal = "Phys. Rev. D",
	volume = "103",
	pages = "114502",
	year = "2021"
}

@article{Fritzsch:2018zym,
	author = "Fritzsch, Patrick",
	title = "{Mass-improvement of the vector current in three-flavor QCD}",
	eprint = "1805.07401",
	archivePrefix = "arXiv",
	primaryClass = "hep-lat",
	reportNumber = "CERN-TH-2018-119",
	doi = "10.1007/JHEP06(2018)015",
	journal = "JHEP",
	volume = "06",
	pages = "015",
	year = "2018"
}

@article{Luscher:1996sc,
    author = "{L\"uscher}, Martin and Sint, Stefan and Sommer, Rainer and Weisz, Peter",
    title = "{Chiral symmetry and O(a) improvement in lattice QCD}",
    eprint = "hep-lat/9605038",
    archivePrefix = "arXiv",
    reportNumber = "DESY-96-086, CERN-TH-96-138, MPI-PHT-96-38",
    doi = "10.1016/0550-3213(96)00378-1",
    journal = "Nucl. Phys. B",
    volume = "478",
    pages = "365--400",
    year = "1996"
}

@article{Bruno:2017gxd,
    author = "Bruno, Mattia and Dalla Brida, Mattia and Fritzsch, Patrick and Korzec, Tomasz and Ramos, Alberto and Schaefer, Stefan and Simma, Hubert and Sint, Stefan and Sommer, Rainer",
    collaboration = "ALPHA",
    title = "{QCD Coupling from a Nonperturbative Determination of the Three-Flavor $\Lambda$ Parameter}",
    eprint = "1706.03821",
    archivePrefix = "arXiv",
    primaryClass = "hep-lat",
    reportNumber = "CERN-TH-2017-129, DESY-17-088, WUB-17-03",
    doi = "10.1103/PhysRevLett.119.102001",
    journal = "Phys. Rev. Lett.",
    volume = "119",
    number = "10",
    pages = "102001",
    year = "2017"
}

@article{Bruno:2016plf,
	author = "Bruno, Mattia and Korzec, Tomasz and Schaefer, Stefan",
	title = "{Setting the scale for the CLS $2 + 1$ flavor ensembles}",
	eprint = "1608.08900",
	archivePrefix = "arXiv",
	primaryClass = "hep-lat",
	reportNumber = "DESY-16-162, WUB-16-05",
	doi = "10.1103/PhysRevD.95.074504",
	journal = "Phys. Rev. D",
	volume = "95",
	number = "7",
	pages = "074504",
	year = "2017"
}

@article{Mohler:2017wnb,
	author = "Mohler, Daniel and Schaefer, Stefan and Simeth, Jakob",
	editor = "Della Morte, M. and Fritzsch, P. and G\'amiz S\'anchez, E. and Pena Ruano, C.",
	title = "{CLS 2+1 flavor simulations at physical light- and strange-quark masses}",
	eprint = "1712.04884",
	archivePrefix = "arXiv",
	primaryClass = "hep-lat",
	reportNumber = "DESY-17-166, HIM-2017-08, MITP-17-093",
	doi = "10.1051/epjconf/201817502010",
	journal = "EPJ Web Conf.",
	volume = "175",
	pages = "02010",
	year = "2018"
}

@article{Bruno:2014jqa,
    author = "Bruno, Mattia and others",
    title = "{Simulation of QCD with N$_{f} =$ 2 $+$ 1 flavors of non-perturbatively improved Wilson fermions}",
    eprint = "1411.3982",
    archivePrefix = "arXiv",
    primaryClass = "hep-lat",
    reportNumber = "DESY-14-216, FTUAM-14-48, HIM-2014-01, HU-EP-14-51, MITP-14-091, SFB-CPP-14-89, IFT-UAM-CSIC-14-117",
    doi = "10.1007/JHEP02(2015)043",
    journal = "JHEP",
    volume = "02",
    pages = "043",
    year = "2015"
}

@article{ParticleDataGroup:2020ssz,
	author = "Zyla, P. A. and others",
	collaboration = "Particle Data Group",
	title = "{Review of Particle Physics}",
	doi = "10.1093/ptep/ptaa104",
	journal = "PTEP",
	volume = "2020",
	number = "8",
	pages = "083C01",
	year = "2020"
}

@article{Husung:2021mfl,
    author = "Husung, Nikolai and Marquard, Peter and Sommer, Rainer",
    title = "{The asymptotic approach to the continuum of lattice QCD spectral observables}",
    eprint = "2111.02347",
    archivePrefix = "arXiv",
    primaryClass = "hep-lat",
    reportNumber = "DESY-21-177, HU-EP-21/45",
    doi = "10.1016/j.physletb.2022.137069",
    journal = "Phys. Lett. B",
    volume = "829",
    pages = "137069",
    year = "2022"
}

@article{Luscher:2010iy,
    author = {L{\"u}scher, Martin},
    title = "{Properties and uses of the Wilson flow in lattice QCD}",
    eprint = "1006.4518",
    archivePrefix = "arXiv",
    primaryClass = "hep-lat",
    reportNumber = "CERN-PH-TH-2010-143",
    doi = "10.1007/JHEP08(2010)071",
    journal = "JHEP",
    volume = "08",
    pages = "071",
    year = "2010",
    note = "[Erratum: \href{https://doi.org/10.1007/JHEP03(2014)092}{JHEP \textbf{03} (2014) 092}]"
}

@article{Lellouch:2000pv,
	author = "Lellouch, Laurent and L{\"u}scher, Martin",
	title = "{Weak transition matrix elements from finite volume correlation functions}",
	eprint = "hep-lat/0003023",
	archivePrefix = "arXiv",
	reportNumber = "CERN-TH-2000-091, CPT-2000-PE-3984, LAPTH-788-00",
	doi = "10.1007/s002200100410",
	journal = "Commun. Math. Phys.",
	volume = "219",
	pages = "31--44",
	year = "2001"
}

@article{Mohler:2020txx,
    author = "Mohler, Daniel and Schaefer, Stefan",
    title = "{Remarks on strange-quark simulations with Wilson fermions}",
    eprint = "2003.13359",
    archivePrefix = "arXiv",
    primaryClass = "hep-lat",
    reportNumber = "DESY-20-041, MITP/20-010, DESY 20-041 ; MITP/20-010",
    doi = "10.1103/PhysRevD.102.074506",
    journal = "Phys. Rev. D",
    volume = "102",
    number = "7",
    pages = "074506",
    year = "2020"
}

@article{Wolff:2003sm,
    author = "Wolff, Ulli",
    collaboration = "ALPHA",
    title = "{Monte Carlo errors with less errors}",
    eprint = "hep-lat/0306017",
    archivePrefix = "arXiv",
    reportNumber = "HU-EP-03-32, SFB-CPP-03-12",
    doi = "10.1016/S0010-4655(03)00467-3",
    journal = "Comput. Phys. Commun.",
    volume = "156",
    pages = "143--153",
    year = "2004",
    note = "[Erratum: Comput.Phys.Commun. 176, 383 (2007)]"
}

@article{Colquhoun:2014ica,
	author = "Colquhoun, B. and Dowdall, R. J. and Davies, C. T. H. and Hornbostel, K. and Lepage, G. P.",
	title = "{$\Upsilon$ and $\Upsilon^{\prime}$ Leptonic Widths, $a_{\mu}^b$ and $m_b$ from full lattice QCD}",
	eprint = "1408.5768",
	archivePrefix = "arXiv",
	primaryClass = "hep-lat",
	doi = "10.1103/PhysRevD.91.074514",
	journal = "Phys. Rev. D",
	volume = "91",
	number = "7",
	pages = "074514",
	year = "2015"
}

@article{Chakraborty:2014aca,
	author = "Chakraborty, Bipasha and Davies, C. T. H. and Galloway, B. and Knecht, P. and Koponen, J. and Donald, G. C. and Dowdall, R. J. and Lepage, G. P. and McNeile, C.",
	title = "{High-precision quark masses and QCD coupling from $n_f=4$ lattice QCD}",
	eprint = "1408.4169",
	archivePrefix = "arXiv",
	primaryClass = "hep-lat",
	doi = "10.1103/PhysRevD.91.054508",
	journal = "Phys. Rev. D",
	volume = "91",
	number = "5",
	pages = "054508",
	year = "2015"
}

@article{Heitger:2021apz,
	author = "Heitger, Jochen and Joswig, Fabian and Kuberski, Simon",
	collaboration = "ALPHA",
	title = "{Determination of the charm quark mass in lattice QCD with $2+1$ flavours on fine lattices}",
	eprint = "2101.02694",
	archivePrefix = "arXiv",
	primaryClass = "hep-lat",
	reportNumber = "MS-TP-21-01",
	doi = "10.1007/JHEP05(2021)288",
	journal = "JHEP",
	volume = "05",
	pages = "288",
	year = "2021"
}

@article{Bussone:2023kag,
	author = "Bussone, Andrea and Conigli, Alessandro and Frison, Julien and Herdo\'\i{}za, Gregorio and Pena, Carlos and Preti, David and S\'aez, Alejandro and Ugarrio, Javier",
	collaboration = "Alpha",
	title = "{Hadronic physics from a Wilson fermion mixed-action approach: charm quark mass and $D_{(s)}$ meson decay constants}",
	eprint = "2309.14154",
	archivePrefix = "arXiv",
	primaryClass = "hep-lat",
	reportNumber = "IFT-UAM/CSIC-23-114",
	doi = "10.1140/epjc/s10052-024-12816-4",
	journal = "Eur. Phys. J. C",
	volume = "84",
	number = "5",
	pages = "506",
	year = "2024"
}

@article{Erb:2025nxk,
	author = "Erb, Dominik and G{\'e}rardin, Antoine and Meyer, Harvey B. and Parrino, Julian and Biloshytskyi, Volodymyr and Pascalutsa, Vladimir",
	title = "{Isospin-violating vacuum polarization in the muon (g {\ensuremath{-}} 2) with SU(3) flavour symmetry from lattice QCD}",
	eprint = "2505.24344",
	archivePrefix = "arXiv",
	primaryClass = "hep-lat",
	doi = "10.1007/JHEP10(2025)157",
	journal = "JHEP",
	volume = "10",
	pages = "157",
	year = "2025"
}

@article{Parrino:2025afq,
	author = "Parrino, Julian and Biloshytskyi, Volodymyr and Chao, En-Hung and Meyer, Harvey B. and Pascalutsa, Vladimir",
	title = "{Computing the UV-finite electromagnetic corrections to the hadronic vacuum polarization in the muon (g {\ensuremath{-}} 2) from lattice QCD}",
	eprint = "2501.03192",
	archivePrefix = "arXiv",
	primaryClass = "hep-lat",
	reportNumber = "MITP-24-088",
	doi = "10.1007/JHEP07(2025)201",
	journal = "JHEP",
	volume = "07",
	pages = "201",
	year = "2025"
}

@article{Biloshytskyi:2022ets,
	author = "Biloshytskyi, Volodymyr and Chao, En-Hung and G{\'e}rardin, Antoine and Green, Jeremy R. and Hagelstein, Franziska and Meyer, Harvey B. and Parrino, Julian and Pascalutsa, Vladimir",
	title = "{Forward light-by-light scattering and electromagnetic correction to hadronic vacuum polarization}",
	eprint = "2209.02149",
	archivePrefix = "arXiv",
	primaryClass = "hep-lat",
	doi = "10.1007/JHEP03(2023)194",
	journal = "JHEP",
	volume = "03",
	pages = "194",
	year = "2023"
}

@article{Kataev:1992dg,
	author = "Kataev, A. L.",
	title = "{Higher order O($\alpha^2$) and O($\alpha alpha_s$) corrections to $sigma_{\mathrm{tot}}(e^+e^- \rightarrow$ hadrons) and Z boson decay rate}",
	reportNumber = "CERN-TH-6465-92",
	doi = "10.1016/0370-2693(92)91901-K",
	journal = "Phys. Lett. B",
	volume = "287",
	pages = "209--212",
	year = "1992"
}

@article{Chetyrkin:1997qi,
	author = "Chetyrkin, K. G. and Harlander, R. and K{\"u}hn, Johann H. and Steinhauser, M.",
	title = "{Mass corrections to the vector current correlator}",
	eprint = "hep-ph/9704222",
	archivePrefix = "arXiv",
	reportNumber = "MPI-PHT-97-012, TTP-97-11",
	doi = "10.1016/S0550-3213(97)00383-0",
	journal = "Nucl. Phys. B",
	volume = "503",
	pages = "339--353",
	year = "1997"
}

@inproceedings{CEPCPhysicsStudyGroup:2022uwl,
    author = "Cheng, Huajie and others",
    collaboration = "CEPC Physics Study Group",
    title = "{The Physics potential of the CEPC. Prepared for the US Snowmass Community Planning Exercise (Snowmass 2021)}",
    booktitle = "{Snowmass 2021}",
    eprint = "2205.08553",
    archivePrefix = "arXiv",
    primaryClass = "hep-ph",
    month = "5",
    year = "2022"
}

@article{FCC:2018evy,
    author = "Abada, A. and others",
    collaboration = "FCC",
    title = "{FCC-ee: The Lepton Collider}: {Future Circular Collider Conceptual Design Report Volume 2}",
    reportNumber = "CERN-ACC-2018-0057",
    doi = "10.1140/epjst/e2019-900045-4",
    journal = "Eur. Phys. J. ST",
    volume = "228",
    number = "2",
    pages = "261--623",
    year = "2019"
}

@article{Riembau:2025ppc,
    author = "Riembau, Marc",
    title = "{Precise Extraction of {\ensuremath{\alpha_{em}(m_Z^2)}} at the Tera-Z Stage of a Future Circular Collider}",
    eprint = "2501.05508",
    archivePrefix = "arXiv",
    primaryClass = "hep-ph",
    reportNumber = "CERN-TH-2025-005",
    doi = "10.1103/PhysRevLett.134.221802",
    journal = "Phys. Rev. Lett.",
    volume = "134",
    number = "22",
    pages = "221802",
    year = "2025"
}

@article{Erler:2024lds,
    author = "Erler, Jens and Ferro-Hernandez, Rodolfo and Kuberski, Simon",
    title = "{Theory-Driven Evolution of the Weak Mixing Angle}",
    eprint = "2406.16691",
    archivePrefix = "arXiv",
    primaryClass = "hep-ph",
    reportNumber = "MITP-24-051, CERN-TH-2024-071",
    doi = "10.1103/PhysRevLett.133.171801",
    journal = "Phys. Rev. Lett.",
    volume = "133",
    number = "17",
    pages = "171801",
    year = "2024"
}

@article{Davier:2023cyp,
    author = "Davier, Michel and Fodor, Zoltan and G{\'e}rardin, Antoine and Lellouch, Laurent and Malaescu, Bogdan and Stokes, Finn M. and Szabo, Kalman K. and Toth, Balint C. and Varnhorst, Lukas and Zhang, Zhiqing",
    title = "{Hadronic vacuum polarization: Comparing lattice QCD and data-driven results in systematically improvable ways}",
    eprint = "2308.04221",
    archivePrefix = "arXiv",
    primaryClass = "hep-ph",
    doi = "10.1103/PhysRevD.109.076019",
    journal = "Phys. Rev. D",
    volume = "109",
    number = "7",
    pages = "076019",
    year = "2024"
}

\end{document}